\documentstyle{BSAXwork}

\input epsf

\def \AAP #1 #2 {{\em Astron. Astrophys.\/} {\bf #1}, #2}
\def \AAPs #1  {{\em Astron. Astrophys.\/} {\bf #1}}
\def \AAL #1 #2 {{\em Astron. Astrophys. Lett.\/} {\bf #1}, L#2}
\def \AAR #1 #2 {{\em Astron. Astrophys. Rev.\/} {\bf #1}, #2}
\def \AAS #1 #2 {{\em Astron. Astrophys. Suppl. Ser.\/} {\bf #1}, #2}
\def \AJ #1 #2 {{\em Astron. J.\/} {\bf #1}, #2}
\def \ANNREV #1 #2 {{\em Ann. Rev. Astron. Astrophys.\/} {\bf #1}, #2}
\def \APJ #1 #2 {{\em Astrophys. J.\/} {\bf #1}, #2}
\def \APJs #1 {{\em Astrophys. J.\/} {\bf #1}}
\def \APJL #1 #2 {{\em Astrophys. J. Lett.\/} {\bf #1}, L#2}
\def \APJS #1 #2 {{\em Astrophys. J. Suppl.\/} {\bf #1}, #2}
\def \APSS #1 #2 {{\em Astrophys. Space Sci.\/} {\bf #1}, #2}
\def \ASR #1 #2 {{\em Adv. Space Res.\/} {\bf #1}, #2}
\def \BAIC #1 #2 {{\em Bull. Astron. Inst. Czechosl.\/} {\bf #1}, #2}
\def \JSQRT #1 #2 {{\em J. Quant. Spectrosc. Radiat. Transfer\/} {\bf #1}, #2}
\def \MN #1 #2 {{\em Mon. Not. R. Astr. Soc.\/} {\bf #1}, #2}
\def \MNs #1 {{\em Mon. Not. R. Astr. Soc.\/} {\bf #1}}
\def \MEM #1 #2 {{\em Mem. R. Astr. Soc.\/} {\bf #1}, #2}
\def \PLR #1 #2 {{\em Phys. Lett. Rev.\/} {\bf #1}, #2}
\def \PASJ #1 #2 {{\em Publ. Astron. Soc. Japan\/} {\bf #1}, #2}
\def \PASP #1 #2 {{\em Publ. Astr. Soc. Pacific\/} {\bf #1}, #2}
\def \NAT #1 #2 {{\em Nature\/} {\bf #1}, #2}
\def \SAIT #1 #2 {{\em Mem.\ Soc.\ Astron.\ It.\/} {\bf #1}, #2}
\def \MESS #1 #2 {{\em The Messenger\/} {\bf #1}, #2}
\def \ASTRNACH #1 #2 {{\em Astron. Nach.\/} {\bf #1}, #2}
\def \ergs {erg~cm$^{-2}~$s$^{-1}~$}
\def \sax {{\it Beppo}SAX$~$}
\def\nupeak{$\nu_{peak}$~}

\def\aut{Giommi et al.}
\epsfverbosetrue
\begin{opening}

\title{A Catalog of 157 X-ray Spectra and 84 Spectral Energy Distributions of Blazars 
Observed with \sax} 
\author{P.Giommi$^1$, M. Capalbi$^1$, M. Fiocchi$^1$,  E. Memola$^1$, 
M. Perri$^{1,2}$, \\ S. Piranomonte$^{1,3}$, S. Rebecchi$^1$ and E. Massaro$^{2,4}$}
\institute{$^1$ASI Science Data Center, c/o ESA-ESRIN, Frascati, Italy\\
$^2$Dipartimento di Fisica, Universit\'a la Sapienza, P.le A. Moro 2, Roma, Italy\\
$^3$Dipartimento di Astronomia, Universit\'a di Padova, Vicolo dell'Osservatorio 2, Padova, Italy\\
$^4$Istituto di Astrofisica Spaziale, CNR, Roma, Italy}
\date{} % DO NOT INSERT ANY DATE HERE !!!
\end{opening}

\begin{document}

\oddpagefooter{}{}{} % LEAVE AS IT IS !
\evenpagefooter{}{}{} % LEAVE AS IT IS !
\medskip  % LEAVE AS IT IS !

\begin{abstract} % LEAVE THIS COMMAND AS IT IS AND WRITE THE ABSTRACT IN THE FOLLOWING!
As a special contribution to the proceedings of the \sax workshop dedicated to 
blazar astrophysics we present a catalog of 157 X-ray spectra and the 
broad-band Spectral Energy Distribution (SED) of 84 blazars observed by  
\sax during its first five years of operations. 
The SEDs have been built by combining \sax LECS, MECS and PDS data with 
(mostly) non-simultaneous multi-frequency photometric data, obtained from 
NED and from other large databases, including the GSC2 and the 2MASS surveys. 
All \sax data have been taken from the public archive and have been analysed 
in a uniform way. For each source we present a $\nu f(\nu)~vs~\nu$ plot, 
and for every \sax observation we give the best fit parameters of the 
spectral model that best describes the data. 
The energy where the maximum of the synchrotron power is emitted spans at least six orders 
of magnitudes ranging from $\approx 0.1~$eV to over $100~$keV. 
A wide variety of X-ray spectral slopes have been seen depending on whether the 
synchrotron or inverse Compton component, or both, are present in the X-ray band. 
The wide energy bandpass of \sax allowed us to detect, and  
measure with good accuracy, continuous spectral curvature in many objects 
whose synchrotron radiation extends to the X-ray band. This convex curvature, 
which is described by a logarithmic parabola law better than other models, 
may be the spectral signature of a particle acceleration process that becomes less and less 
efficient as the particles energy increases. 
Finally some brief considerations about other statistical properties of the sample 
are presented.
\end{abstract}

\medskip

\section{Introduction}

Blazars emission is known to be dominated by strong and highly variable 
non-thermal radiation across the entire electromagnetic spectrum. 
Multi-frequency ground based observations, combined with data from 
high energy astronomy satellites, have often been used to derive the broad-band Spectral 
Energy Distribution (SED) of blazars, that is the source intensity as a function 
of energy, usually represented in the $\nu f(\nu)~vs~\nu$ or $\nu L(\nu)~vs~\nu$ space.
These measurements are consistent with the widely accepted scenario where blazar emission is 
due to synchrotron radiation whose power 
increases with energy up to a peak value above which it drops sharply.
At higher energies the spectrum is dominated by inverse Compton emission which 
also smoothly raises until it reaches a second luminosity peak. 
The often extreme observational characteristics of blazars are thought to be the result 
of the emission from a relativistic jet seen at a very small angle with respect to the line of 
sight (e.g. Urry \& Padovani 1995), an interpretation first proposed by Blandford \& Rees (1978).  
According to this scenario the position and the relative power of the synchrotron 
and inverse Compton peaks directly depend on important physical parameters such 
as the intensity of the magnetic field, the maximum energy at which electrons can be 
accelerated, and the relativistic motion and orientation of the emitting plasma. 
The synchrotron peak is located at energies ranging from less than $\approx 0.1~$eV 
(or $ \nu \approx 10^{13}~$Hz) to well over $ 10~$keV (or $\nu \approx 
10^{18}~$Hz) or even 100 keV in flaring states, demonstrating the existence 
of a wide variety of physical and geometric conditions in blazars.
For these reasons the Spectral Energy Distribution of blazars has been and still 
is the subject of intense research activity. 
Figure 1 shows the expected emission from Synchrotron Self Compton models (SSC) 
tracing a hypothetical sequence of blazar SEDs that ranges from LBL sources 
where the synchrotron peak frequency (\nupeak ) occurs at low energies to 
HBL objects where \nupeak reaches the X-ray band, and up to the extremely large 
\nupeak energies of the, possibly existing but still unseen, Ultra High energy 
peaked BL Lacs (UHBLs).
As shown in Figure 1, within the broad-band energy spectrum of blazars the X-ray region is 
particularly important since at these energies a variety of different spectral 
components can be (and have been) seen. 
These include the flat and rising Compton component, the transition between the two regimes, and 
the high energy end of the synchrotron spectrum which is produced by very, sometimes extremely, 
energetic electrons.  
These crucial observations, in combination with other multi-frequency data allow the determination 
of the overall spectral shape and therefore the estimation of important physical parameters. 

With its very wide X-ray band pass, good sensitivity and spectral 
capabilities {\it Beppo\-}SAX has provided a very important opportunity to study blazars 
astrophysics, especially when simultaneous multi-frequency observations could be
arranged.

As a special contribution to the proceedings of the \sax workshop dedicated 
to blazar Astrophysics we present the catalog of X-ray spectral 
fits and broad-band Spectral Energy Distribution of all the blazars 
observed with \sax whose data are currently public.

\setcounter{figure}{0}
\begin{figure}[!ht]
\vspace*{-3.0cm}
\centering
\epsfxsize=10.0cm\epsfbox{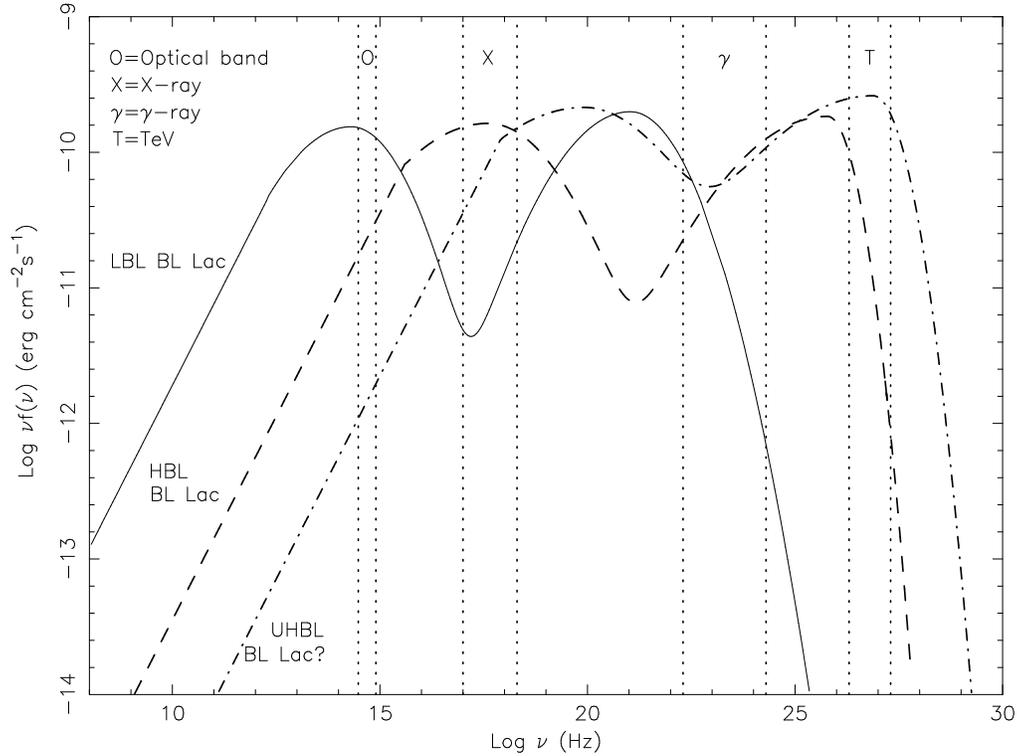}
\caption[h]{The Spectral Energy Distribution of BL Lacs is shown as a sequence 
of Synchrotron Self Compton spectra peaking at different energies. Objects
whose synchrotron component peaks at low energy are called LBLs, the maximum 
of their synchrotron power output occurs in the IR-Optical band, while 
High energy peaked BL Lacs (HBLs) peak in the UV or X-ray band. It is not known
how far the sequence of \nupeak goes on. If it continues to very high energies
it could be that in some very extreme objects (UHBLs) the synchrotron component 
might even reach the gamma-ray band. Note that for the same peak luminosity, the 
radio power decreases by orders of magnitudes in going form LBLs to HBLs and possibly
to UHBLs.}
\label{fig1}
\end{figure}

\section{\sax Blazars: the sample}

A complete description of the \sax X-ray astronomy satellite is given in 
Boella et al. (1997a). 
The scientific instrumentation is composed of six co-aligned Narrow Field 
Instruments (NFI) and two  Wide Field Cameras (WFC, Jager et al.  1997) 
pointing in opposite directions and perpendicularly to the pointing direction 
of the NFI. 
The NFI include the Low Energy Concentrator Spectrometer, 
(LECS, Parmar et. al 1997), three identical Medium Energy Concentrator 
Spectrometers, (MECS, Boella et al. 1997b), one of which (MECS1) failed 
in May 1997, a High Pressure Gas Scintillation Proportional Counter 
(HPGSPC, Manzo et al. 1997), and a Phoswich Detector System (PDS, Frontera 
et al. 1997). Collectively the NFI cover  a very wide spectral range 
(0.1--200 keV) providing an unprecedented opportunity to study the broad 
band SED of blazars.
During its lifetime \sax dedicated approximately 15\% of its scientific program 
to the study of blazars performing about 200 observations of nearly 
100 blazars.

We have carried out a systematic analysis of all blazar observations made 
with the LECS, MECS and PDS (for sufficiently bright and unconfused sources) 
instruments  and for which the corresponding data were available in the \sax public archive 
(Giommi \& Fiore 1997) at the date of March 2002. As \sax data are subject to the usual 
one year proprietary period, this approximately corresponds to observations 
performed in the period July 1996 - March 2001. 
The blazar class is usually divided in BL Lacs and Flat Spectrum Radio 
Quasars (FSRQs), these last also including GigaHertz Peaked Spectrum (GPS) QSOs. 

Our list  includes a total of 84 blazars, 58 of which are BL Lacs, 22 are FSRQs, and 4 
(1JY 2149$-$3006, PKS 2126$-$15, OX 57 and PKS 2243$-$123) are GPS QSOs.
These sources were observed as part of several \sax projects  
and were discovered both in the radio and the X-ray band, sometimes in surveys 
with widely different flux limits. In addition the \sax selection process clearly favoured 
objects that were known to be X-ray bright and possibly detectable by the 
high energy instruments (PDS).
The sample is therefore highly heterogeneous and cannot be used for detailed 
statistical analyses requiring strict completeness criteria. The size of the sample 
and its heterogeneity nevertheless provide an unprecedented opportunity to explore 
a very wide portion of the blazar parameter space. 

The lists of BL Lacs and FSRQs considered in this paper are presented in Tables I 
and II, respectively.
Column 1 gives the name of the object, columns 2 and 3
give the Right Ascension and  Declination for the equinox 2000.0,
column 4 gives the redshift when available, column 5 gives the number
of \sax observations available in the public archive, and column 6 gives 
the references to previous \sax publications. 

In general, sources brighter than $\approx 5\times 10^{-12}~$ \ergs can be detected 
in all three \sax NFI 
instruments considered in this paper (LECS, MECS, and PDS), fainter sources 
can only be detected in the imaging instruments (LECS and MECS) up to 10 keV. 
A total of 157 spectra have been analysed.

\section{Data analysis}

All LECS, MECS and PDS cleaned and calibrated spectral data (pha files) have 
been taken from the \sax public archive (Giommi \& Fiore 1997). 
The standard {\it Beppo\-}SAX procedure that generated these archival spectra
used photons located within a circular region centered on the source
with a typical radius of 4 arcminutes for the MECS data and of
6 arcminutes for the LECS data. For the case of weak sources, however, this procedure 
used a radius of 4 and sometimes even 2 arcminutes to extract the LECS spectra, a value 
that turned out to be too small to ensure a good signal-to-noise ratio.
All LECS spectral files that were originally produced using a 2 arcminutes radius have been 
re-generated by us using a more adequate 4 arcminutes radius.
In a few cases non-standard extraction radii had to be used to avoid 
contamination from a nearby source. 
Background subtraction was carried out using the standard background files 
available from the \sax archive. Since these background data were taken observing ``blank"
areas of the sky (that is fields not including a detectable target) with low 
Galactic absorption ($N_H$), the low energy counts of X-ray sources located in regions 
of high $N_H$ may be underestimated. We have therefore ignored data below 0.3 keV. 
Finally, spectral channels were binned in such a way that each channel includes 
at least 20 counts.

All data were analysed in a homogeneous way following the recommendations 
given in the BeppoSAX handbook for NFI spectral analysis (Fiore et al. 1999).

We have combined LECS data covering the range 0.3--2.0 keV, MECS data 
between 2--10 keV, and, for bright and unconfused sources, PDS data above 
12 keV and up to the maximum energy at which these sources could be detected.

\subsection{Spectral models}

Spectral fits have been done with the XSPEC 11.0 package using one of the 
following spectral models \\

\begin{enumerate}
\item single power law 

\medskip\hspace{3cm} $ dN/dE = ke^{-\sigma(E)N_H}E^{-\gamma} $ \medskip
\item sum of two power laws 

\medskip \hspace{3cm} $ dN/dE =e^{-\sigma(E)N_H}(kE^{-\gamma}+k_1E^{-\gamma_1}) $ 
\medskip
\item broken power law 

\medskip\hspace{3cm} $ dN/dE = \left\{\begin{array}{ll}
ke^{-\sigma(E)N_H}E^{-\gamma} \quad  \quad \quad \,\,\,\,\,\,\, \mbox{if} \quad E < E_{break}\\[0.2cm]
kE_{break}^{\gamma_1-\gamma}e^{-\sigma(E)N_H}E^{-\gamma_1} \quad  \mbox{if} \quad  E \ge E_{break}
\end{array}\right.$
\medskip
\item logarithmic parabola 

\medskip\hspace{3cm}$ dN/dE = ke^{-\sigma(E)N_H} E^{(-\gamma +\beta Log(E))}$ 
\medskip
\end{enumerate}

\noindent 
in this last case  $\gamma$ is the photon spectral index at 1 keV 
and $\beta$ is the coefficient of the quadratic term in the logarithmic parabola 
($Log ( dN/dE ) = log(k) -\gamma Log(E) +\beta (Log(E))^2$),
which is proportional to the slope change in an energy decade. 
This spectral law was first used by Landau et al. (1986) who were able 
to obtain very good fits to the wide band spectra (mm to optical-UV) of a number of BL Lacs. 
More recently, Massaro et al. (2000) and Cusumano et al. (2001) successfully applied
this model to fit the X-ray spectra of young pulsars and developed
the XSPEC routine used in the present work.
With respect to other curved spectra, the log-parabola has the advantage of
describing well the spectral curvature that is often seen in the broad band spectrum
of blazars with only three parameters. Furthermore, as discussed by Massaro (2002),
it is possible to show that a log-parabolic spectrum is naturally obtained
by a statistical acceleration mechanism where the probability for a particle to
remain inside the acceleration region is assumed to decrease as the particles
energy grows.

The amount of $N_H$ in the low energy absorption term ($e^{-\sigma(E)N_H}$) was 
always fixed to the Galactic value as estimated from the 21 cm measurements of 
Dickey \& Lockman (1990). 

The LECS-MECS and MECS-PDS relative normalizations were treated as free 
parameters but constrained to be within 0.6 and 1.1 and 0.8-1.0, respectively. 
Typical values are 0.7 for LECS/MECS and 0.9 for MECS/PDS, as expected 
from the \sax NFI intercalibration (see Fiore et al. 1999 for details).

\subsection{Results of the spectral fitting}

The results of our spectral fits are reported in Tables III-IV, where, for each 
source, we give the best fit parameters for the spectral model that gives the 
lowest $\chi^2$.
For sources that were observed more than once only one model was used 
choosing the one that gives the lowest $\chi^2$ in most of the observations. 

Column 1 gives the source name built with the J2000.0 coordinates (for 
reasons of space we use here a IAU like naming convention rather than 
the full name), column 2
gives the observation date, columns 3, 4, 5, 6 and 7 give the model used and the 
best fit parameters (with one sigma errors), column 8 gives the reduced
$\chi^2$  and the number of degrees of freedom, column 9 gives the 2--10
keV flux in units of \ergs.

To keep the amount of work manageable we have neglected the effects 
of rapid variability always integrating the spectral data over the full observation. 
The high $\chi^2$ values that resulted from the analysis of some rapidly variable 
sources are due to significant spectral changes that usually accompany large 
intesity variations (e.g. Giommi et al. 1990, Tanihata et al. 2001).
The best estimates of the $\beta$ parameter in model 4 are mostly negative 
and between $-$0.5 and $-$0.15; the few positive values attest the presence of an 
upward spectral curvature due to presence of both the synchrotron and the Compton 
component in the \sax band.  

As all of the data used in this work are publicly available, a large fraction of the \sax 
spectra has already been analysed and published by the original 
investigators. We have therefore compared the results of our 
fits with published results and we have found that, when the spectral model used
is the same, our best fit parameters agree well with the published values.

\subsection{Spectral Energy Distributions} 

In order to build the broad band (radio to hard X-ray) Spectral Energy
Distributions we have first converted all \sax spectra into flux using 
the best model and best fit parameters listed in Tables III-IV. We have then 
de-reddened the soft X-ray fluxes using the cross sections of 
Morrison \& McCammon (1983) setting the amount of absorbing material ($N_H$)
equal to the Galactic value, consistently with the fitting procedure. 
We have then combined these X-ray fluxes with (mostly non-simultaneous) 
multi-frequency photometric measurements available from NED or from 
specific BeppoSAX publications. We have also 
added infrared and optical fluxes converting magnitudes from the blazars
counterparts in the 2MASS and GSC2 catalogs (when available) using the 
prescriptions given in Cardelli et al. (1989).  
In several nearby objects (e.g. Mkn~421, Mkn~501, 3C~66A, PKS~0548$-$322) the 
optical emission is dominated by the host galaxy and therefore our fluxes 
should be considered as an upper limit to the non-thermal nuclear emission. 

All SEDs are plotted in Figures 2a through 2o for BL Lacs and in Figures 3a through 3g  
for FSRQs. Fluxes have not been corrected for redshift and therefore are in the 
observer's frame.

All the SEDs, including the data files, and spectral fitting details are also 
available on-line from the ASI Science Data Center web server at the following 
address \par 
\begin{center}
http://www.asdc.asi.it/blazars/
\end{center}

\setcounter{table}{0}
\begin{table}[*t]
\caption{BL Lacs available in the \sax public archive} 

\begin{center}
\begin{tabular}{lrrrcc}
\hline
\multicolumn{1}{c}{Blazar name} & \multicolumn{1}{r}{Ra(J2000.0)} & \multicolumn{1}{r}{Dec(J2000.0)} & \multicolumn{1}{c}{Redshift} & 
 \multicolumn{1}{c}{No. of} &Ref.\\ 
& & & & observ. &\\
\hline
%1ES 0033+595    &00 35 52.7&59 50 03.8&0.06& 1 & a\\
1ES 0033+595    &00 35 52.7&59 50 03.8&$-$& 1 & a\\
PKS 0048$-$097   &00 50 41.3&$-$09 29 04.9&$-$ & 2 & b \\
1ES 0120+340    &01 23 08.6&34 20 48.8&0.272& 2 &a\\
RGB J0136+39     &01 36 32.7&39 06 00.0&$-$& 1 &\\
1ES 0145+138    &01 48 29.7&14 02 17.8&0.125& 1 &c\\
MS 0158.5+0019  &02 01 06.1&00 34 00.1&0.299& 1&d\\
3C 66A         &02 22 39.6&43 02 08.1&0.444& 2 &\\
%1ES 0229+200    &02 32 48.5&20 17 17.1&0.139& 1 &\\ Private
AO 0235+164    &02 38 38.8&16 36 59.0&0.94& 1 &e\\
MS 0317.0+1834  &03 19 51.9&18 45 33.8&0.19& 1 &d\\
1H 0323+022    &03 26 13.8&02 25 14.8&0.147& 1 &c\\
1ES 0347$-$121    &03 49 23.2&$-$11 59 26.8&0.188& 1 &d\\
1H 0414+009    &04 16 52.3&01 05 24.0&0.287& 1 &d\\
%EXO 0423.4$-$0840 &04 25 50.7&$-$08 33 42.8&0.039& 1 &\\
1ES 0502+675    &05 07 56.2&67 37 23.8&$0.314$& 1 &d\\
1H 0506$-$039    &05 09 38.0&$-$04 00 46.0&0.304& 1 &c\\
PKS 0521$-$365   &05 22 57.7&$-$36 27 02.8&0.055& 1 &m\\
PKS 0537$-$441   &05 38 50.3&$-$44 05 08.8&0.896& 1 &n\\
PKS 0548$-$322   &05 50 41.9&$-$32 16 10.9&0.069& 3 &a\\  
S5 0716+714    &07 21 53.4&71 20 35.8&$-$& 3 &g\\
MS 0737.9+7441  &07 44 05.4&74 33 57.9&0.315& 1 &d\\
OJ 287         &08 54 48.9&20 06 30.9&0.306& 1 & b,f\\
B2 0912+29     &09 15 52.2&29 33 23.0&$-$& 1 & b\\
1ES 0927+500    &09 30 37.6&49 50 26.1&0.188& 1 &c\\
1ES 1028+511  &10 31 18.5&50 53 34.0&0.361& 1 &c\\
RGB J1037+57   &10 37 44.3&57 11 54.9&$-$& 1 &\\
RGB J1058+56   &10 58 37.7&56 28 10.9&$-$& 2 &\\
1H 1100$-$230    &11 03 37.7&$-$23 29 30.8&0.186& 2 &d\\
Mkn 421        &11 04 27.2&38 12 32.0&0.031& 15 &h,i,j\\
1RXS J111706.3+20141&11 17 06.3&20 14 09.9&0.137& 1 &\\
EXO 1118.0+4228 &11 20 48.0&42 12 11.8&0.124& 1 &c\\
Mkn 180        &11 36 26.4&70 09 28.0&0.046& 1 &d\\
PKS 1144$-$379   &11 47 01.3&$-$38 12 11.1&1.048& 1 &b\\
1RXS J121158.1+22423&12 11 58.0&22 42 36.0&0.77& 3 &\\
EXO 1215.3+3022 &12 17 52.1&30 07 00.1&0.237& 2 &\\
1H 1219+301    &12 21 22.0&30 10 36.8&0.13& 1 &\\
ON 231         &12 21 31.7&28 13 58.0&0.102& 2 &k\\
1RXS J123511.1$-$14033&12 35 11.1&$-$14 03 32.0&0.406& 3 &\\
1ES 1255+244    &12 57 31.8&24 12 39.9&0.141& 1 &c\\
MS 1312.1$-$4221  &13 15 03.4&$-$42 36 50.0&0.105& 1 &\\
EXO 1415.6+2557 &14 17 56.2&25 43 55.9&0.237& 3 &\\
PG 1418+546    &14 19 46.6&54 23 15.0&0.152& 3 &e\\
1H 1430+423    &14 28 32.4&42 40 23.8&0.13& 1 &\\
MS 1458.8+2249 &15 01 01.8&22 38 03.8&0.235& 1 &f\\
1H 1515+660    &15 17 47.4&65 25 23.1&0.702& 1 &d\\
PKS 1519$-$273   &15 22 37.6&$-$27 30 11.1&$-$& 1 &\\
1ES 1533+535   &15 35 00.7&53 20 36.9&0.89& 1 &c\\
1ES 1544+820   &15 40 15.6&81 55 05.8&$-$& 1 &c\\
1ES 1553+113    &15 55 43.1&11 11 21.1&0.36& 1 &c\\
Mkn 501        &16 53 52.2&39 45 37.0&0.033& 11 &l\\
% 1H 1720+117   &17 25 04.2&11 52 14.8&0.1& 1 &\\ Private
\hline
\end{tabular}
\end{center}
\noindent $^{a}$ Costamante et al. 2001, $^{b}$ Padovani et al. 2001, $^{c}$ Beckmann et al. 2002,
$^{d}$ Wolter et al. 1998,  $^{e}$ Padovani et al. 2002a, $^{f}$ Massaro et al. 2002, 
$^{g}$ Giommi et al. 1999,  $^{h}$ Fossati et al. 2000a, \\ 
\noindent $^{i}$ Fossati et al. 2000b, $^{j}$Malizia et al. 2000, $^{k}$
Tagliaferri et al. 2000, $^{l}$Pian et al. 1998, \\ 
\noindent $^{m}$Tavecchio et al. 2002, $^{n}$Pian et al. 2002
\end{table}

\clearpage
\setcounter{table}{0}
\begin{table}[*t]
\caption{BL Lacs available in the \sax public archive - Continued } 

\begin{center}
\begin{tabular}{lrrrcc}
\hline
\multicolumn{1}{c}{Blazar name} & \multicolumn{1}{r}{Ra(J2000.0)} & \multicolumn{1}{r}{Dec(J2000.0)} & \multicolumn{1}{c}{Redshift} & 
 \multicolumn{1}{c}{No. of} &Ref.\\ 
& & & & observ. &\\
\hline
1ES 1741+196    &17 43 57.5&19 35 09.9&0.083& 1 &a\\
S5 1803+784    &18 00 45.6&78 28 04.0&0.684& 1 &b\\
3C 371         &18 06 50.7&69 49 27.8&0.051& 1 &b\\
S4 1823+568    &18 24 07.1&56 51 01.0&0.664& 1 &c\\
1ES 1959+650   &19 59 59.9&65 08 54.9&0.048& 3 &d\\
PKS 2005$-$489   &20 09 25.3&$-$48 49 54.1&0.071& 2 & c\\
PKS 2155$-$304   &21 58 51.9&$-$30 13 31.0&0.116& 3 & e,f,g\\
BL Lacertae    &22 02 43.2&42 16 40.0&0.069& 5 &c,h\\
1ES 2344+514   &23 47 04.7&51 42 17.9&0.044& 7 &i\\
1H 2354$-$315    &23 59 07.8&$-$30 37 39.0&0.165& 1 &\\
\hline
\end{tabular}
\end{center}
\noindent $^{a}$ Siebert et al. 1999, $^{b}$ Padovani et al. 2002a, $^{c}$ Padovani et al. 2001, 
$^{d}$ Beckmann et al. 2002, \\
\noindent $^{e}$ Giommi et al. 1998,  
$^{f}$ Chiappetti et al. 1999, $^{g}$ Zhang et al. 1999, $^{h}$ Ravasio et al. 2002, \\
\noindent $^{i}$ Giommi et al. 2000  
\end{table}

\setcounter{table}{1}
\begin{table}[*t]
\caption{Flat Spectrum Radio Quasars available in the \sax public archive} 
\begin{center}
\begin{tabular}{lrrccc}
\hline
\multicolumn{1}{c}{Blazar name} & \multicolumn{1}{r}{Ra(J2000.0)} & \multicolumn{1}{r}{Dec(J2000.0)} & \multicolumn{1}{c}{Redshift} & 
\multicolumn{1}{c}{No. of}& Ref.\\ 
& & & & observ.& \\
\hline
PKS 0208$-$512         &02 10 46.1&$-$51 01 01.9&1.003& 1& a \\
NRAO 140             &03 36 30.0&32 18 29.8&1.259& 1 &\\
PKS 0523$-$33           &05 25 06.1&$-$33 43 05.0&4.401& 1 &b\\
OG 147               &05 30 56.3&13 31 54.8&2.06& 8 &c\\
WGA J0546.6$-$6415      &05 46 42.4&$-$64 15 21.9&0.323& 1 &d\\
1Jy 0836+710          &08 41 24.4&70 53 40.9&2.19& 1 &e\\
RGB J0909+039         &09 09 15.8&03 54 42.1&3.2& 1 &f\\
1Jy 1127$-$145          &11 30 07.0&$-$14 49 27.1&1.187& 1 &\\
3C 273               &12 29 06.6&02 03 07.9&0.158& 9 &g\\
3C 279          &12 56 11.1&$-$05 47 21.1&0.538& 5 &h\\
B3 1428+422          &14 30 23.5&42 04 36.1&4.715& 1 &i\\
GB 1508+5714         &15 10 02.8&57 02 47.0&4.301& 2 &f\\
PKS 1510$-$08          &15 12 50.4&$-$09 05 59.9&0.361& 1 &e\\
1ES 1627+402         &16 29 01.3&40 07 58.0&0.272& 1 &d\\
3C 345                &16 42 58.7&39 48 36.0&0.594& 1 &a\\
V 396 HER            &17 22 41.3&24 36 19.0&0.175& 1 &d\\
4C 62.29             &17 46 13.9&62 26 53.8&3.886& 1 &f\\
S5 2116+81           &21 14 01.6&82 04 46.9&0.086& 2 &d\\
PKS 2126$-$15          &21 29 12.0&$-$15 38 41.9&3.266& 1& \\
OX 57                 &21 36 38.4&00 41 53.8&1.936& 1 &\\
1Jy 2149$-$306          &21 51 55.3&$-$30 27 53.9&2.34& 1 &j\\
PKS 2223+21          &22 25 38.0&21 18 06.1&1.949& 1 &a,f\\
3C 446                &22 25 47.2&$-$04 57 01.0&1.404& 1 &\\
CTA 102              &22 32 36.4&11 43 50.8&1.037& 5 &e\\
PKS 2243$-$123         &22 46 18.1&$-$12 06 51.8&0.63& 1 &a\\
3C 454.3             &22 53 57.6&16 08 53.1&0.86& 1 &a\\
\hline
\end{tabular}
\end{center}
\noindent $^{a}$ Tavecchio et al. 2002, $^{b}$ Fabian et al. 2001a, 
$^{c}$ Ghisellini et al. 1999,  $^{d}$ Padovani et al. 2002b, \\
\noindent $^{e}$ Tavecchio et al. 2000, 
$^{f}$  Costantini et al. 1999, $^{g}$ Mineo et al. 2000, 
$^{h}$ Ballo et al. 2002, \\ $^{i}$~Fabian et al. 2001b, 
$^{j}$ Elvis et al. 2000 
\end{table}
\clearpage
\setcounter{table}{2}
\begin{table}[*t]
\scriptsize
\caption{Results of spectral fits for the sample of BL Lacertae Objects} 
\begin{center}
\begin{tabular}{lcccccrrc}
\hline
\\
Blazar name & Observ.&\multicolumn{2}{l}{model$^{(a)}$}&
\multicolumn{3}{l}{Best fit parameters} & 
\multicolumn{1}{c}{$\chi^2/d.o.f.$ }&\multicolumn{1}{c}{flux$^{(b)}$} \\ 
 & date  & & $\gamma$ &$\gamma_1$ &$\beta$ &$E_{break}$ & (d.o.f.) & \\
\hline
0035+5950 &18-12-99 &4&$1.42\pm0.08$& &$-0.44\pm0.06$&
        &0.95(85) & 5.83 \\ 
0050$-$0929 &19-12-97 &1&$1.9\pm0.2$& & & 
        &0.52(9) & 0.13 \\
0123+3420 &03-01-99 &4&$1.8\pm0.1$& &$-0.2\pm0.1$&
        &1.1(45)& 1.71 \\
          &02-02-99 &4&$1.9\pm0.2$& &$-0.4\pm0.2$&
        &0.9(45)& 1.31 \\
0136+3906 &09-01-01 &4&$2.0\pm 0.1$& &$-0.5\pm0.1$&
        &1.39(44) &1.05  \\
0148+1402 &30-12-97 &1&$2.2\pm0.4$& & &
        & 0.57(8) &0.05  \\
0201+0034 &16-08-96 &4&$2.8\pm0.3$& &$0.5\pm0.3$&
        &1.21(22) &0.3  \\
0222+4302& 31-01-99& 1  & $2.2\pm0.1$ & &  &  & 0.75(24)& 0.20 \\  
0238+1636&28-01-99 &1&$1.9\pm0.2$& & &
        & 0.9(22) &0.08  \\
0319+1845&15-01-97 &4&$1.5\pm0.3$& &$-0.5\pm0.3$&
        & 0.74(39) &0.7  \\
0326+0225 &20-01-98 &1&$2.2\pm0.1$& & &
        & 1.3(22) &0.3  \\
0349$-$1159 &10-01-97 &4&$1.8\pm0.2 $& &$-0.4\pm0.2 $&
        & 0.92(42) &0.6  \\
0416+0105 &21-09-96 &1&$2.6\pm0.06 $& & &
        & 0.75(34) &0.85  \\
0507+6737 & 06-10-96 & 1 &$ 2.31\pm0.04 $& &  &  
        & 1.34(37) &  1.94     \\
0509$-$0400   & 11-02-99 & 1 &$ 2.00\pm0.07 $& &  &  
        & 1.20(37) &  0.56     \\
0522$-$3627  & 02-10-98 & 1 &$ 1.75\pm0.04 $& &  &  
        & 0.83(54) &  0.88     \\
0538$-$4405  & 28-11-98 & 1 &$ 1.75\pm0.07 $& &  &  
        & 0.68(42) &  0.24     \\
0550$-$3216  & 20-02-99 & 4 &$ 1.53\pm0.13 $& &$-0.49\pm0.11$&     
        & 0.96(56) &  2.28 \\ 
        & 07-04-99 & 4 &$ 1.74\pm 0.15 $& &$-0.45\pm0.14$&     
        & 0.82(43) &  1.52  \\ 
0721+7120 & 14-11-96 &3&$2.3\pm0.1$&$1.8\pm0.2$&&$3.7\pm0.6$&1.11(41)&0.14\\
 & 07-11-98 &3&$2.1\pm0.2$&$1.7\pm0.2$&&$2.9\pm0.9$&1.46(36)&0.26\\
 & 30-10-00 &3&$2.6\pm0.1$&$1.7\pm0.1$&&$3.0\pm0.3$&0.99(48)&0.33\\
0744+7433 & 29-10-96 & 1 &$ 2.62\pm0.15 $& &  &  
           & 1.32(18) &  0.15     \\
0854+2006  & 24-11-97    & 1 &$ 1.62\pm0.12 $& &  &  
           & 0.99(17) &  0.21     \\ 
0915+2933 & 14-11-97    & 1 &$ 2.36\pm0.09 $& &  &  
           & 1.36(42) &  0.26     \\
0930+4950&25-11-98&4&$1.7\pm0.1$&&$-0.4\pm 0.1$&&1.08(42)&0.67 \\
1031+5053&01-05-97&1&$2.40\pm0.05$&&&&1.05(40) &1.03 \\
1037+5711&02-12-98&1&$2.5\pm0.1$&&&&0.98(22) & 0.09 \\
1058+5628&21-05-98&1&$2.7\pm0.3$&&&&0.68(14) &0.03\\
         &23-11-98&1&$2.8\pm0.1$&&&&0.91(16) &0.17\\
1103$-$2329&04-01-97&1&$2.07\pm0.03$&&&&1.10(242) &3.81 \\
         &19-06-98&1&$2.25\pm0.02$&&&&1.32(85) &2.56 \\
1104+3812 &29-04-97&4&$2.23\pm 0.03$&&$-0.46\pm 0.03$ & 
        &1.40(108)& 8.5$^{(c)}$ \\ 
 &30-04-97&4&$2.21\pm 0.03$ &&$ -0.51\pm 0.03$  &
        &0.76(102) & 8.3 \\ 
 &01-05-97&4&$2.19\pm 0.03$&&$-0.47\pm 0.03$  &
        &1.03(102) & 10.0 \\ 
 &02-05-97&4&$2.16\pm 0.04$&&$-0.51\pm 0.04$  &
        &0.91(104) & 12.3 \\ 
 &03-05-97&4 &$2.24\pm 0.06$&&$-0.52\pm 0.06$  &
        &0.98(89) & 7.1 \\ 
 &04-05-97 & 4 &$2.49\pm 0.07$&&$-0.47\pm 0.07$  &
        &0.97(84) & 4.1 \\ 
 &05-05-97&4&$2.29\pm 0.04$&&$-0.54\pm 0.04$  &
        &1.17(88) & 6.3 \\ 
 &21-04-98&4&$2.12\pm 0.01$&&$-0.47\pm 0.01$  &
        &1.10(140) & 17.8 \\ 
 &23-04-98&4&$1.99\pm 0.01$&&$-0.43\pm 0.01$  &
        &1.50(136) & 31.0$^{(c)}$ \\ 
 &22-06-98&4&$2.03\pm 0.02$&&$-0.39\pm 0.02$  &
        &1.27(123) & 23.9 \\ 
 &04-05-99&4&$2.30\pm 0.01$&&$-0.53\pm 0.01$  &
        &1.44(137) & 11.1 \\ 
 &26-04-00&4&$1.75\pm 0.01$&&$-0.26\pm 0.01$  &
        &2.72(130) & 61.9$^{(c)}$ \\ 
 &28-04-00&4&$1.72\pm 0.01$&&$-0.26\pm 0.01$  &
        &1.67(130) & 66.8$^{(c)}$ \\ 
 &30-04-00&4&$1.72\pm 0.01$&&$-0.32\pm 0.01$  &
        &1.72(130) & 59.8$^{(c)}$ \\ 
 &09-05-00&4&$1.82\pm 0.01$&&$-0.23\pm 0.01$  &
        &3.05(130) & 49.5$^{(c)}$ \\ 
1117+2014&13-12-99&4&$2.34\pm0.09$&&$-0.6\pm0.1$&&1.89(41)&0.61\\
1120+4212&01-05-97&1&$2.7\pm0.1$&&&&0.73(19) &0.29 \\
1136+7009&10-12-96&4&$2.1\pm0.2$&&$-0.4\pm0.2$&&0.85(36)&0.51 \\
1147$-$3812&10-01-97&1&$1.6\pm0.2$&&&&1.02(17)&0.09\\
1211+2242&27-12-99&4&$1.6\pm0.3$&&$-0.5\pm0.3$&&1.24(41)&0.24\\
1217+3007&23-12-98&2&$3.0\pm2.1$&$1.0\pm1.5$&&&1.09(17)&0.07\\
1221+3010&12-07-99&4&$2.21\pm0.08$&&$-0.31\pm0.07$&&0.83(88)&1.49\\
1221+2813&11-05-98&2&$2.8\pm0.1$&$0.6\pm0.3$&&&0.97(60)&0.42\\
         &11-06-98&2&$2.9\pm0.3$&$1.1\pm0.4$&&&1.19(46)&0.33\\
\hline
\end{tabular}
\end{center}
\noindent $^{(a)}$ 1: single power law, 2: double power law, 3: broken
power law, 4: logarithmic parabola model \\
 $^{(b)}$ X-ray flux in the 2--10 keV band in units of $10^{-11}$ \ergs \\
 $^{(c)}$ The source flux strongly varied during the observation 
\end{table}
\clearpage

\setcounter{table}{2}
\begin{table}[ht]
\vspace*{0.5truecm} 
\scriptsize
\caption{Results of spectral fits for the sample of BL Lacertae Objects - Continued} 
\begin{center}
\begin{tabular}{lclcccrrc}
\hline
\\
Blazar name & Observ.&\multicolumn{2}{l}{model$^{(a)}$}&
\multicolumn{3}{l}{Best fit parameters} & 
\multicolumn{1}{c}{$\chi^2/d.o.f.$ }&\multicolumn{1}{c}{flux$^{(b)}$} \\ 
 & date  && $\gamma$ &$\gamma_1$ &$\beta$ &$E_{break}$ & (d.o.f.) & \\
\hline
1235$-$1403&27-06-99&4&$1.6\pm0.7$&&$-0.8\pm0.7$&&0.63(15)&0.19\\
         &16-07-99&4&$2.8\pm0.5$&&$0.4\pm0.5$&&0.40(18)&0.16\\
1257+2412&20-06-98&1&$2.14\pm0.08$&&&&0.94(36)&1.12\\
1315$-$4236&21-02-97&1&$2.2\pm0.1$&&&&0.94(20)&0.44\\
1417+2543&13-07-00&4&$1.7\pm0.1$&&$-0.5\pm0.1$&&1.03(49)&1.27\\
         &23-07-00&4&$1.5\pm0.3$&&$-0.8\pm0.3$&&0.96(42)&0.85\\
         &27-07-00&4&$1.8\pm0.1$&&$-0.3\pm0.1$&&1.11(49)&0.93\\
1419+5423&12-02-99&1&$1.7\pm0.2$&&&&1.28(15)&0.06\\
         &03-03-00&1&$1.3\pm0.2$&&&&0.94(19)&0.11\\
         &26-03-00&1&$1.9\pm0.2$&&&&0.93(15)&0.08\\
1428+4240&08-02-99&1& $1.90\pm0.02$         &
         &              &           & 1.04(89) & 2.0  \\   
1501+2238&19-02-01&1& $2.7 \pm0.1 $         &
        &              &           & 1.01(22) & 0.07 \\ 
1517+6525&05-03-97&1& $2.51 \pm0.06 $ & & & & 1.20(38) & 0.99 \\
1522$-$2730&01-02-98&1& $2.0 \pm0.2 $         &
        &              &           & 0.73(17) & 0.06 \\ 
1535+5320&13-02-99&1& $2.42\pm0.09$         &
        &              &           & 0.88(40) & 0.26 \\
1540+8155 & 13-02-99 &1&$2.6\pm0.1$& & & &0.71(23) &0.15\\
1555+1111 & 05-02-98 &4&$2.2\pm0.2$& &$-0.6\pm 0.1$ & &1.40(45) &1.30 \\
1653+3945 &07-04-97&4&$1.69\pm 0.01$&&$-0.19\pm 0.01$  &
        &1.39(122) & 21.5 \\ 
 &11-04-97&4&$1.67\pm 0.01$&&$-0.11\pm 0.01$  &
        &1.38(139) & 23.9 \\ 
 &16-04-97&4&$1.43\pm 0.01$&&$-0.14\pm 0.01$  &
        &1.06(139) & 52.4 \\ 
 &28-04-98&4&$1.64\pm 0.02$&&$-0.17\pm 0.02$  &
        &1.11(122) & 17.6 \\ 
 &29-04-98&4&$1.62\pm 0.02$&&$-0.18\pm 0.01$  &
        &1.18(122) & 20.9 \\ 
 &01-05-98&4&$1.71\pm 0.02$&&$-0.26\pm 0.02$  &
        &1.06(122) &14.7 \\ 
 &20-06-98&4&$1.81\pm 0.02$&&$-0.22\pm 0.02$  &
        &1.10(122) & 4.95 \\ 
 &29-06-98&4&$1.70\pm 0.02$&&$-0.22\pm 0.02$  &
        &1.07(122) & 10.3 \\ 
 &16-07-98&4&$1.73\pm 0.03$&&$-0.33\pm 0.02$  &
        &1.07(122) & 9.0 \\ 
 &25-07-98&4&$1.78\pm 0.02$&&$-0.27\pm 0.02$  &
        &1.06(122) & 9.3 \\ 
 &10-06-99&4&$2.16\pm 0.01$&&$-0.24\pm 0.01$  &
        &1.37(122) & 3.0 \\ 
1743+1935 & 26-09-98 &1&$2.08\pm0.04$& & & &0.97(41) &0.69 \\
1800+7828 & 28-09-98 &1&$1.51\pm0.08$& & & &0.90(40)&0.22 \\
1806+6949 &22-09-98&4& $2.5\pm0.5$&   &$0.7\pm0.4$&   & 0.62(19)& 0.19 \\  
1824+5651 & 11-10-97 &1&$1.9\pm0.2$& & & &0.85(12)&0.10\\
1959+6508 & 04-05-97 &4&$2.1\pm0.2$& &$-0.4\pm0.2$&&0.89(40)&1.35\\
2009$-$4849 &29-09-96 &4&$2.3\pm0.1$& &$-0.05\pm0.11 $& &1.23(75)&6.01\\
            & 01-11-98 &4&$2.05\pm0.02$& &$-0.12\pm0.02$&&1.00(124)&17.6\\
2158$-$3013 & 20-11-96 &4&$2.38\pm0.02$& &$-0.27\pm0.02$& &1.27(121)&5.59\\
 & 22-11-97 &4&$2.28\pm0.02$& &$-0.46\pm0.02$& &1.22(102)&8.28\\
 & 04-11-99 &4&$2.57\pm0.02$& &$-0.30\pm0.02$& &1.30(95)&2.47\\
2202+4216 & 08-11-97 &1&$1.84\pm0.06$& & & &1.02(46)&1.11\\
          & 05-06-99 &2&$2.4\pm0.2$&$1.2\pm0.4$ & & &1.07(59)&0.66\\
          & 05-12-99 &1&$1.56\pm0.03$& & & &0.97(53)&1.26\\
          & 26-07-00 &1&$1.84\pm0.06$& & & &0.65(41)&0.59\\
          & 31-10-00 &1&$2.63\pm0.03$& & & &0.98(96)&2.03\\
2347+5142 &03-12-96&4& $1.7\pm0.1$   & &$-0.3\pm0.1$  &   & 0.77(40) & 1.8  \\
          &04-12-96&4& $1.8\pm0.1$   & &$-0.15\pm0.09$&   & 1.23(45) & 2.0  \\
          &05-12-96&4& $2.0\pm0.1$   & &$-0.14\pm0.09$&   & 1.03(45) & 2.9  \\
          &07-12-96&4& $1.62\pm0.07$ & &$-0.15\pm0.07$&   & 0.99(87) & 3.7  \\
          &11-12-96&4& $1.7\pm0.1$   & &$-0.4\pm0.1$  &   & 0.85(43) & 2.2  \\
          &26-06-98&1& $2.34\pm0.03$ & &              &   & 0.79(45) & 0.8  \\
          &03-12-99&4& $2.0\pm0.1$   & &$-0.3\pm0.1$  &   & 1.08(83) & 0.9  \\
2359$-$3037 &21-06-98&4& $1.82\pm0.07$        &
        &$-0.23\pm0.06$&   & 1.36(91) & 2.9  \\
\hline
\end{tabular}
\end{center}
\noindent $^{(a)}$ 1: single power law, 2: double power law, 3: broken
power law, 4: logarithmic parabola model \\
 $^{(b)}$ X-ray flux in the 2--10 keV band in units of $10^{-11}$ \ergs
\vspace*{6truecm} 
\end{table}
\clearpage

\setcounter{table}{3}
\begin{table}[t]
\vspace*{0.5truecm} 
\scriptsize
\caption{Results of spectral fits for Flat Spectrum Radio Quasars} 
\begin{center}
\begin{tabular}{lcccccrrc}
\hline
Blazar & Observ.&\multicolumn{2}{l}{model$^{(a)}$}&
\multicolumn{3}{l}{Best fit parameters} &
\multicolumn{1}{c}{$\chi^2/d.o.f.$ }&\multicolumn{1}{c}{flux} \\
 & date  && $\gamma$ &$\gamma_1$ &$\beta$ &$E_{break}$ & (d.o.f.) & $^{(b)}$\\
\hline
%3C 57      & 28-01-99 & 1  & $1.76\pm0.10$ &   &      &   & 0.73(24)& 0.27 \\
%      & 27-01-99 & 1  & $1.73\pm0.10$ &    &      &   & 0.60(24)& 0.25 \\
0210$-$5101 & 14-01-01 & 1  & $1.67 \pm0.06$ &         &  & & 1.81(41)& 0.46 \\ 
%0210 con riga
0336+3218  & 05-08-99 & 1  & $1.65\pm0.05$ &      &  &              & 1.16(49)& 0.73 \\
0525$-$3343 &27-02-00 &4&$1.6\pm0.1 $& & &
        & 1.1(17) & 0.07 \\
0530+1331   &21-02-97 &1&$1.3\pm0.1 $& &$ $&
        & 1.35(17) & 0.25 \\    
        &22-02-97 &1&$1.4\pm0.1 $& &$ $&
        & 1.2(16) & 0.26 \\    
        &27-02-97 &1&$1.5\pm0.2 $& &$ $&
        & 0.91(17) & 0.23 \\    
        &01-03-97 &1&$1.3\pm0.1 $& &$ $&
        & 0.84(18) & 0.26 \\    
        &03-03-97 &1&$1.3\pm0.1 $& &$ $&
        & 1.06(19) & 0.23 \\    
        &04-03-97 &1&$1.5\pm0.1 $& &$ $&
        & 0.72(17) & 0.25 \\    
        &06-03-97 &1&$1.5\pm0.1 $& &$ $&
        & 1.13(16) & 0.23 \\    
        &11-03-97 &1&$1.4\pm0.2 $& &$ $&
        & 0.69(14) & 0.23 \\
0546$-$6415 &01-10-98 &4&$2.2\pm0.3 $& &$0.4\pm0.2 $&
        & 0.98(38) & 0.38 \\
0841+7053 & 27-05-98 & 1 &$ 1.30\pm0.02 $&$ $& &$ $
          & 1.04(101)&  2.65     \\
0909+0354 & 27-11-97 & 1 &$1.26\pm0.12$& & &
        & 0.83(17) & 0.18 \\ 
1130$-$1449 & 04-06-99 & 1 &$ 1.42\pm0.04  $& &  &  
            & 0.95(65) &  0.87     \\
1229+0203 & 18-07-96 & 3  & $1.2\pm0.3$ & $1.58\pm0.01$    &  &$1.0\pm0.3$  &1.36(116)&6.99\\
          & 13-01-97 & 3  & $1.9\pm0.4$ & $1.58\pm0.03$    &  &$1.0\pm0.3$  &0.87(112)&11.88\\
          & 15-01-97 & 3  & $1.3\pm0.4$ & $1.57\pm0.03$  &  &  $1.0\pm0.4$  &1.34(112)&11.43\\
          & 17-01-97 & 3  & $1.3\pm0.2$ & $1.62\pm0.01$    &  &$1.0\pm0.2$  &1.14(112)&10.80\\
          & 22-01-97 & 3  & $1.6\pm0.3$ & $1.56\pm0.04$  &  &  $1.6\pm44$   &1.09(99) &10.34\\
          & 24-06-98 & 3  & $1.8\pm0.3$ & $1.61\pm0.02$  &  &  $1.0\pm0.3$  &1.21(126)&10.77\\
          & 09-01-00 & 3  & $2.2\pm0.3$ & $1.63\pm0.02$    &  &$1.18\pm0.08$&1.46(125)&11.59\\
          & 13-06-00 & 3  & $1.7\pm0.3$ & $1.62\pm0.02$  &  &  $0.8\pm0.6$  &1.03(110)&8.67\\
          & 12-06-01 & 3  & $1.9\pm0.4$ & $1.58\pm0.03$  &  &  $1.0\pm0.3$  &0.87(112)&11.94\\
1256$-$0547 & 13-01-97 & 1 &$ 1.65\pm0.06  $& &  &  
            & 0.92(43) &  0.56     \\
            & 15-01-97 & 1 &$ 1.66\pm0.05  $& &  &  
            & 1.00(42) &  0.57     \\
            & 21-01-97 & 1 &$ 1.76\pm0.07  $& &  &  
            & 1.04(41) &  0.55     \\
            & 23-01-97 & 1 &$ 1.69\pm0.05  $& &  &  
            & 1.47(42) &  0.57     \\
            & 19-06-98 &1&$2.25\pm0.02$&&&&1.32(85) &2.56 \\ 
1430+4204   & 04-02-99 &1&$1.38\pm0.05$& & & &0.98(40)&0.28 \\
1510+5702   & 29-03-97 &1&$1.3\pm0.5$& & & &0.47(1) &0.05 \\
            & 01-02-98 &1&$1.2\pm0.4$& & & &1.99(4) &0.04 \\
1512$-$0905 & 03-08-98 &1&$1.36\pm0.05$& & & &0.77(41) & 0.50 \\
1629+4007   & 11-08-99 &1&$2.5\pm0.1$& & & &1.27(21) &0.12 \\
1642+3948   & 19-02-99 & 1  & $1.59\pm0.06$ &               &  &     & 1.29(57)&0.51 \\ 
1722+2436   & 13-02-00&1& $1.7 \pm0.1 $ &             &  &           &
1.08(17)    & 0.10 \\
1746+6226   & 29-03-97&1& $1.60\pm0.09$ &             &              &                     & 0.44(39) & 0.06 \\
2114+8204   & 29-04-98&4& $1.4 \pm0.1 $ &             &$-0.3\pm0.1 $&          & 1.29(45)  & 1.6  \\
            &12-10-98&4& $1.5 \pm0.2 $ &             &$-0.2\pm0.1 $&          &
1.12(39) & 1.2  \\
2129$-$1538 &24-05-99&4& $1.08\pm0.09$ &             &$-0.42\pm0.07$&           &
1.08(95) & 1.1  \\
2136+0041  &25-11-00&1& $1.57\pm0.06$ &             &              &           &
0.85(36) & 0.22 \\
2151$-$3027  &31-10-97&1& $1.39\pm0.04$ &             &              &           &
0.69(46) & 0.79 \\
2225+2118 & 12-11-97 & 1  & $1.3\pm0.1$ &         &  &           & 1.28(18)& 0.21\\ 
2225$-$0457 & 10-11-97 & 1  &$1.9\pm0.2$&               &  &     & 1.53(16)& 0.12 \\
2232+1143    &11-11-97&1& $1.51\pm0.06$ &             &              &           &
1.06(40) & 0.55 \\
              &13-11-97&1& $1.61\pm0.06$ &             &              &           &
0.74(40) & 0.59 \\
              &16-11-97&1& $1.55\pm0.06$ &             &              &           &
0.81(37) & 0.61 \\
              &18-11-97&1& $1.53\pm0.08$ &             &              &           &
0.99(40) & 0.59 \\
              &21-11-97&1& $1.58\pm0.07$ &             &              &           &
1.71(36) & 0.59 \\
2246$-$1206  &18-11-98&1& $1.6 \pm0.1 $ &             &              &           &
0.85(17) & 0.19 \\
2253+1608 & 05-06-00 & 1  & $1.40\pm0.04$ &         & & & 1.07(62)& 1.14 \\ 
\hline
\end{tabular}
\end{center}
\noindent $^{(a)}$ 1: single power law, 2: double power law, 3: broken
power law, 4: logarithmic parabola model \\
 $^{(b)}$ X-ray flux in the 2--10 keV band in units of $10^{-11}$ \ergs
\vspace{7truecm} 
\end{table}
%---------------------------------------------------
\clearpage 
\setcounter{figure}{1} 
\begin{figure}[t]
\centering
\includegraphics{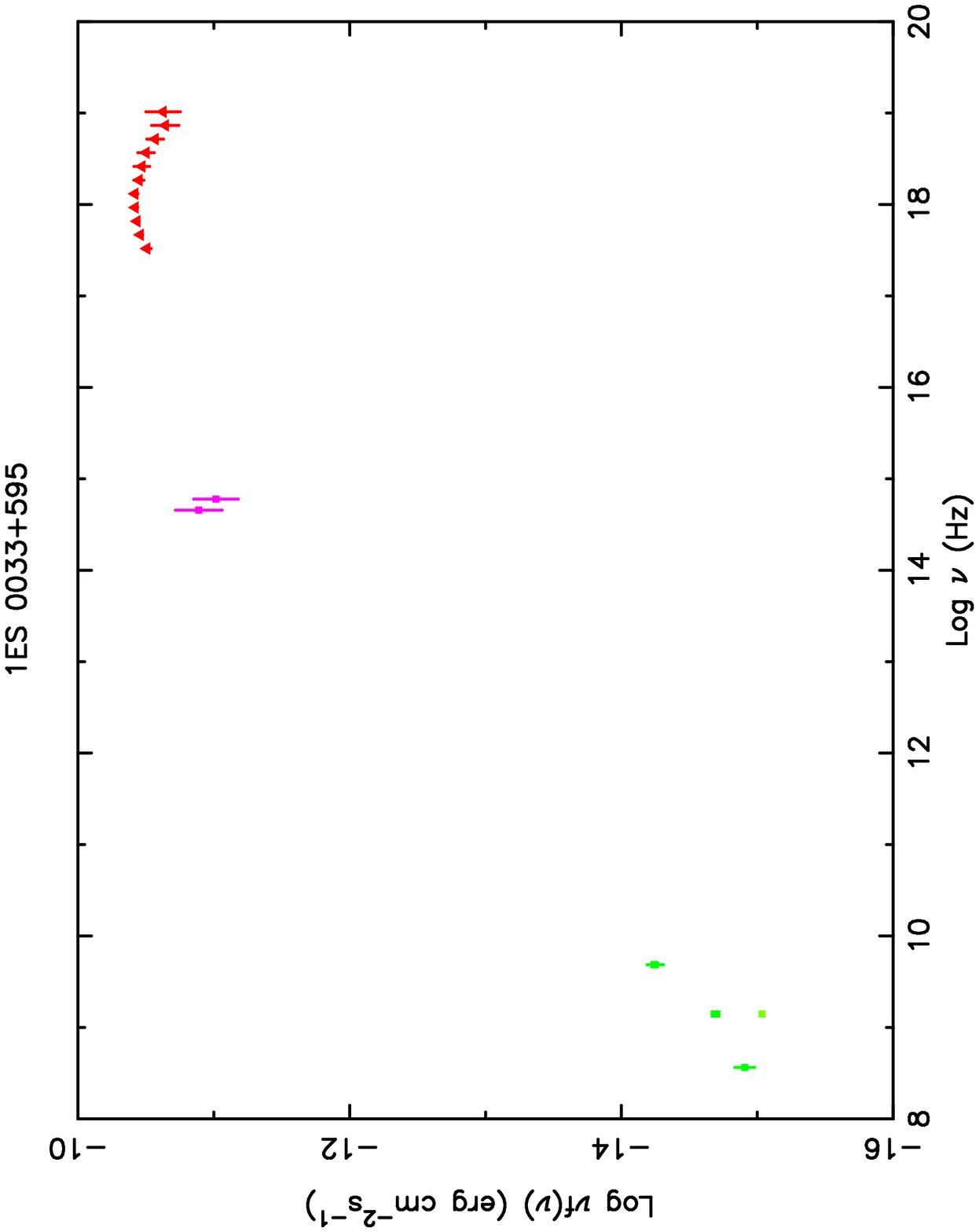} 
\includegraphics{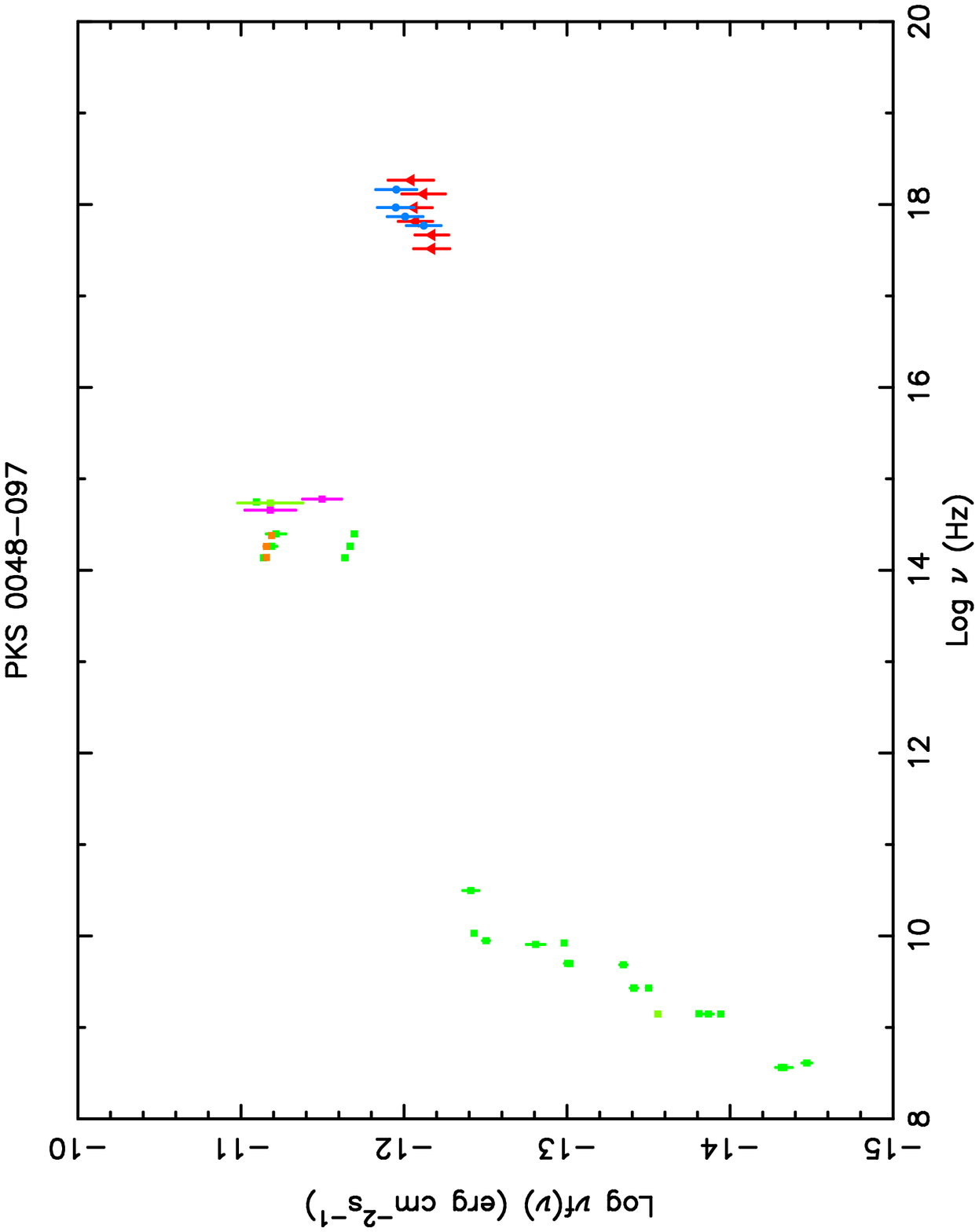} 
\includegraphics{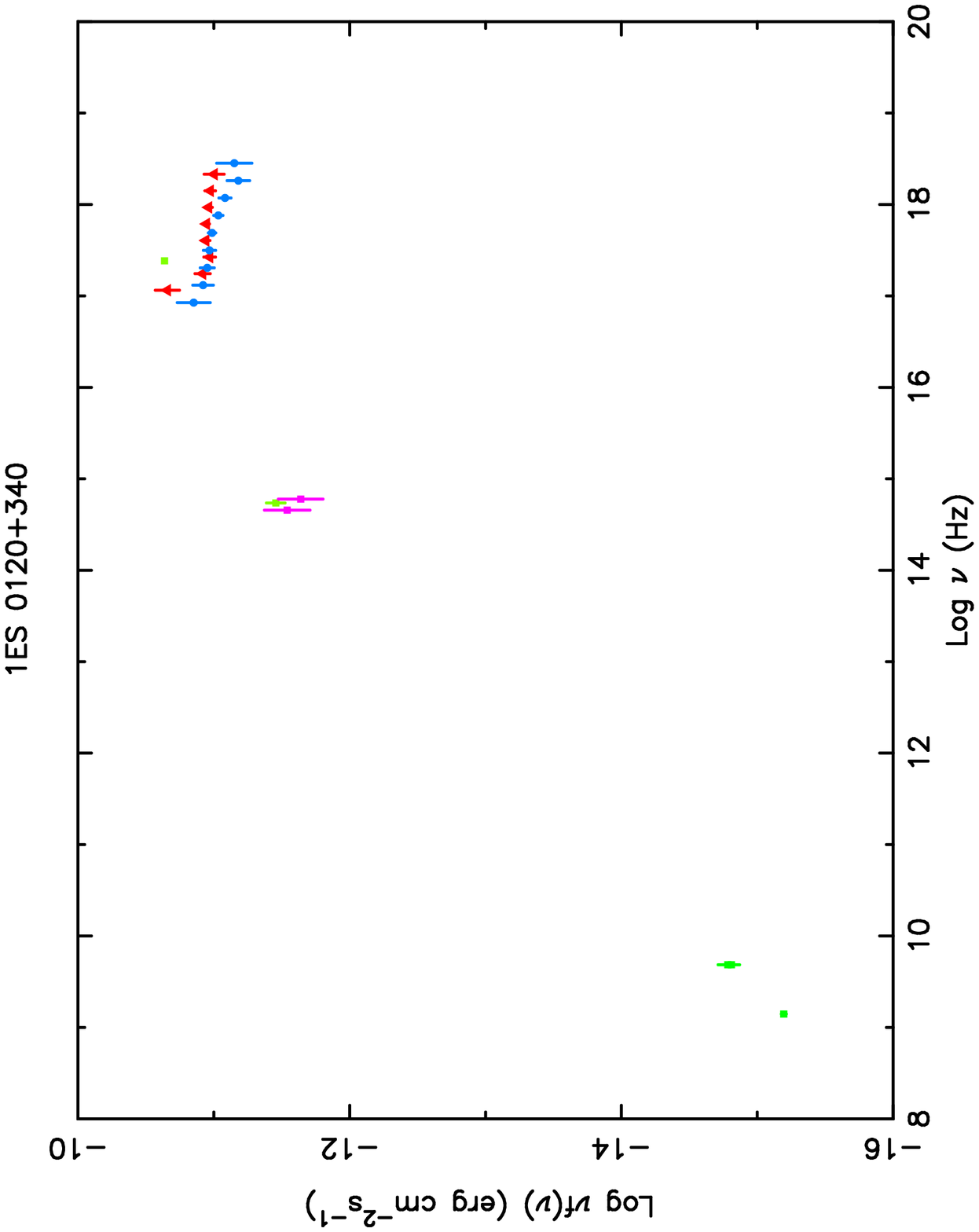} 
\includegraphics{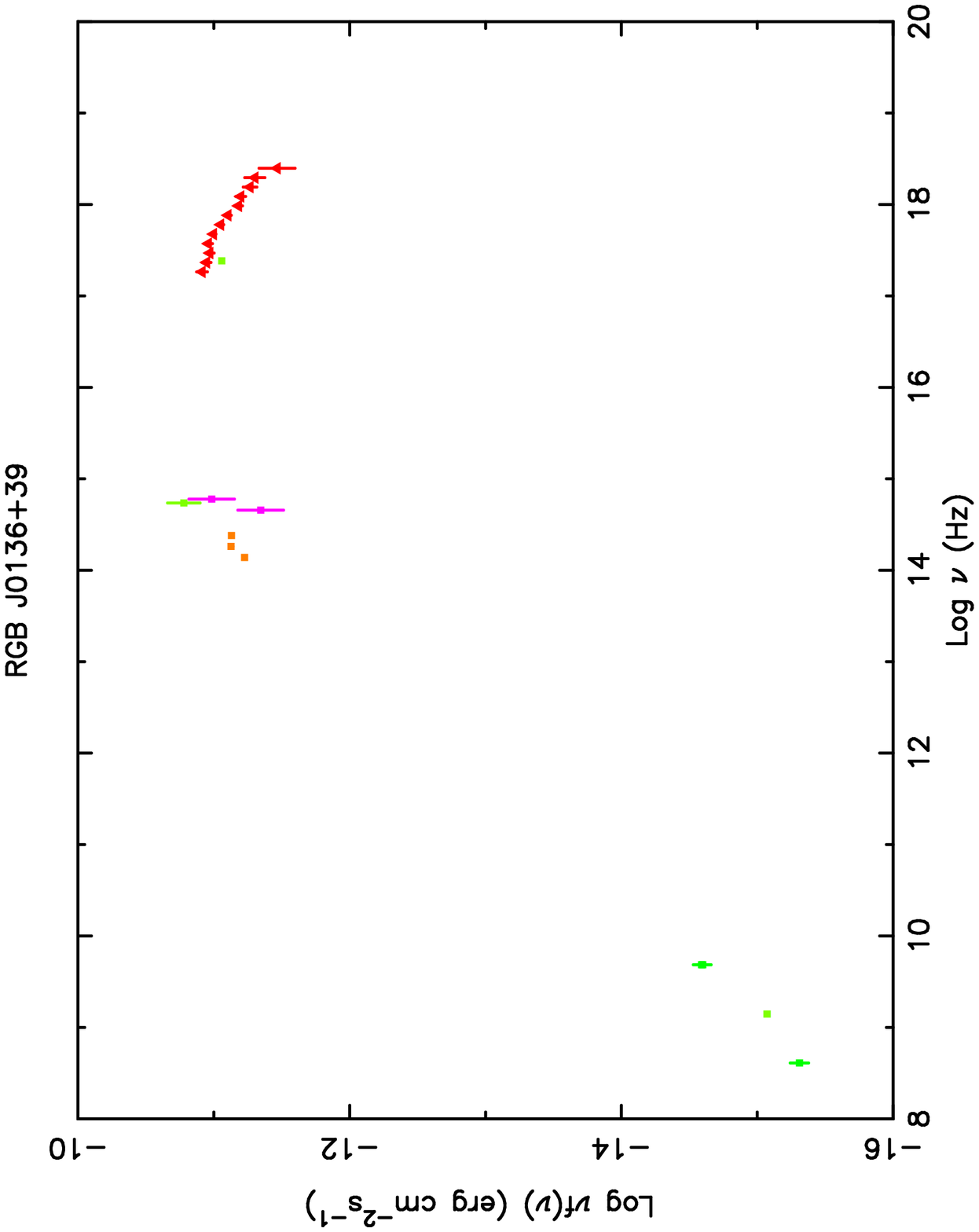} 
\vspace{19.0cm} 
\caption[t]{a- Spectral Energy Distribution of the BL Lacs 1ES 0033+595, PKS 0048$-$097,
1ES 0120+340 and RGB J0136+39} 
\label{fig1a} 
\end{figure} 
\clearpage 
\setcounter{figure}{1} 
\begin{figure}[t]
\centering
\includegraphics{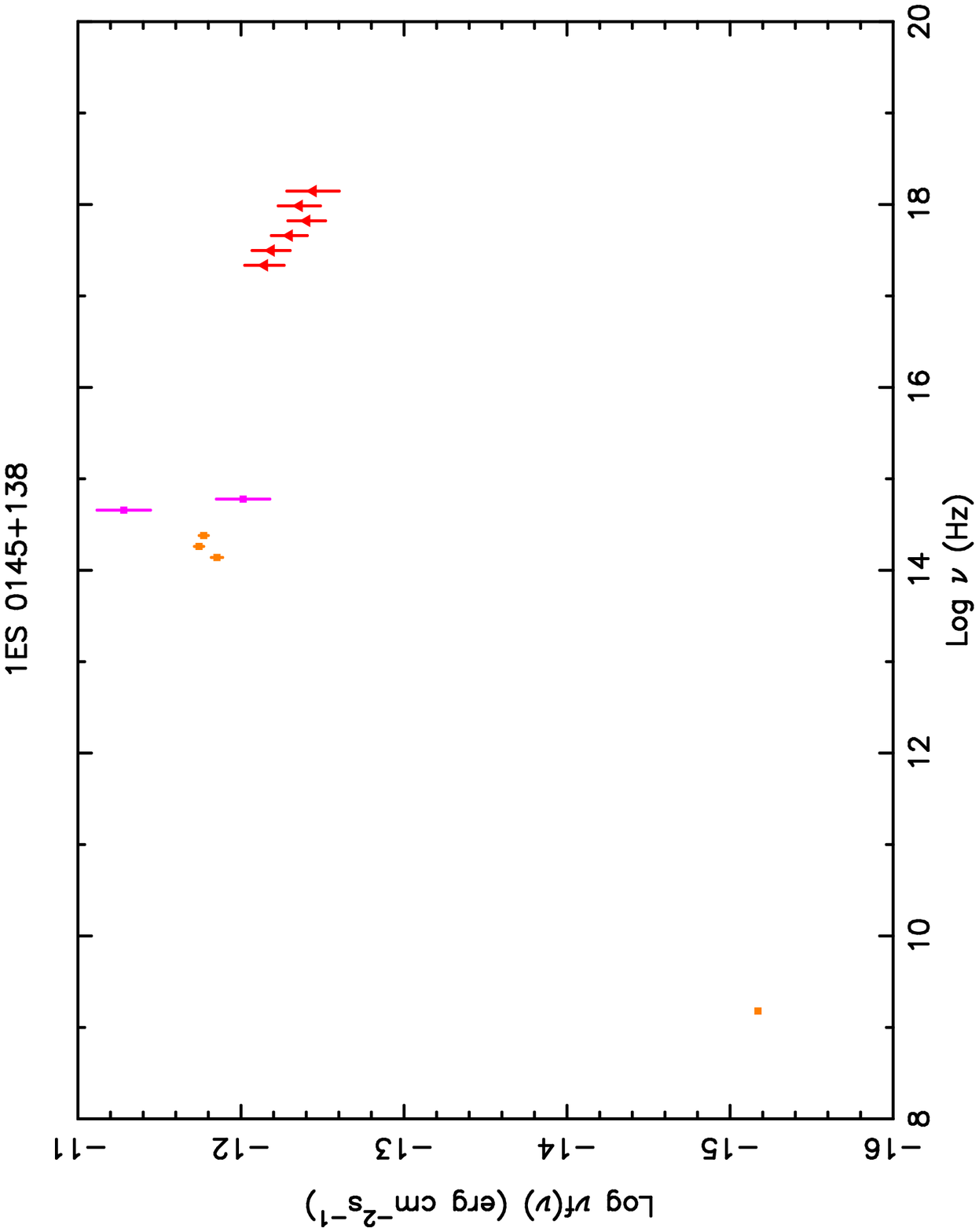} 
\includegraphics{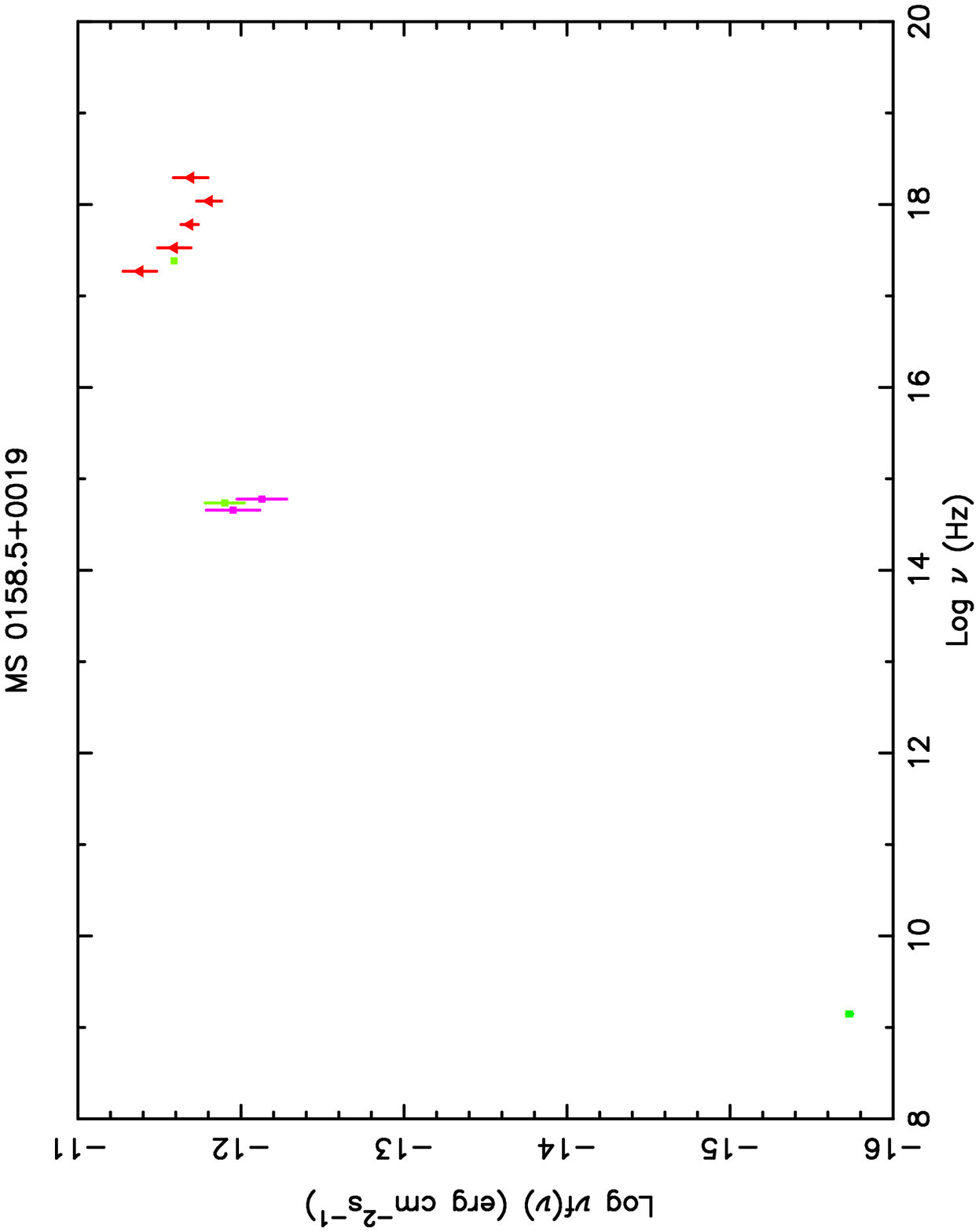} 
\includegraphics{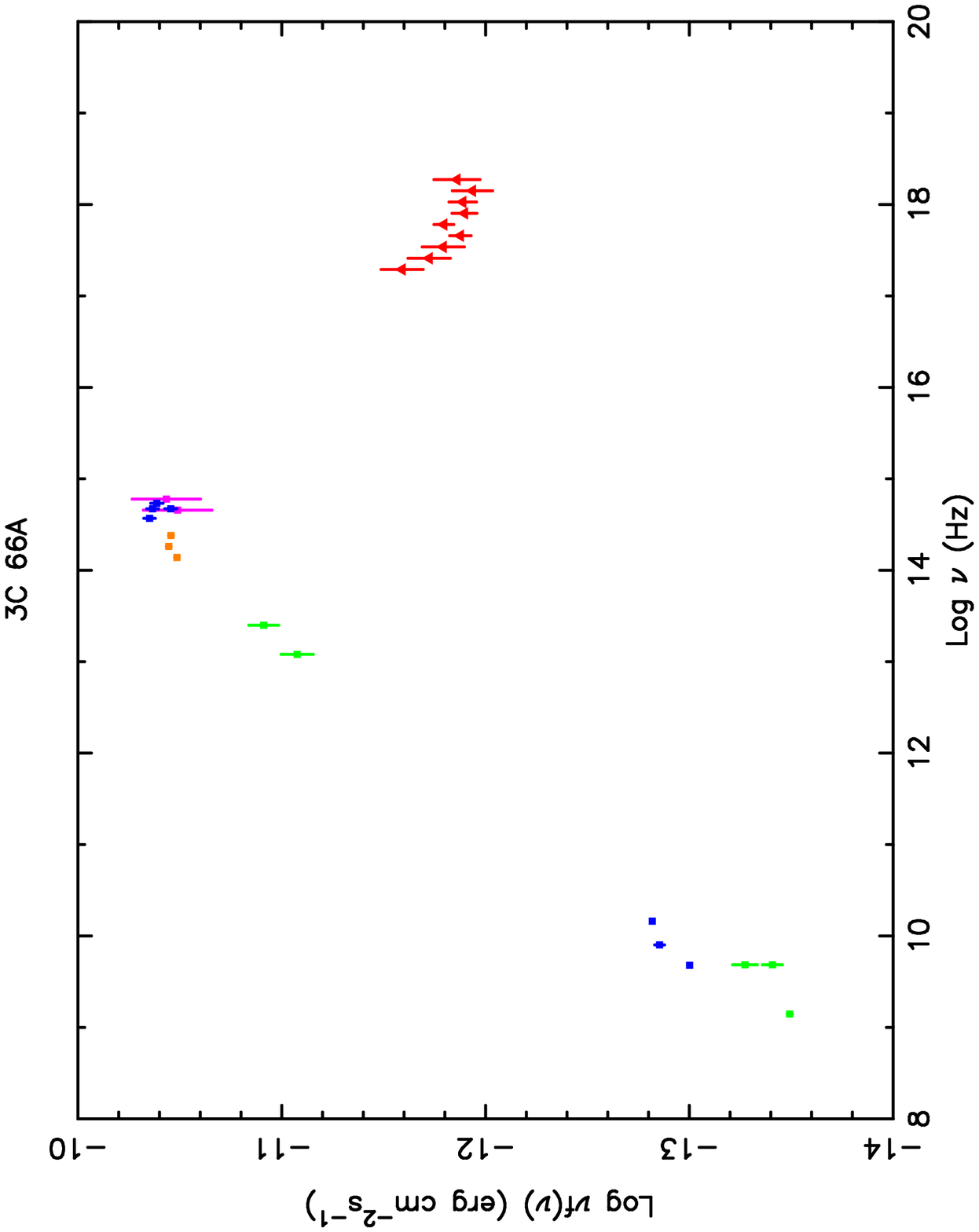} 
\includegraphics{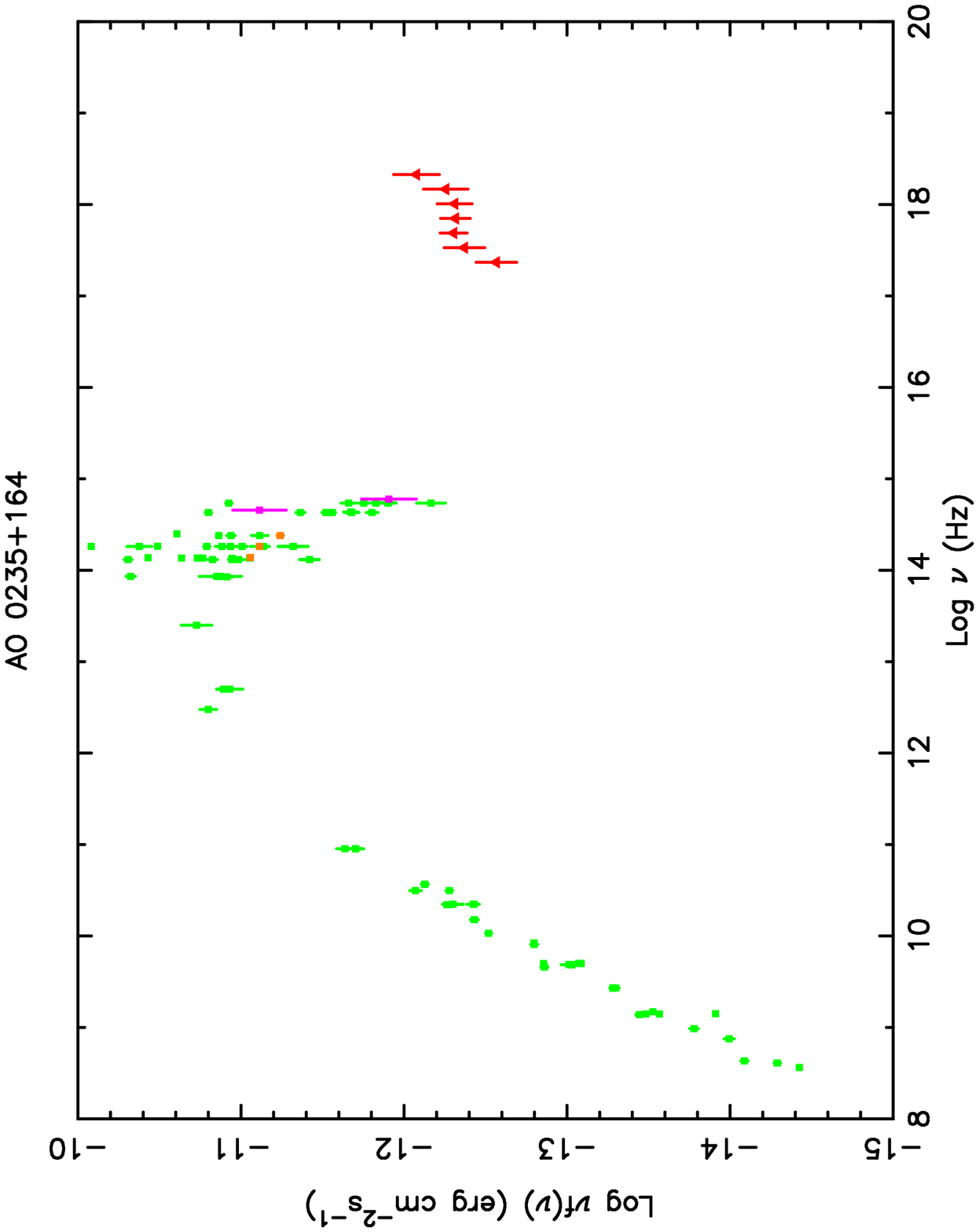} 
\vspace{19.0cm} 
\caption[t]{b- Spectral Energy Distribution of the BL Lacs 1ES 0145+138, 
MS 0158.5+0019, 3C 66A and AO 0235+164} 
\label{fig1b} 
\end{figure} 
\clearpage 
\setcounter{figure}{1} 
\begin{figure}[t]
\centering
\includegraphics{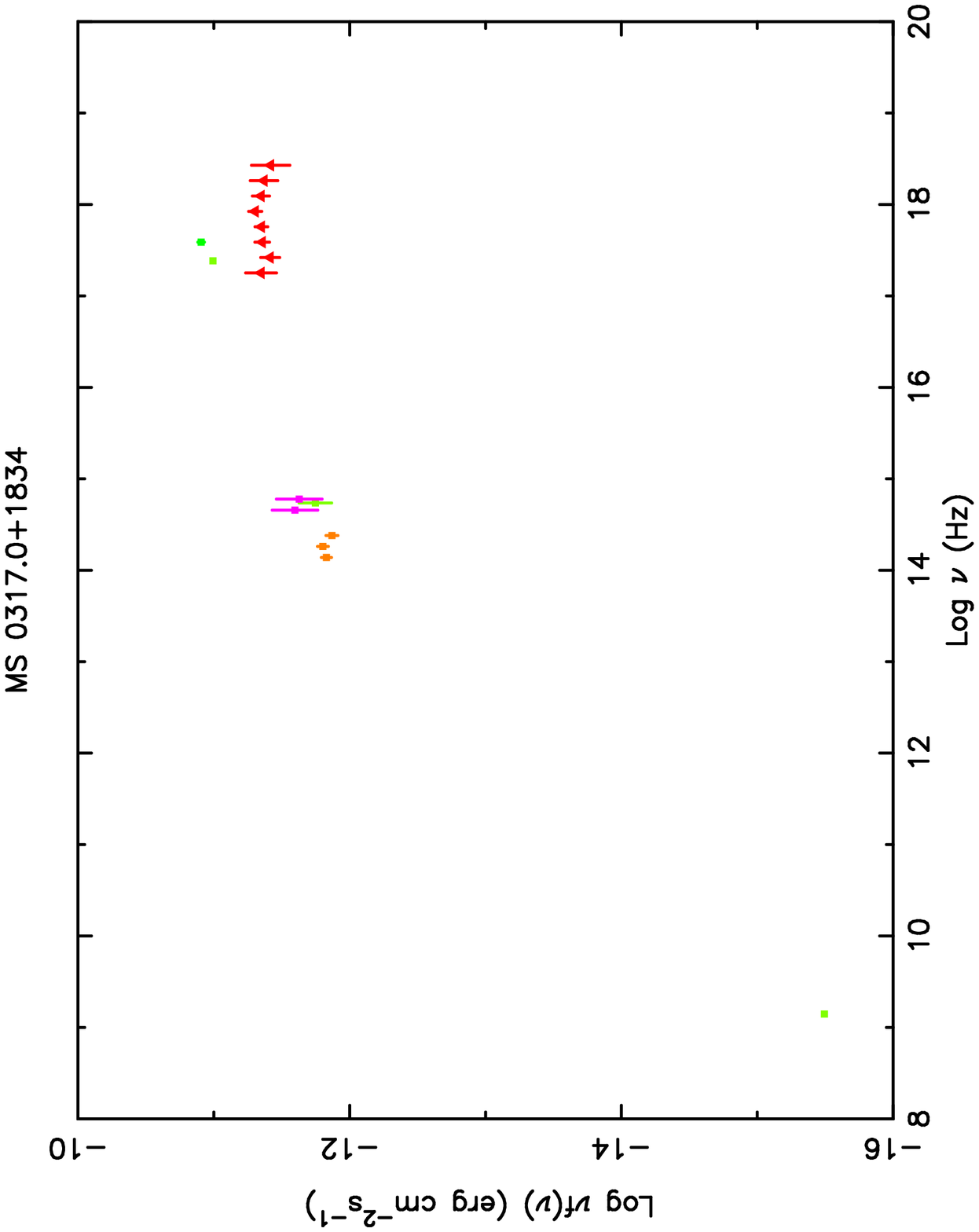} 
\includegraphics{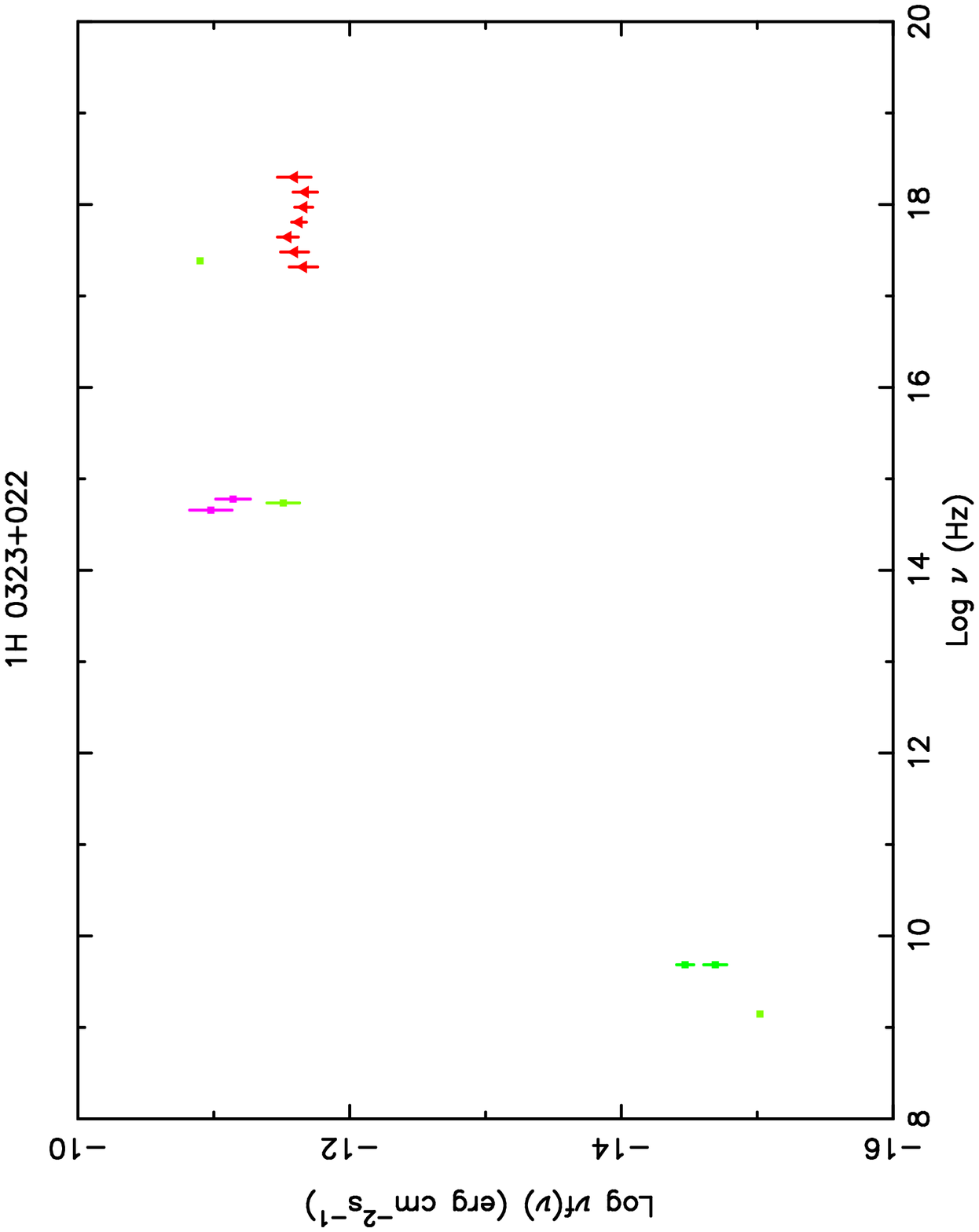} 
\includegraphics{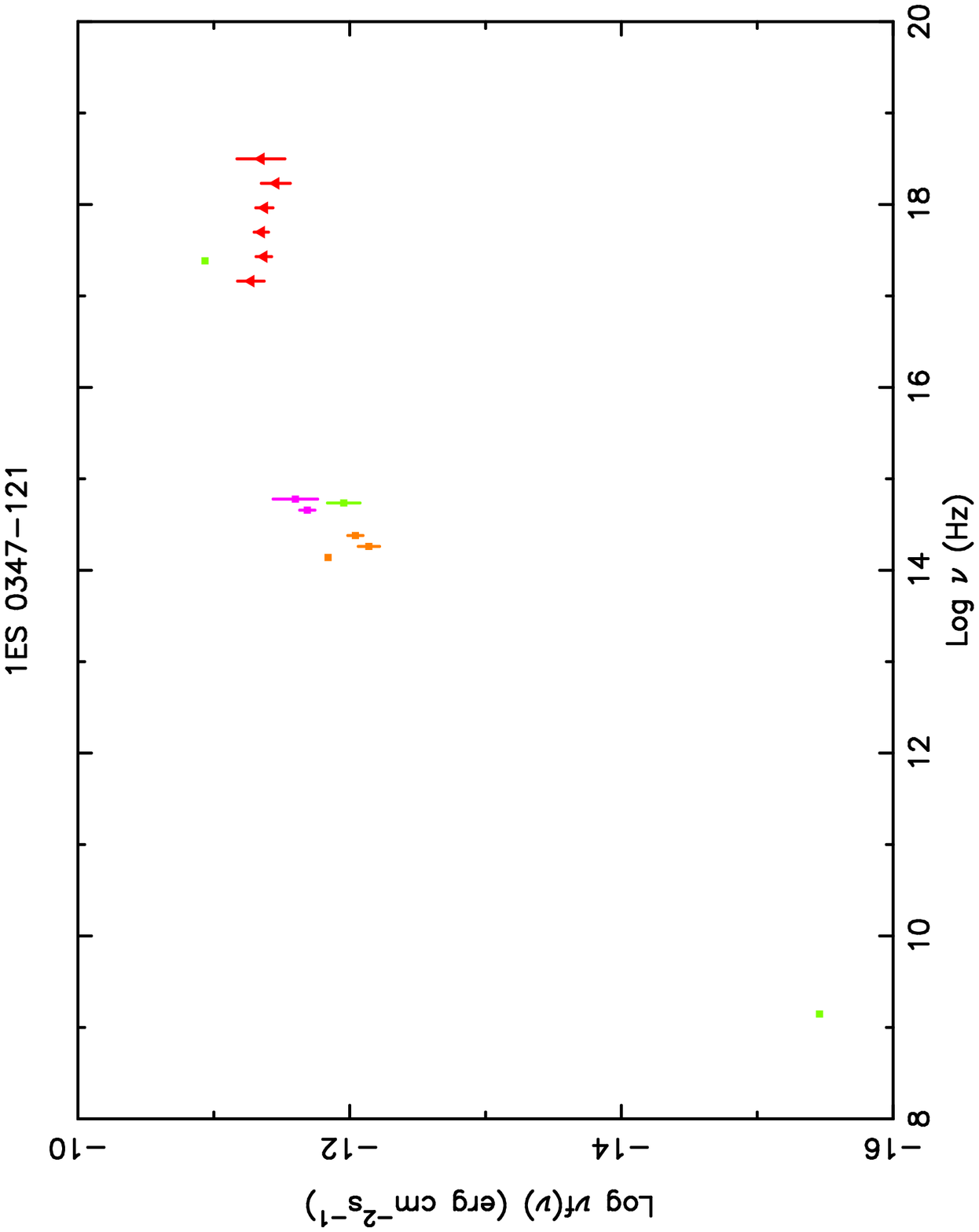} 
\includegraphics{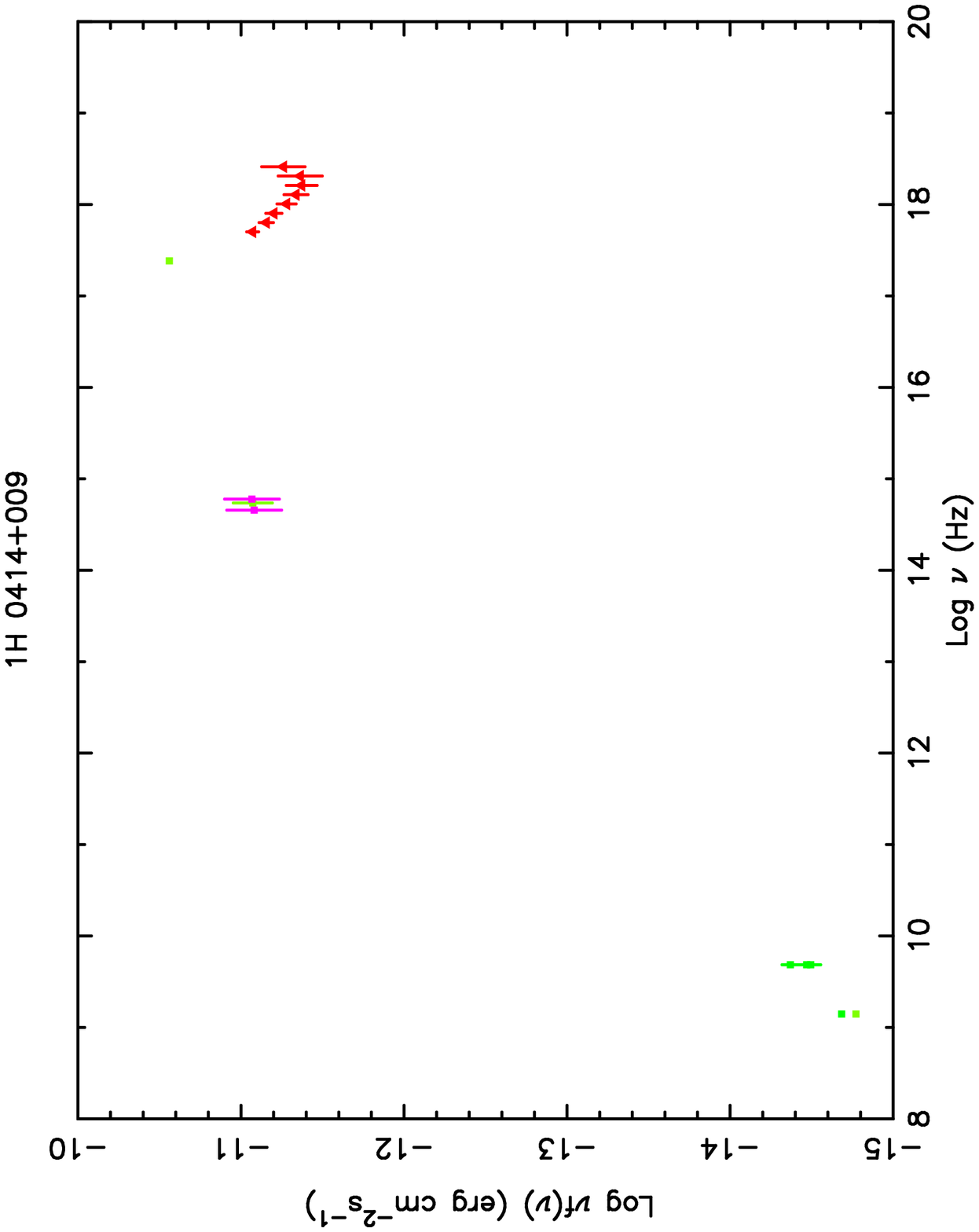} 
\vspace{19.0cm} 
\caption[t]{c- Spectral Energy Distribution of the BL Lacs MS 0317.0+1834,
1H 0323+022, 1ES 0347$-$121 and 1H 0414+009} 
\label{fig1c} 
\end{figure} 
\clearpage 
\setcounter{figure}{1} 
\begin{figure}[t]
\centering
\includegraphics{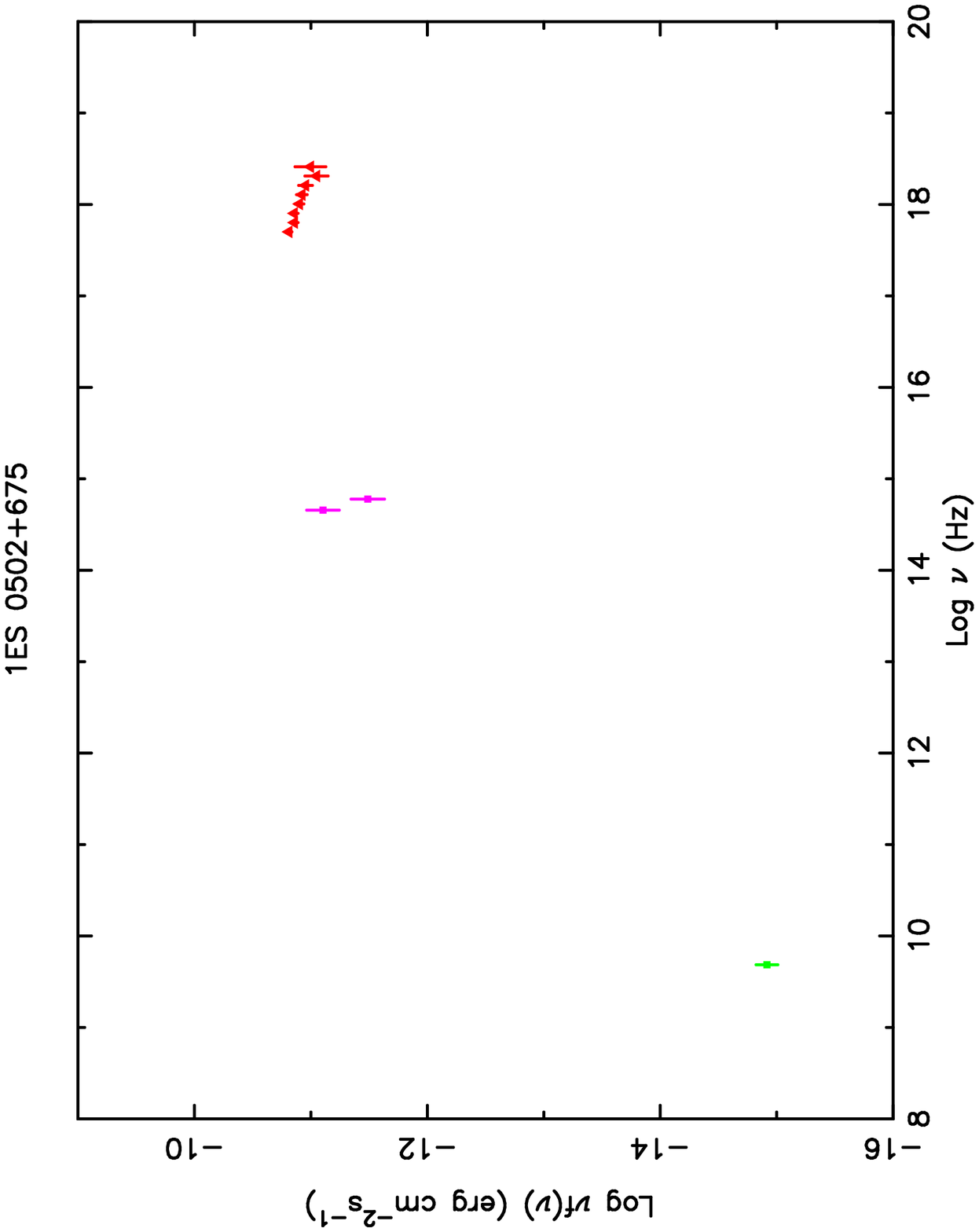} 
\includegraphics{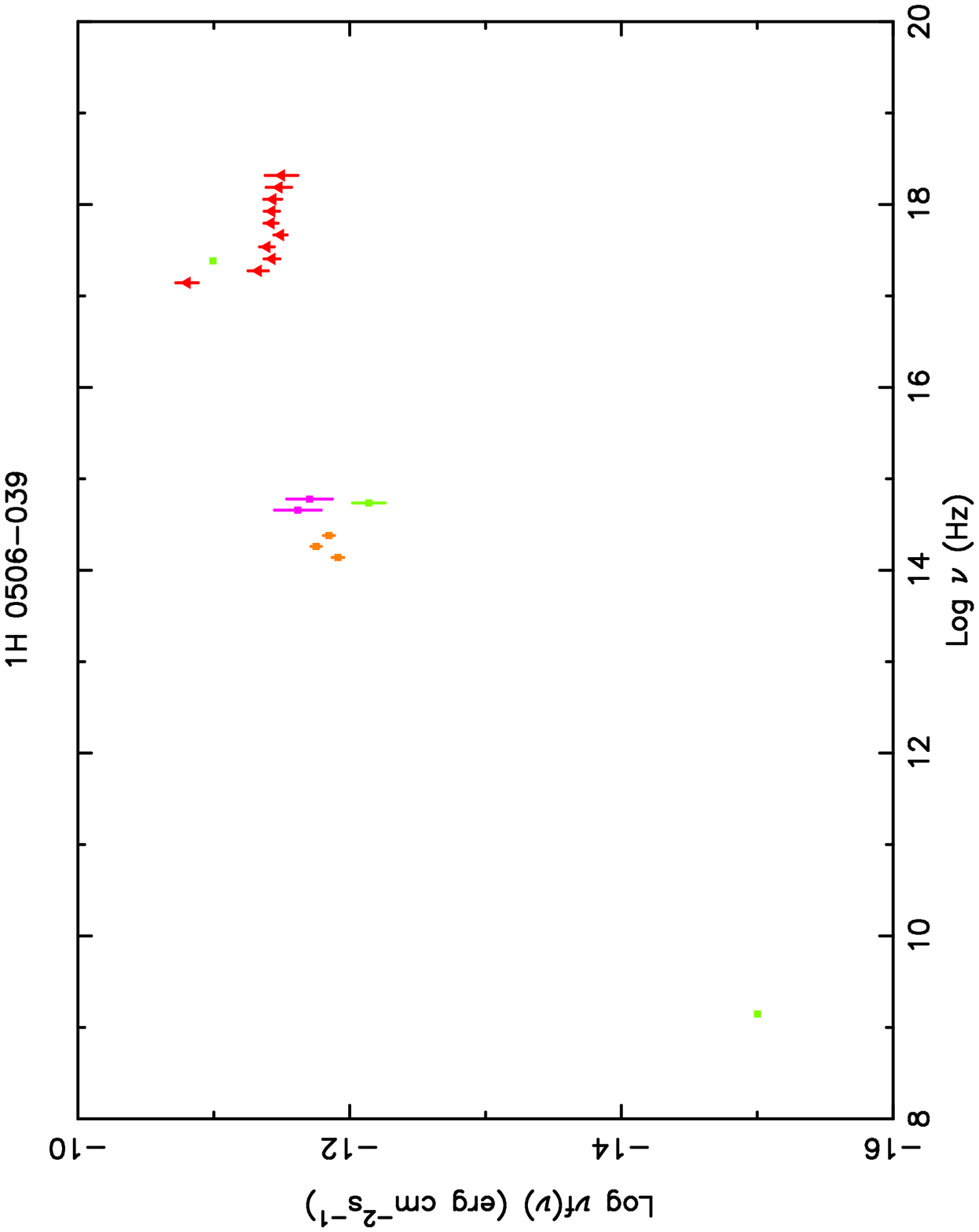} 
\includegraphics{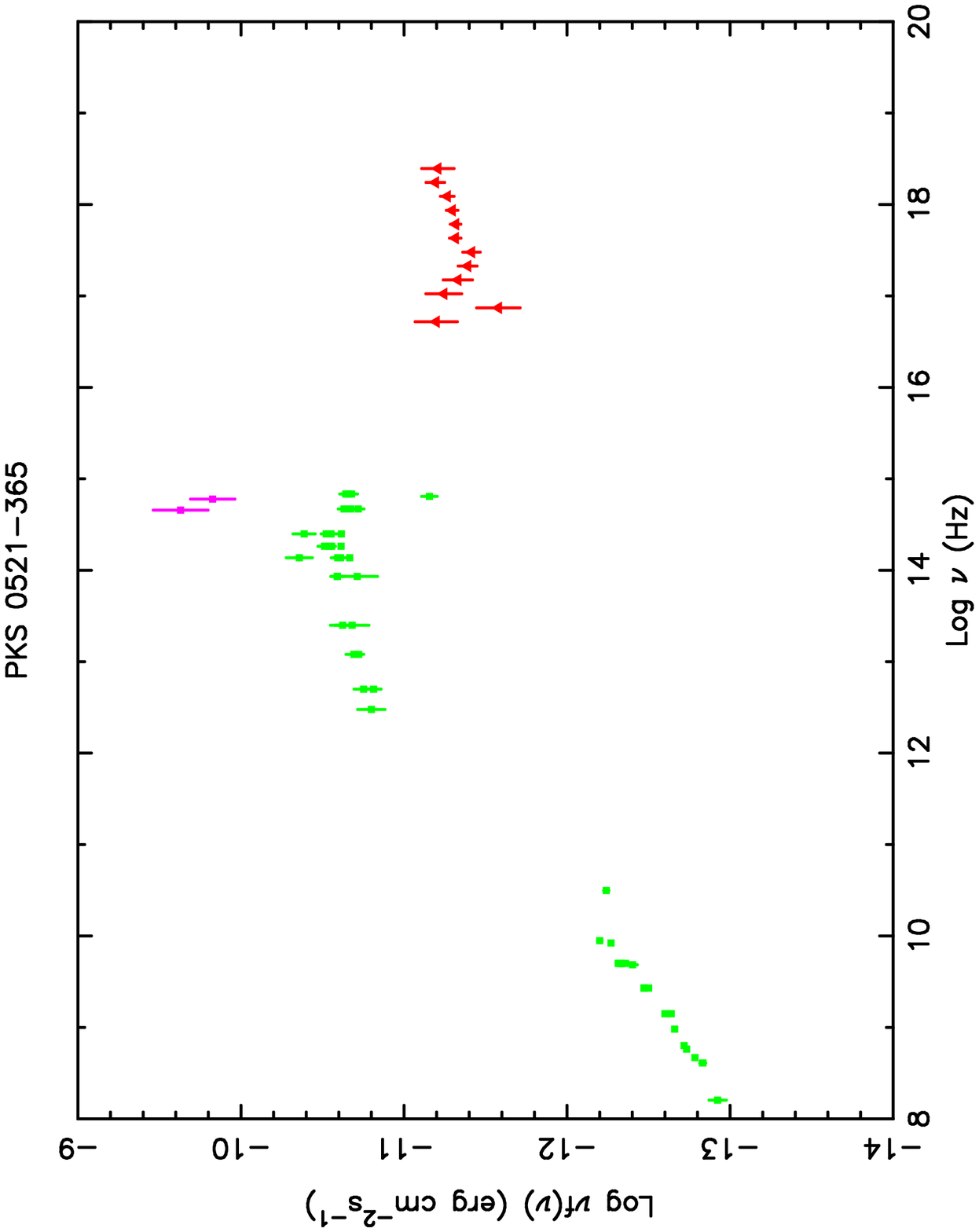} 
\includegraphics{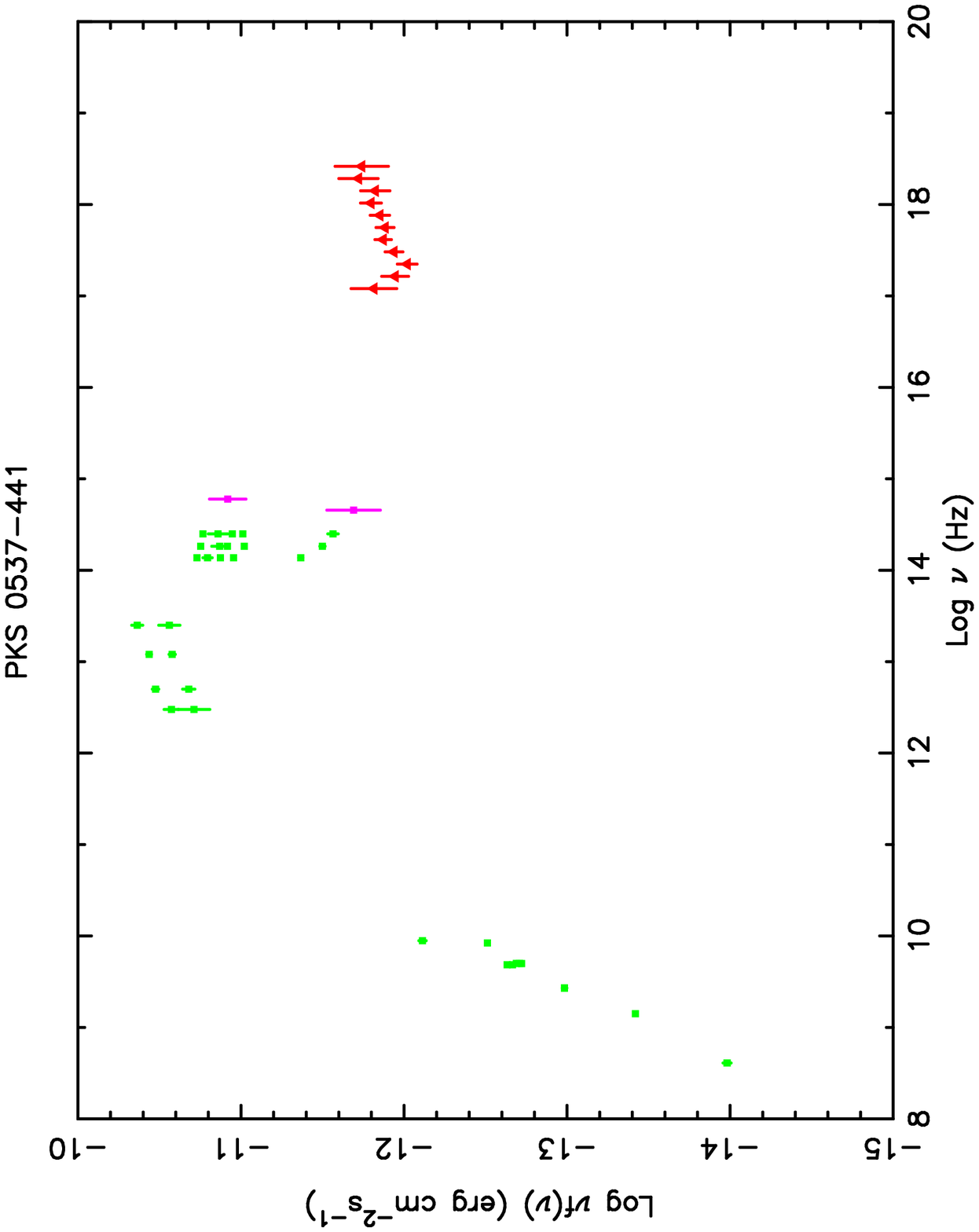} 
\vspace{19.0cm} 
\caption[t]{d- Spectral Energy Distribution of the BL Lacs 1ES 0502+675, 1H 0506$-$039,
PKS 0521$-$365 and PKS 0537$-$441} 
\label{fig1d} 
\end{figure} 
\clearpage 
\setcounter{figure}{1} 
\begin{figure}[t]
\centering
\includegraphics{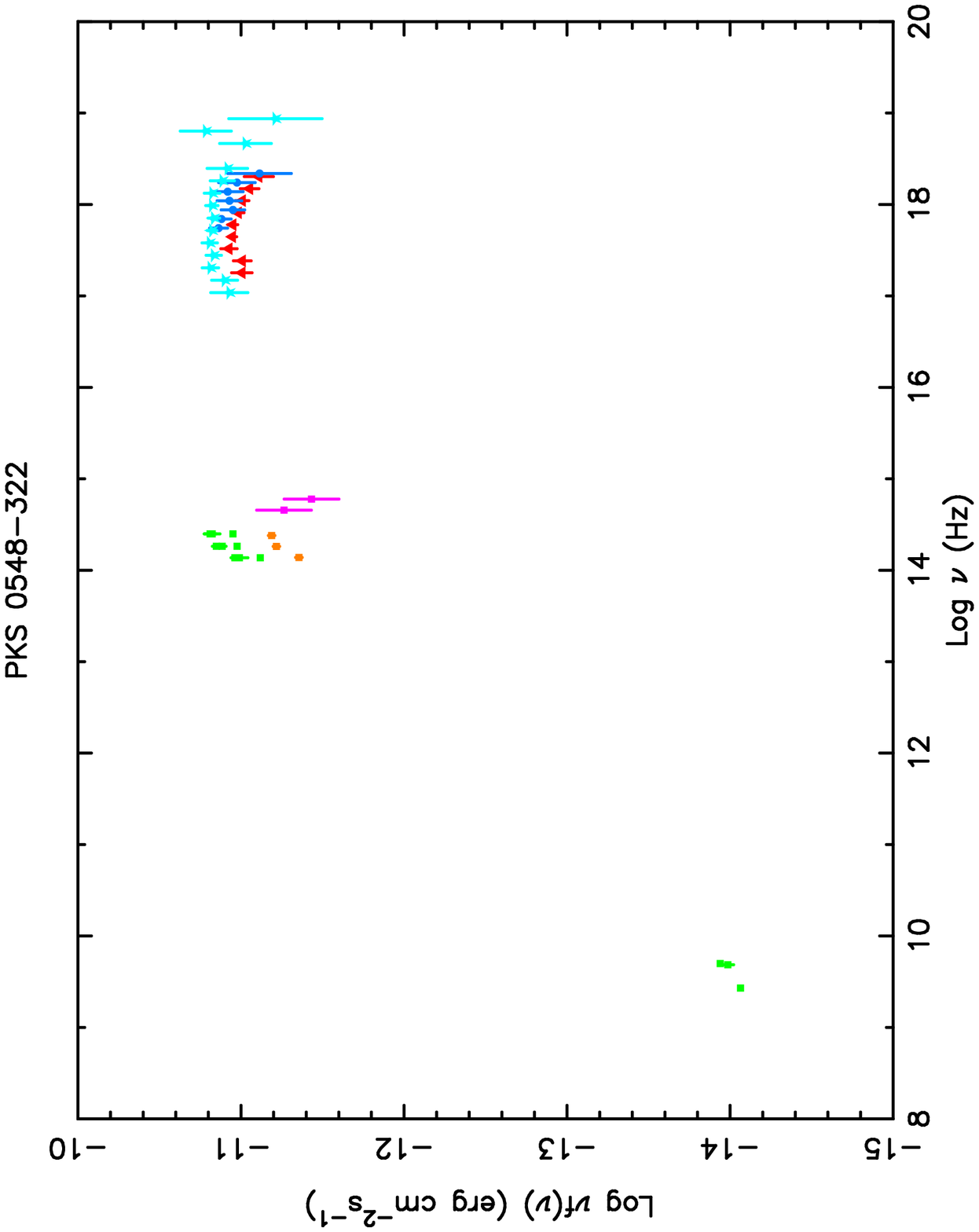} 
\includegraphics{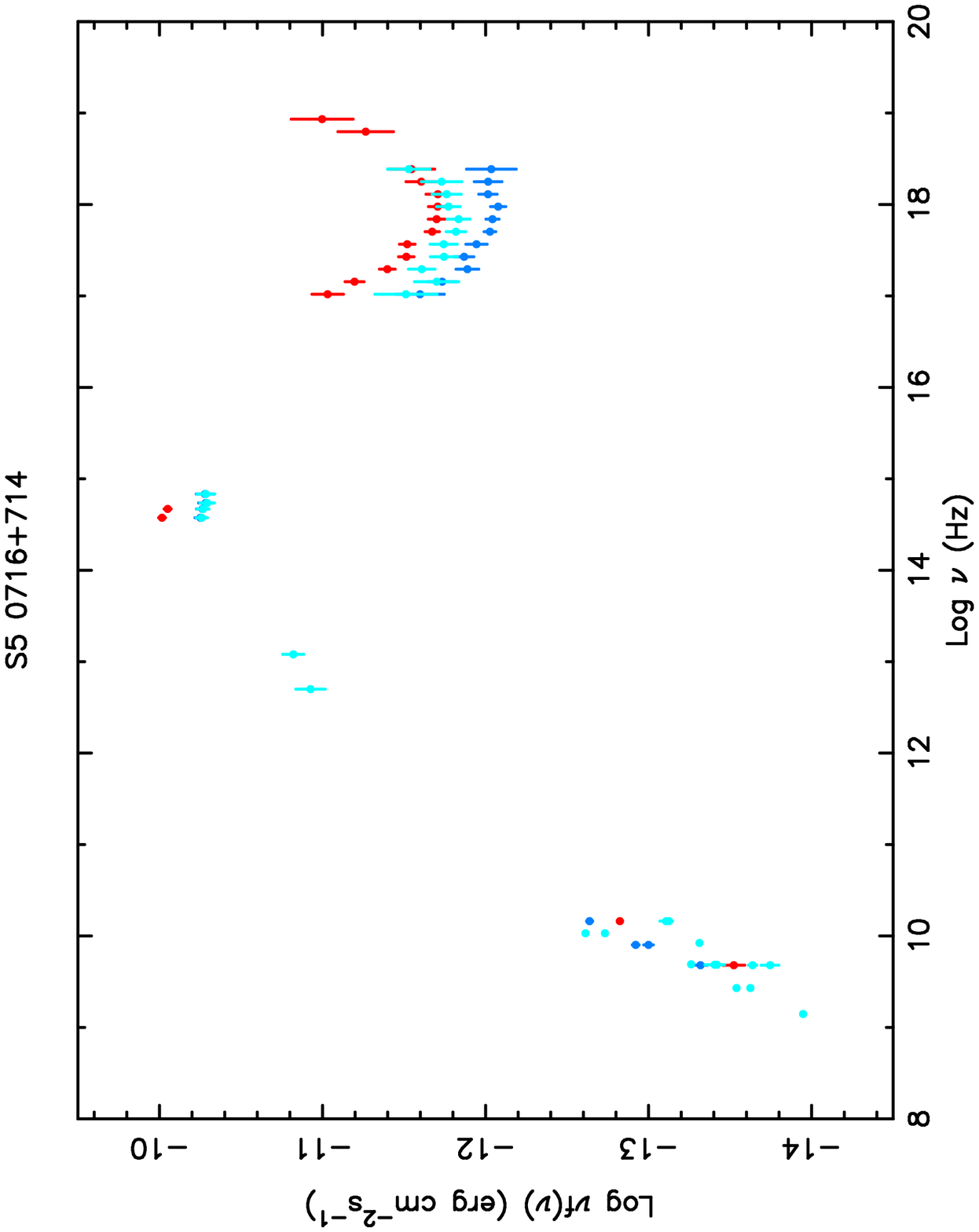} 
\includegraphics{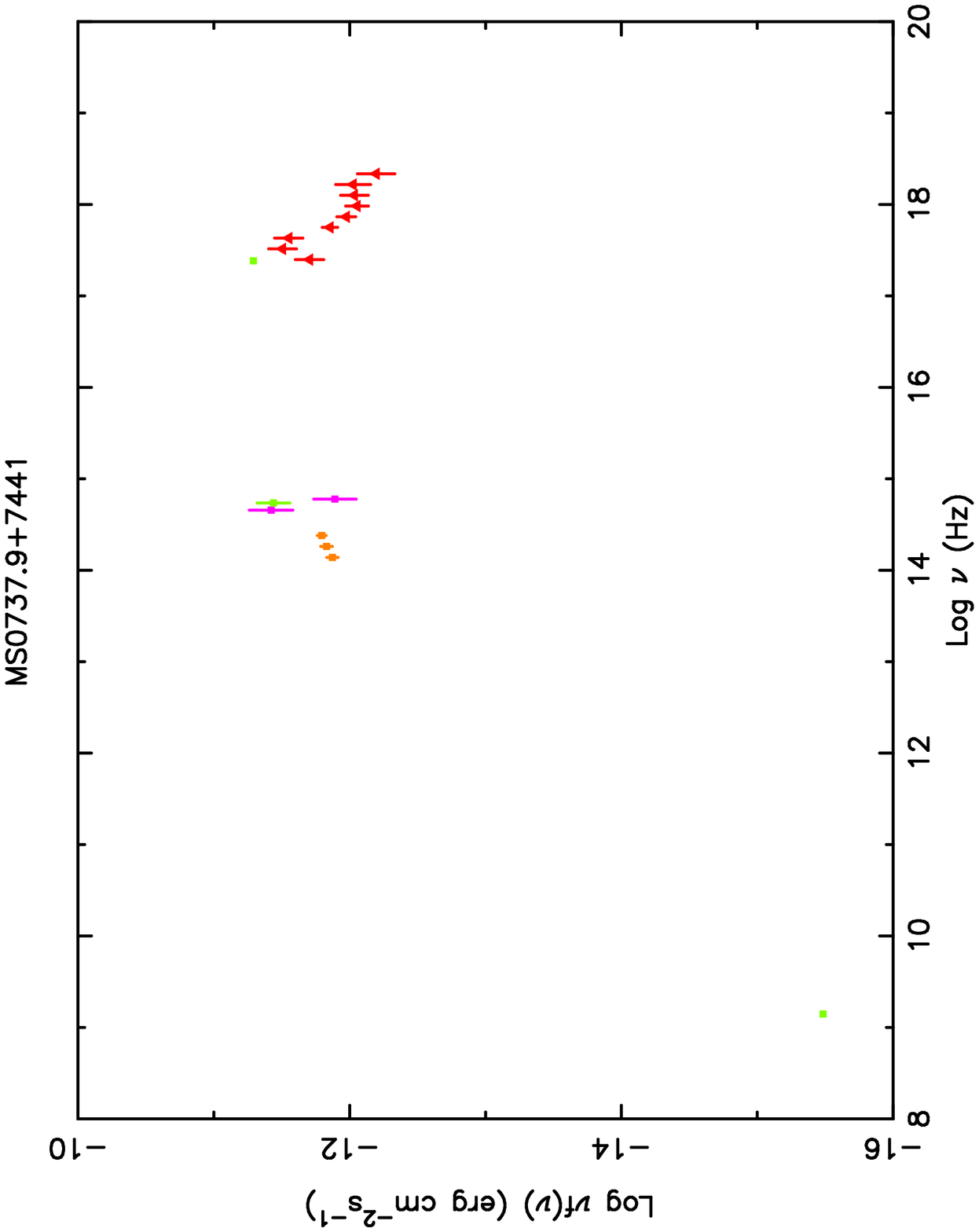} 
\includegraphics{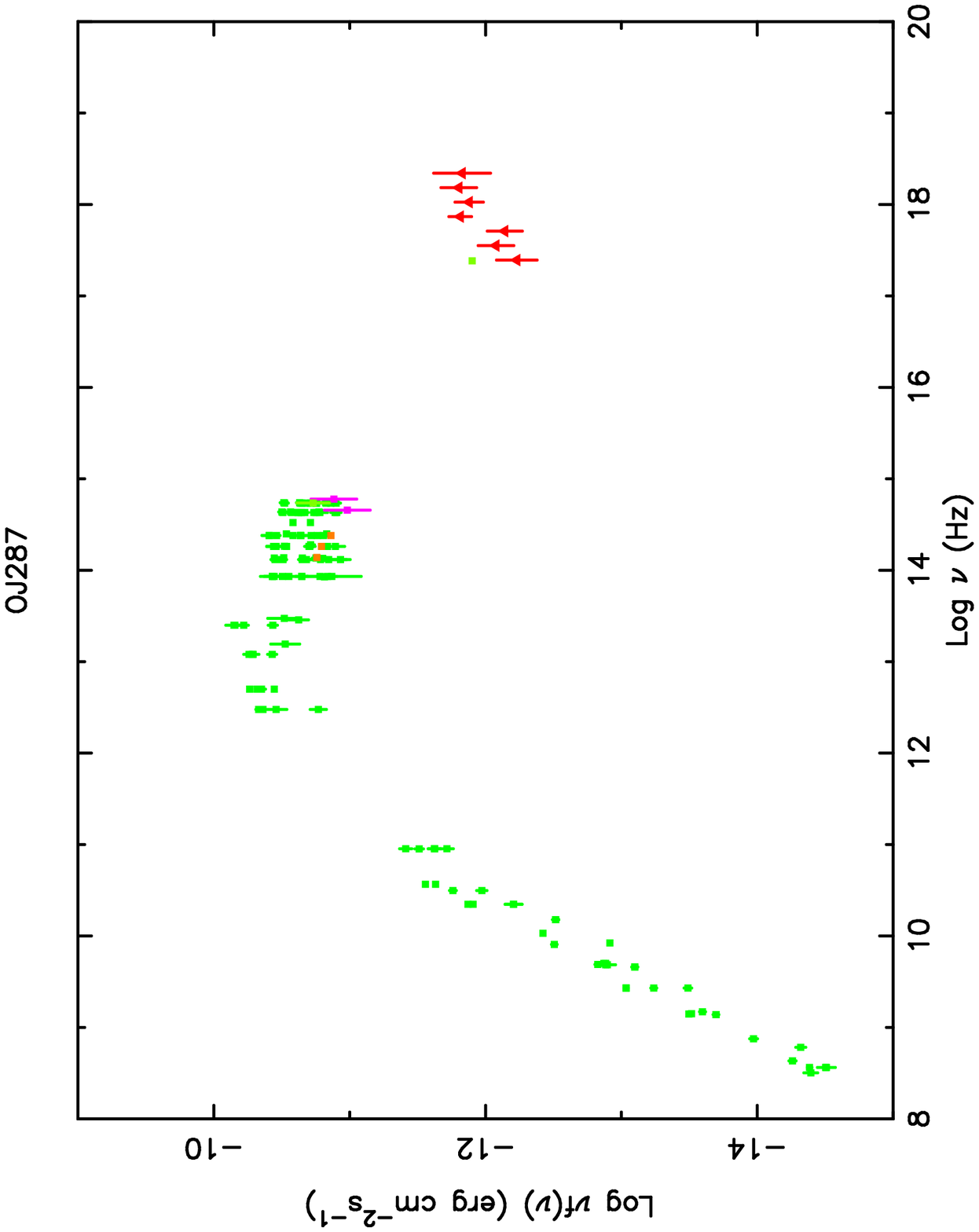} 
\vspace{19.0cm} 
\caption[t]{e- Spectral Energy Distribution of the BL Lacs PKS 0548$-$322, 
S5 0716+714, MS 0737.9+7441 and OJ 287} 
\label{fig1e} 
\end{figure} 
\clearpage 
\setcounter{figure}{1} 
\begin{figure}[t]
\centering
\includegraphics{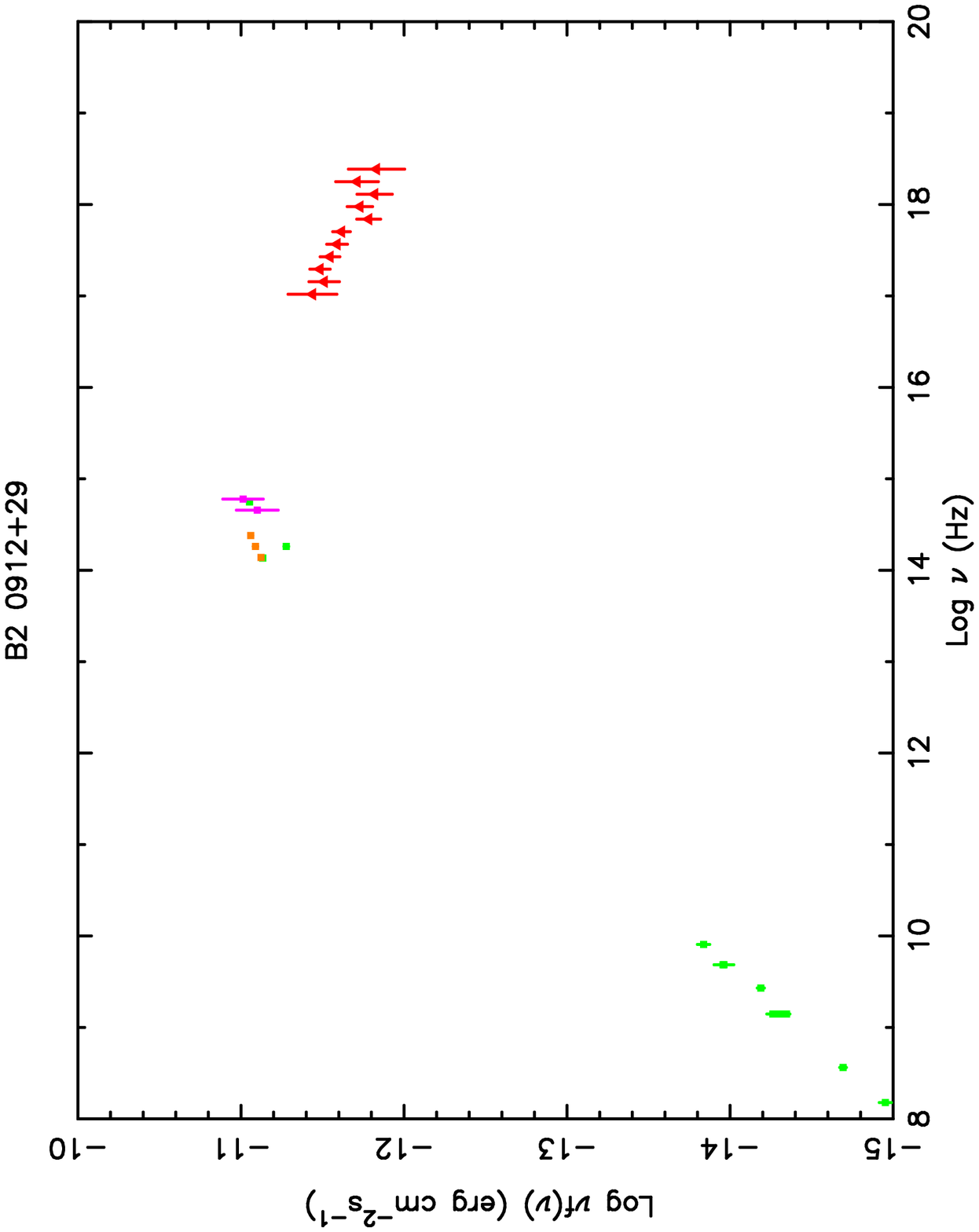} 
\includegraphics{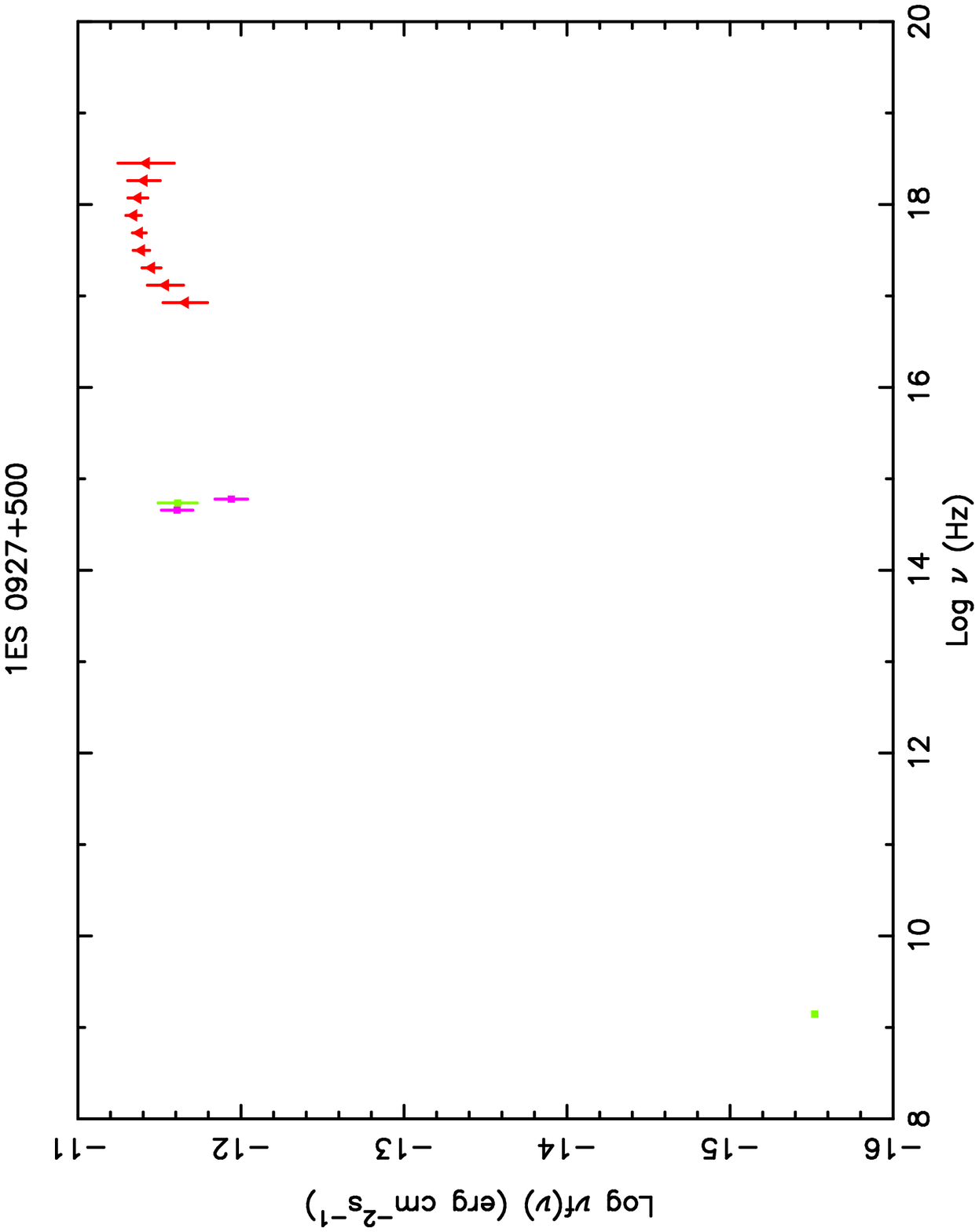} 
\includegraphics{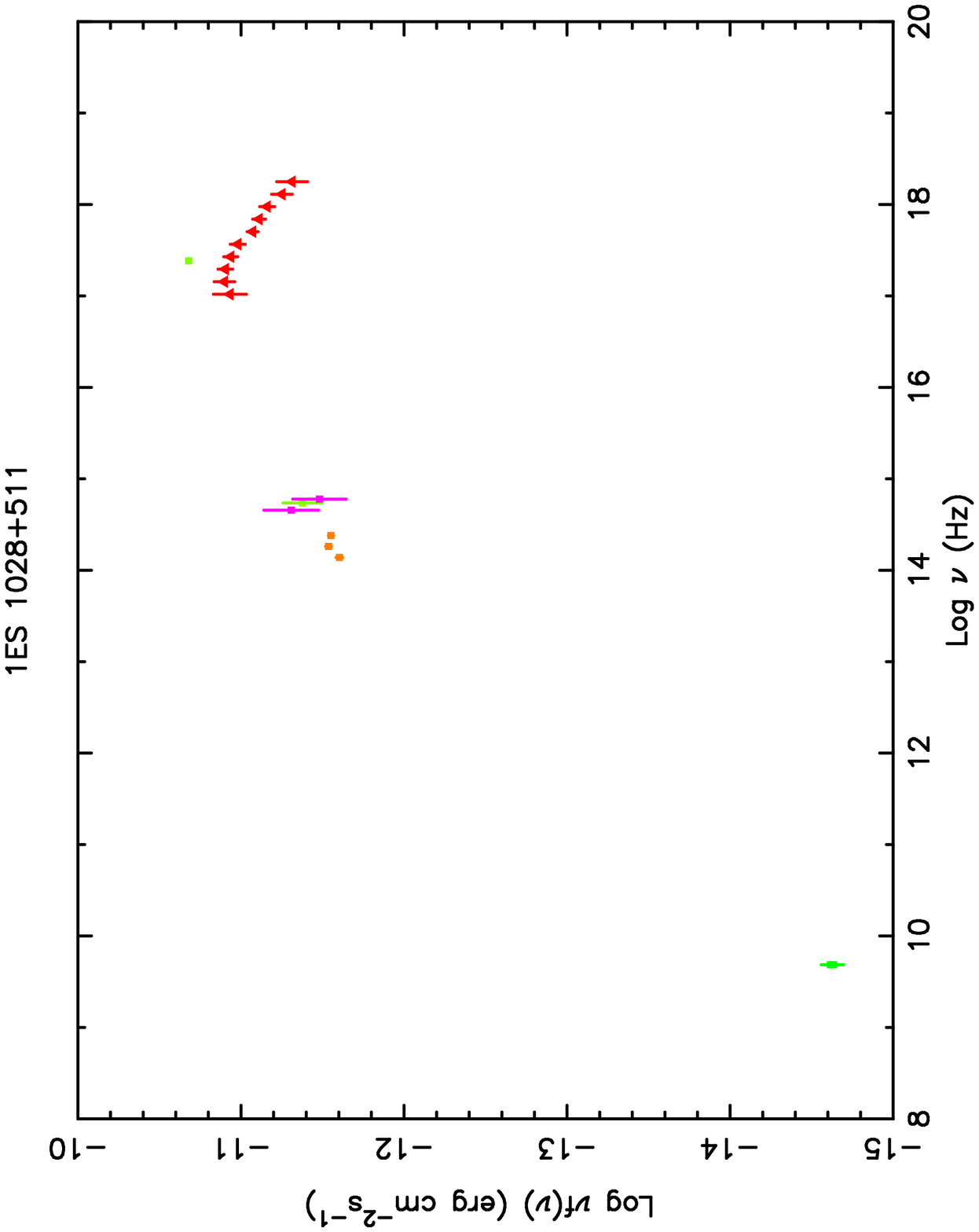} 
\includegraphics{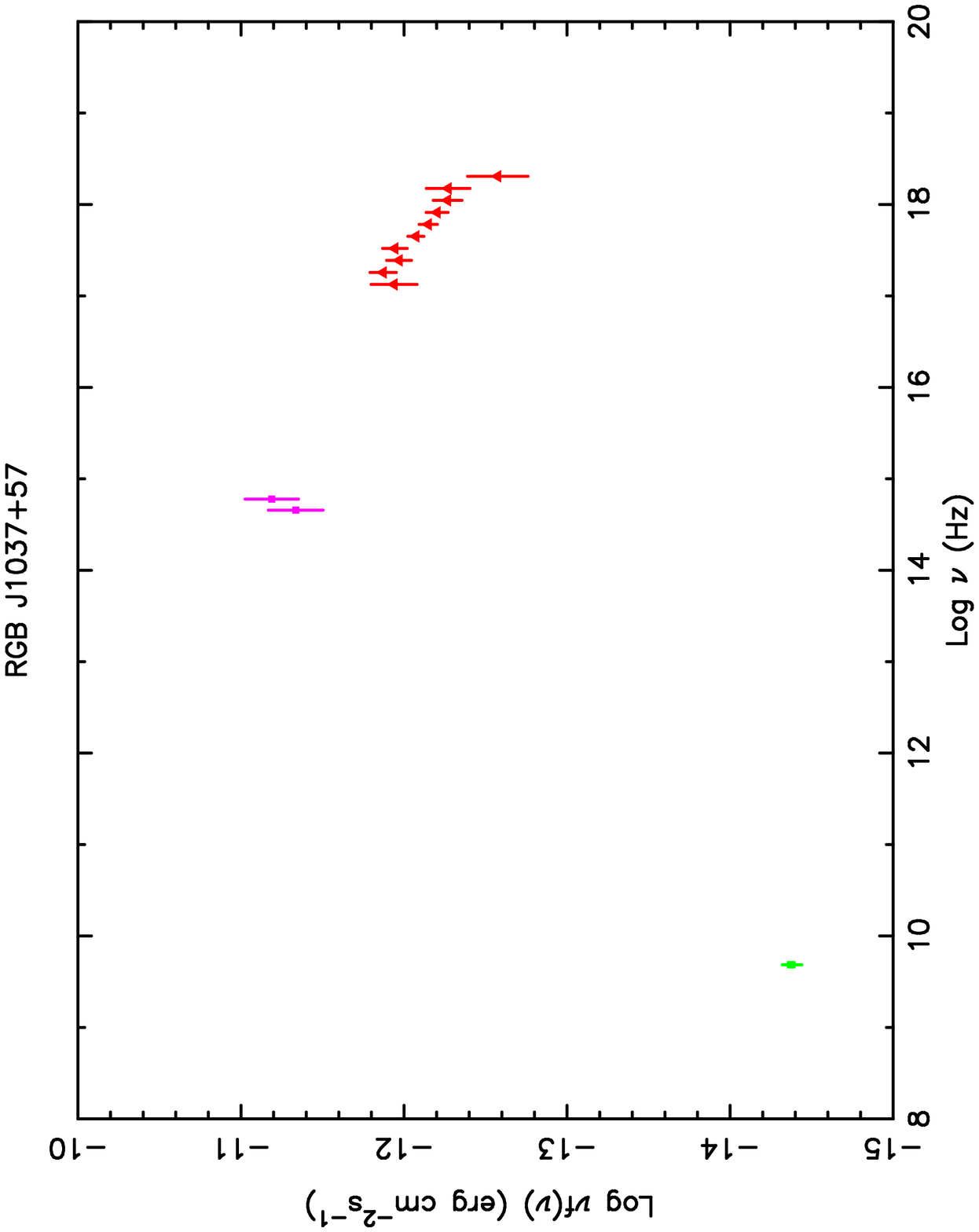} 
\vspace{19.0cm} 
\caption[t]{f- Spectral Energy Distribution of the BL Lacs B2 0912+29, 1ES 0927+500,
1ES 1028+511 and RGB J1037+57} 
\label{fig1f} 
\end{figure} 
\clearpage 
\setcounter{figure}{1} 
\begin{figure}[t]
\centering
\includegraphics{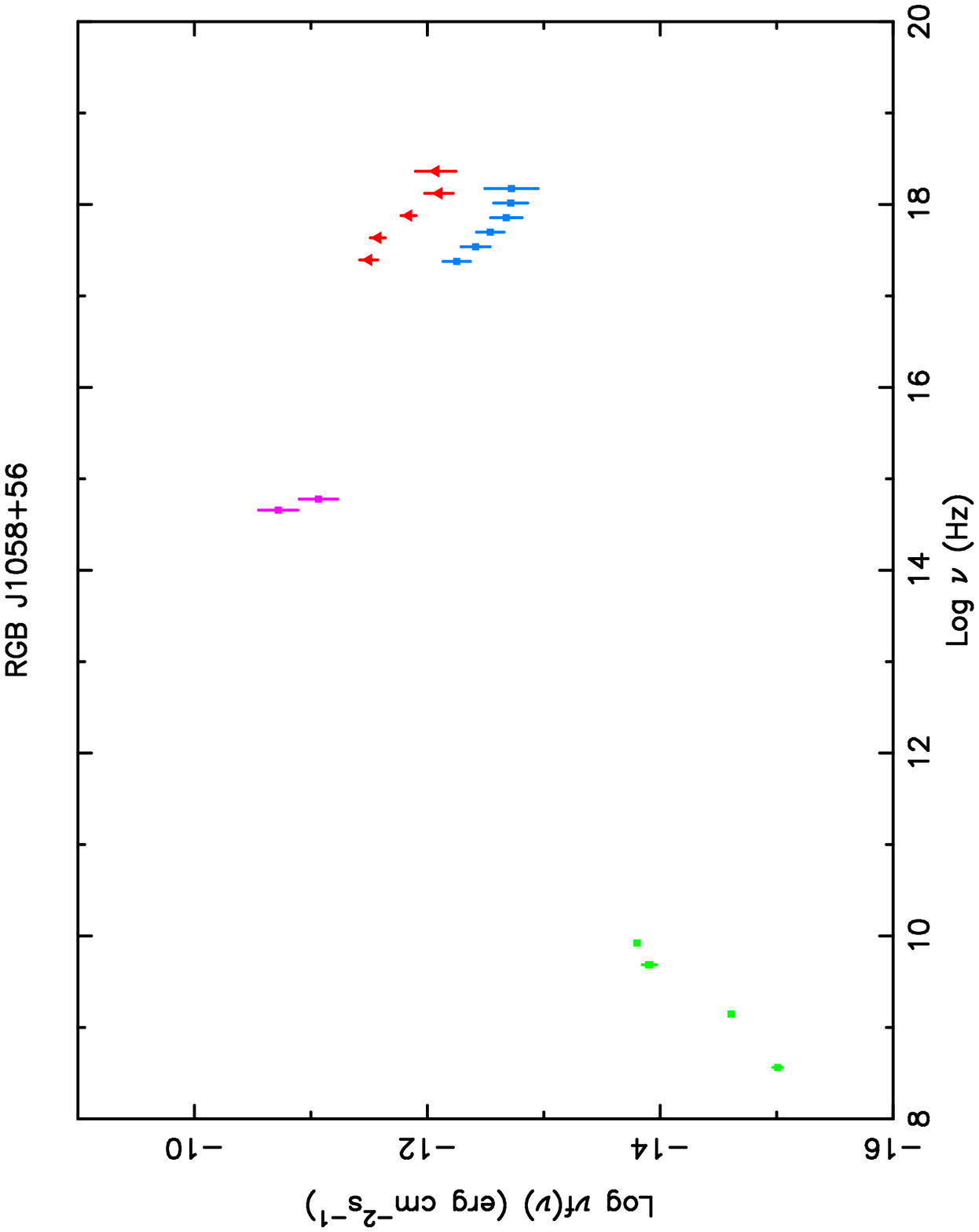} 
\includegraphics{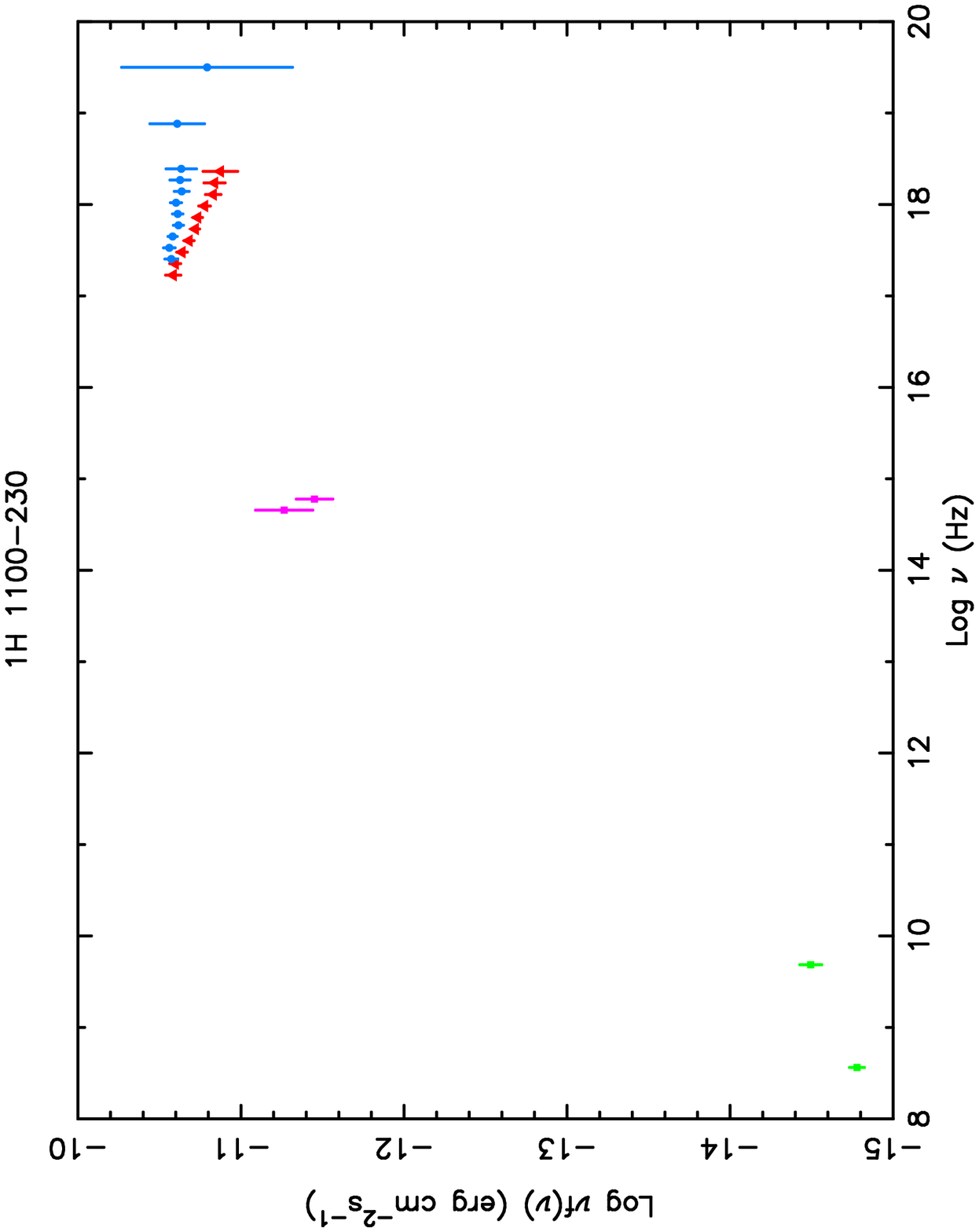} 
\includegraphics{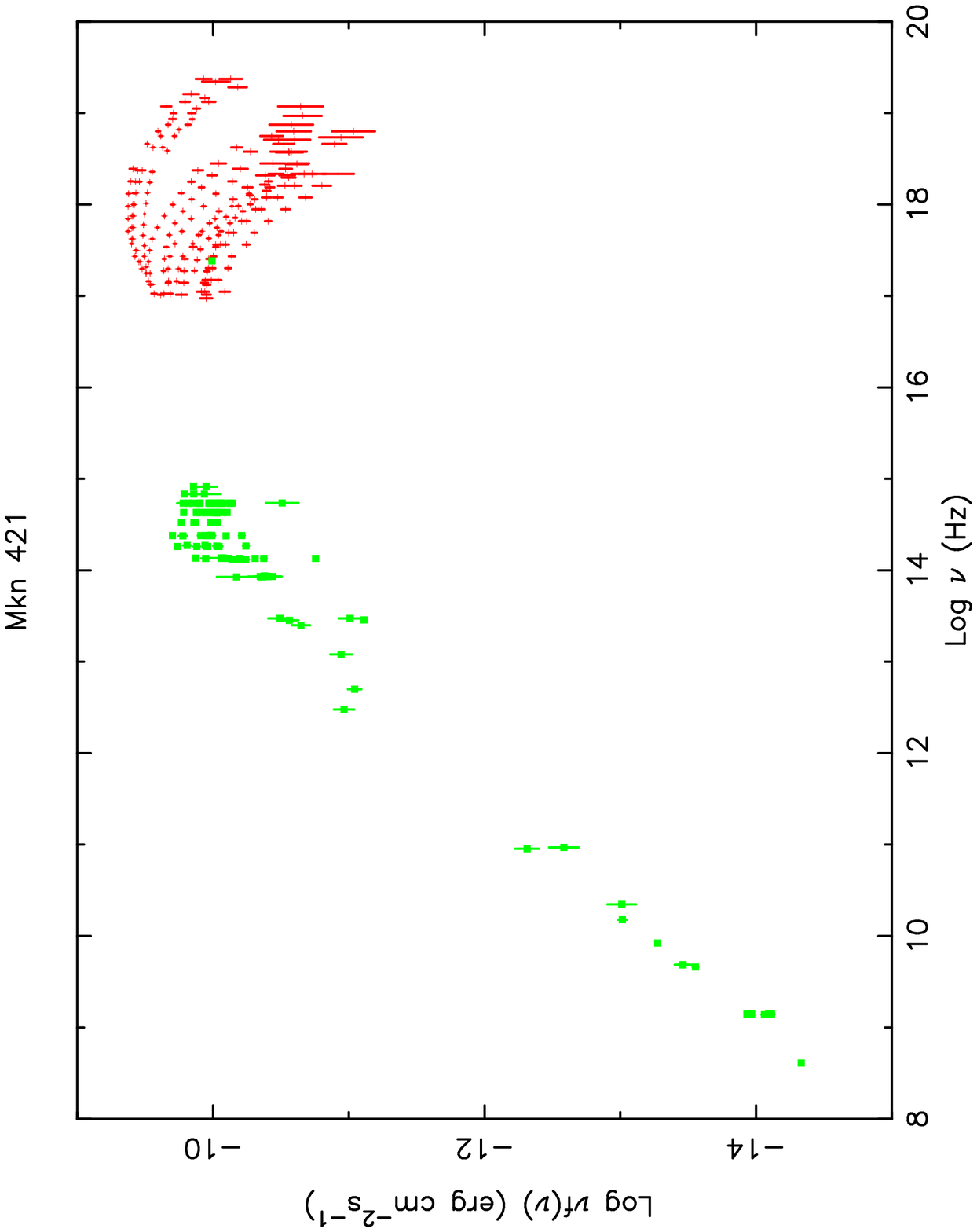} 
\includegraphics{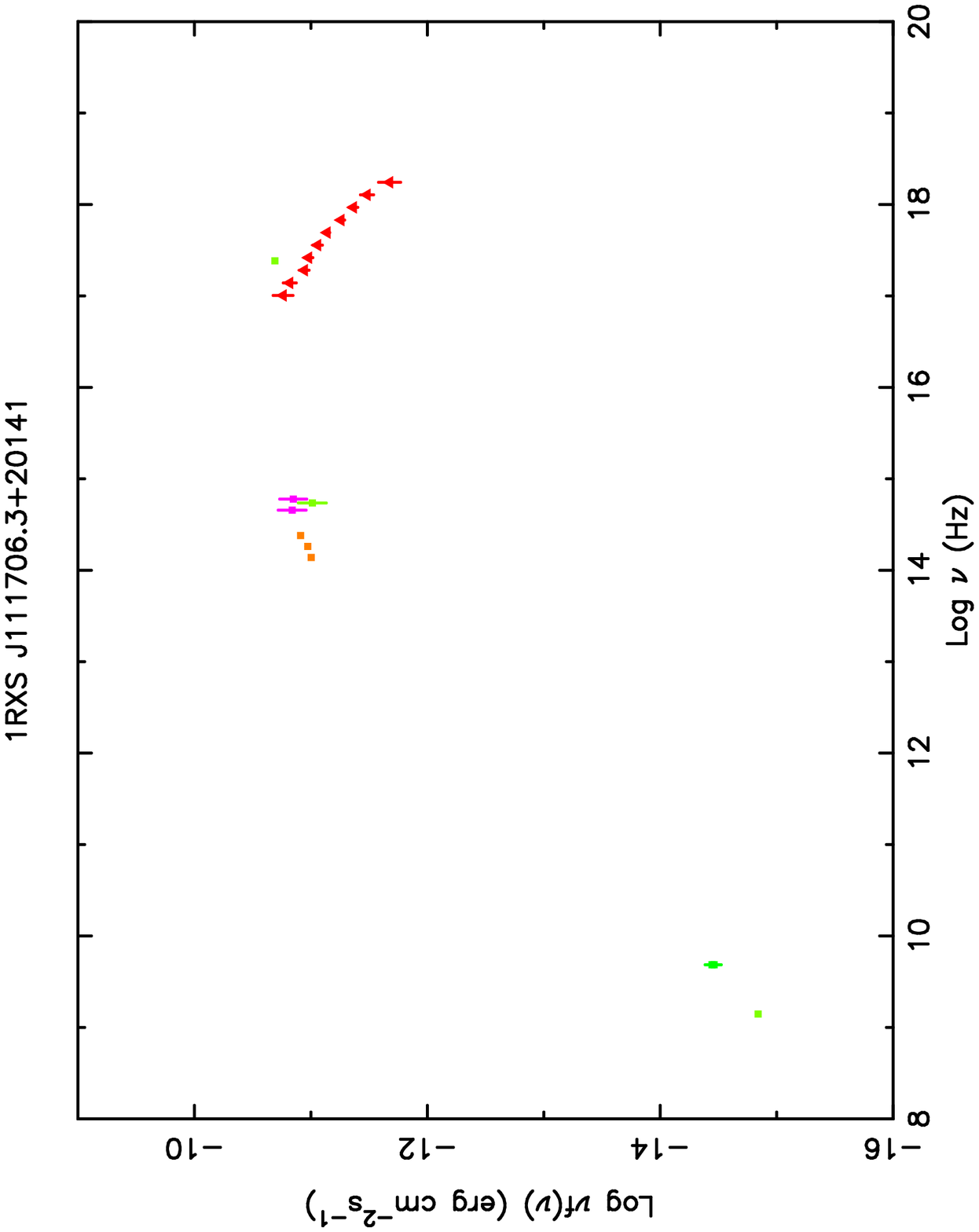} 
\vspace{19.0cm} 
\caption[t]{g- Spectral Energy Distribution of the BL Lacs RGB J1058+56, 1H 1100$-$230,
Mkn 421 and 1RXS J111706.3+20141} 
\label{fig1g} 
\end{figure} 
\clearpage 
\setcounter{figure}{1} 
\begin{figure}[t]
\centering
\includegraphics{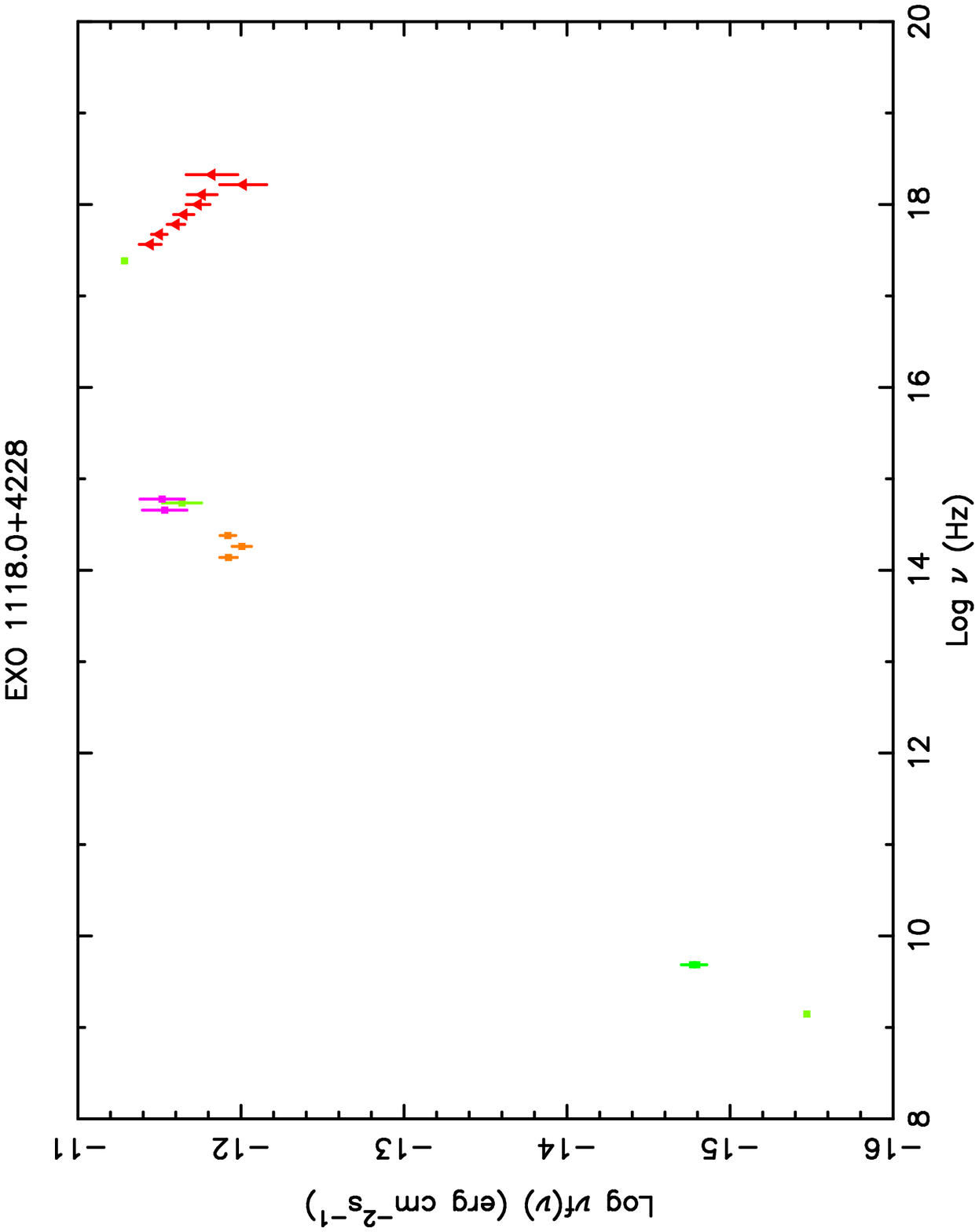} 
\includegraphics{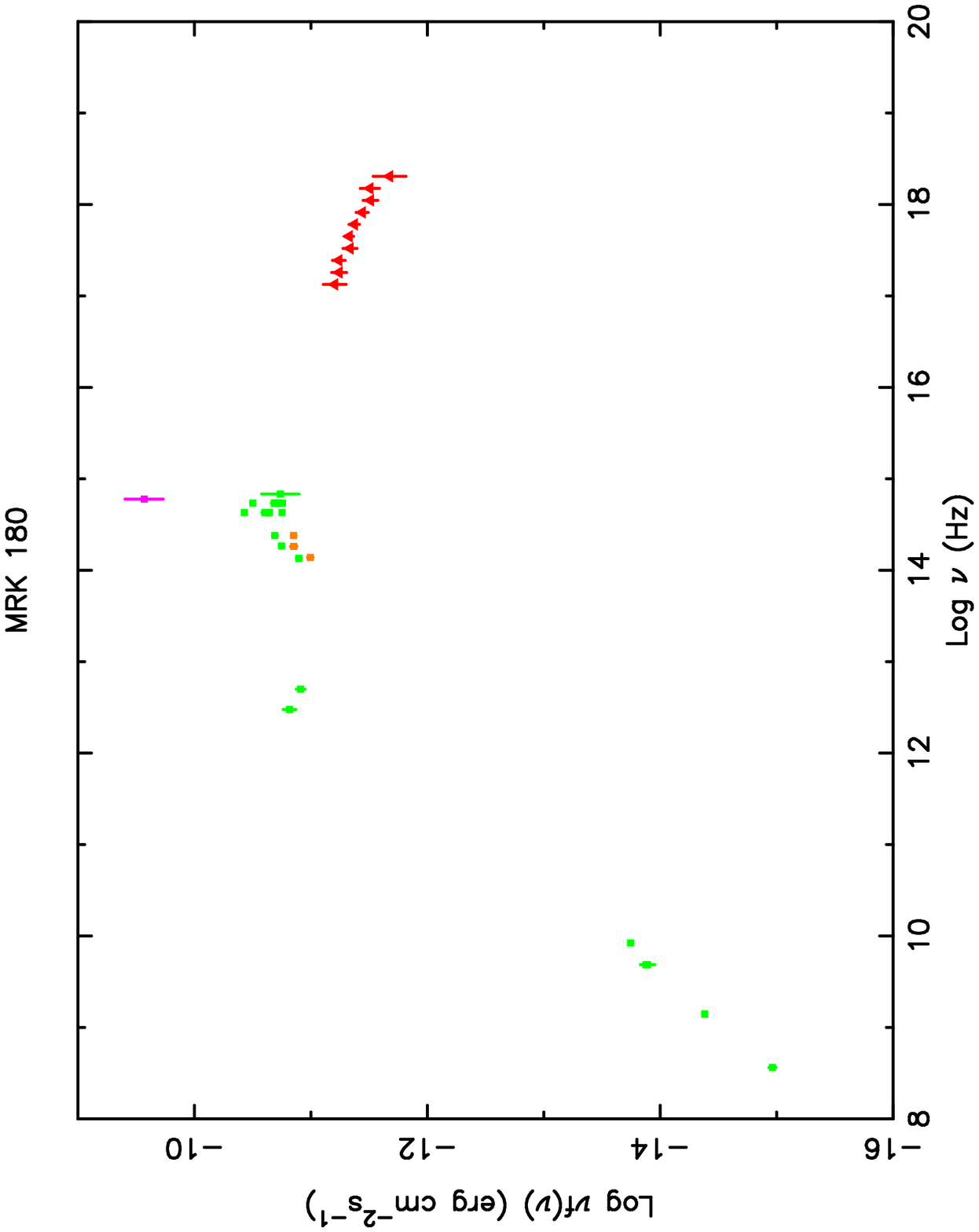} 
\includegraphics{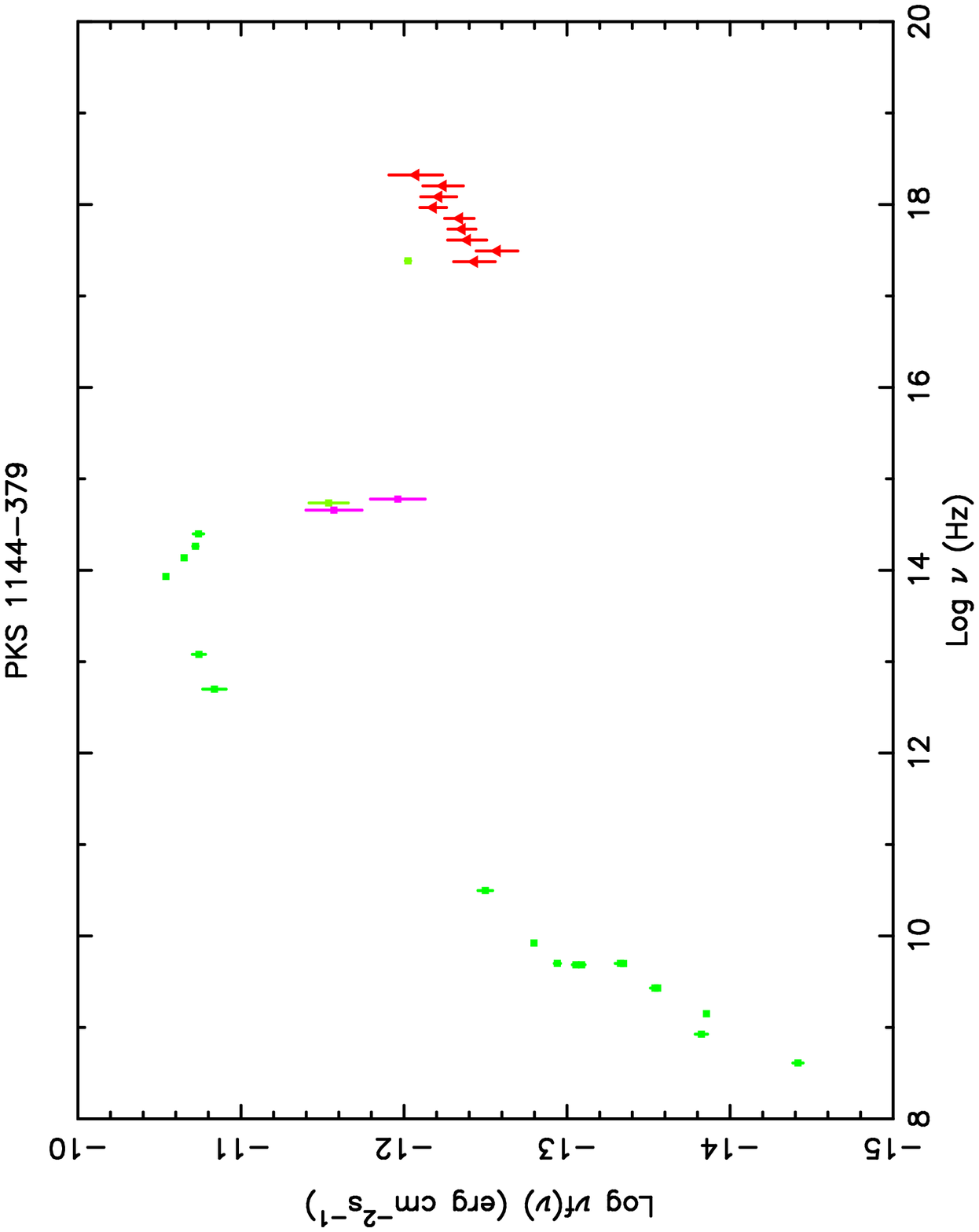} 
\includegraphics{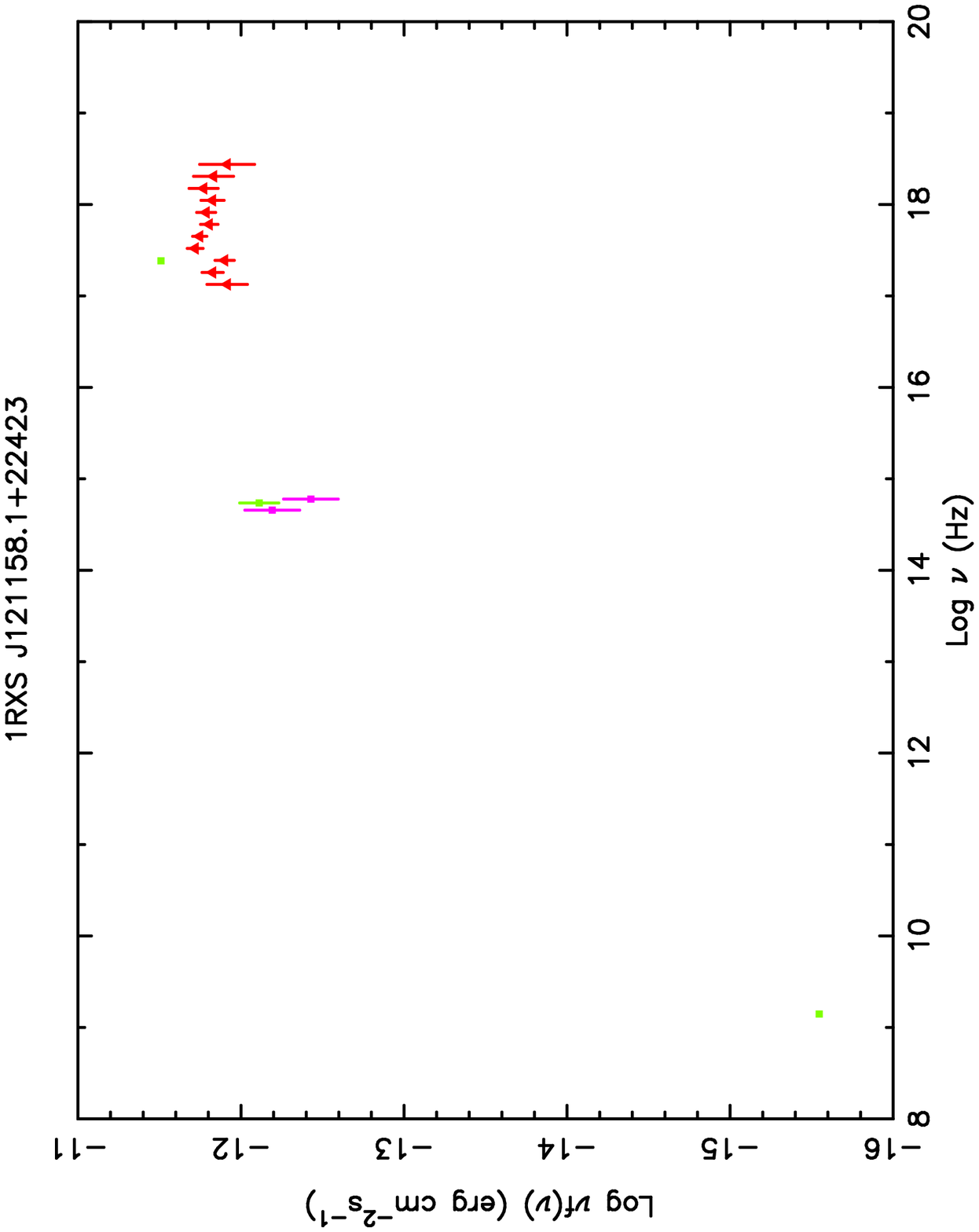} 
\vspace{19.0cm} 
\caption[t]{h- Spectral Energy Distribution of the BL Lacs EXO 1118.0+4228,
Mkn 180, PKS 1144$-$379 and 1RXS J121158.1+22423} 
\label{fig1h} 
\end{figure} 
\clearpage 
\setcounter{figure}{1} 
\begin{figure}[t]
\centering
\includegraphics{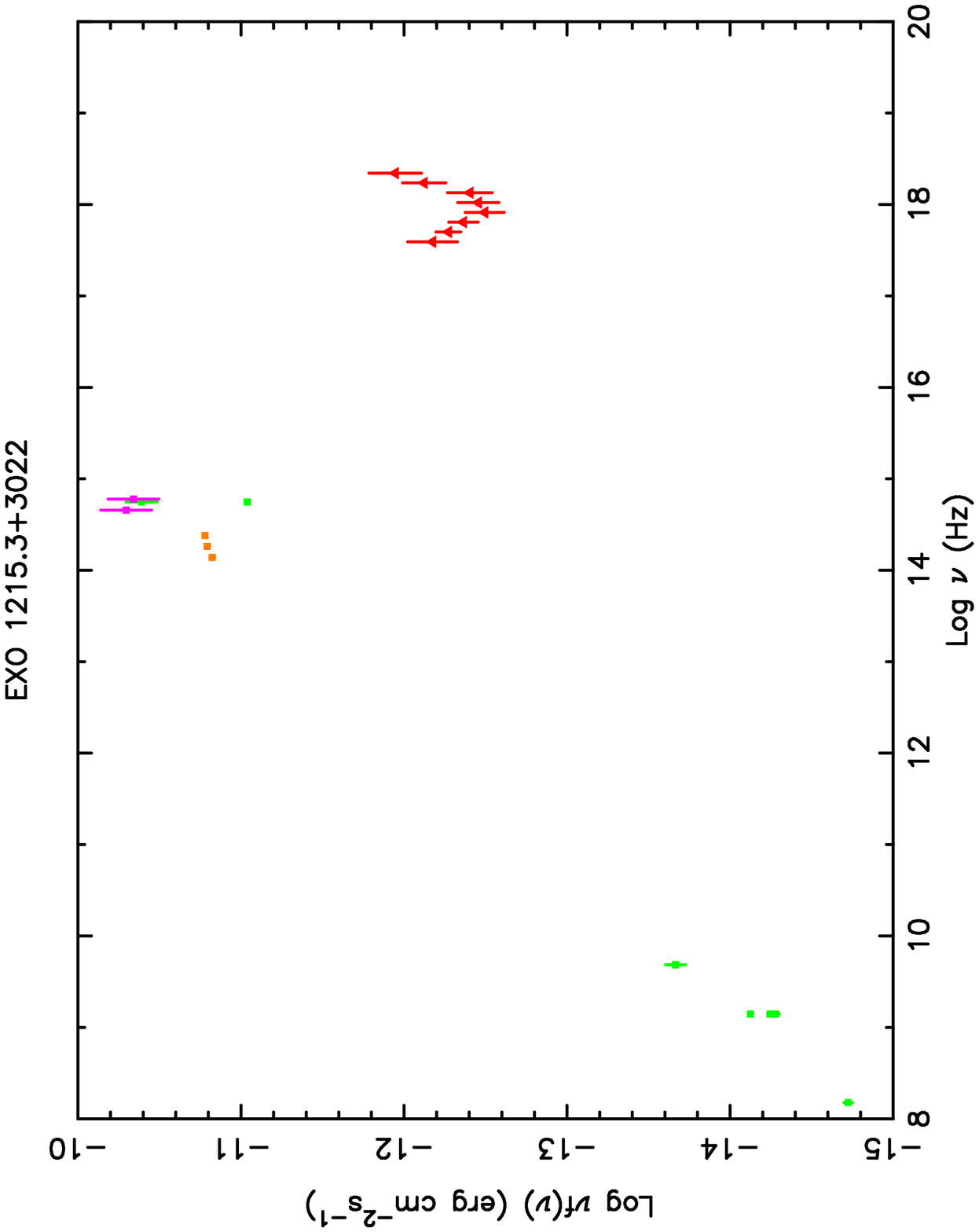} 
\includegraphics{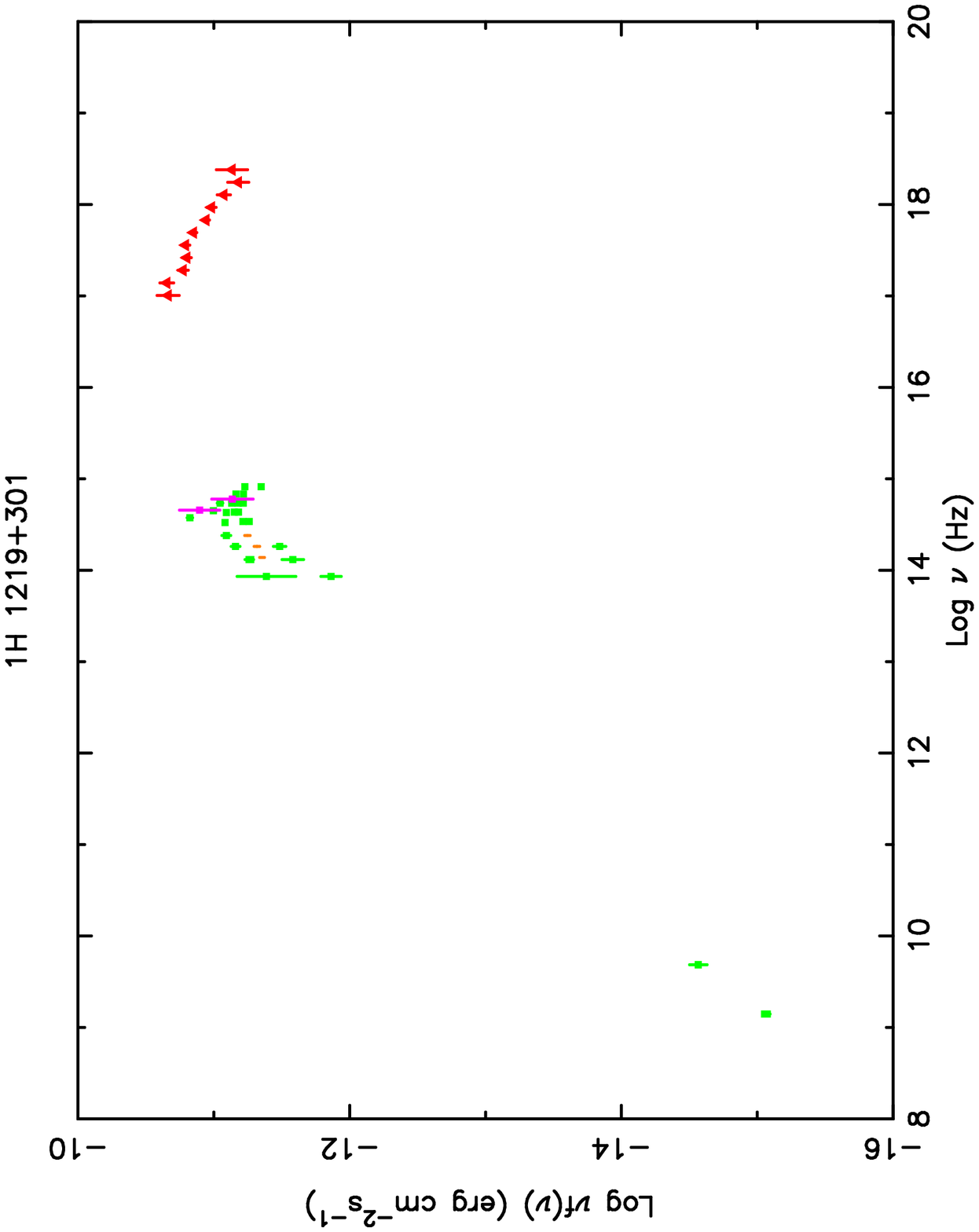} 
\includegraphics{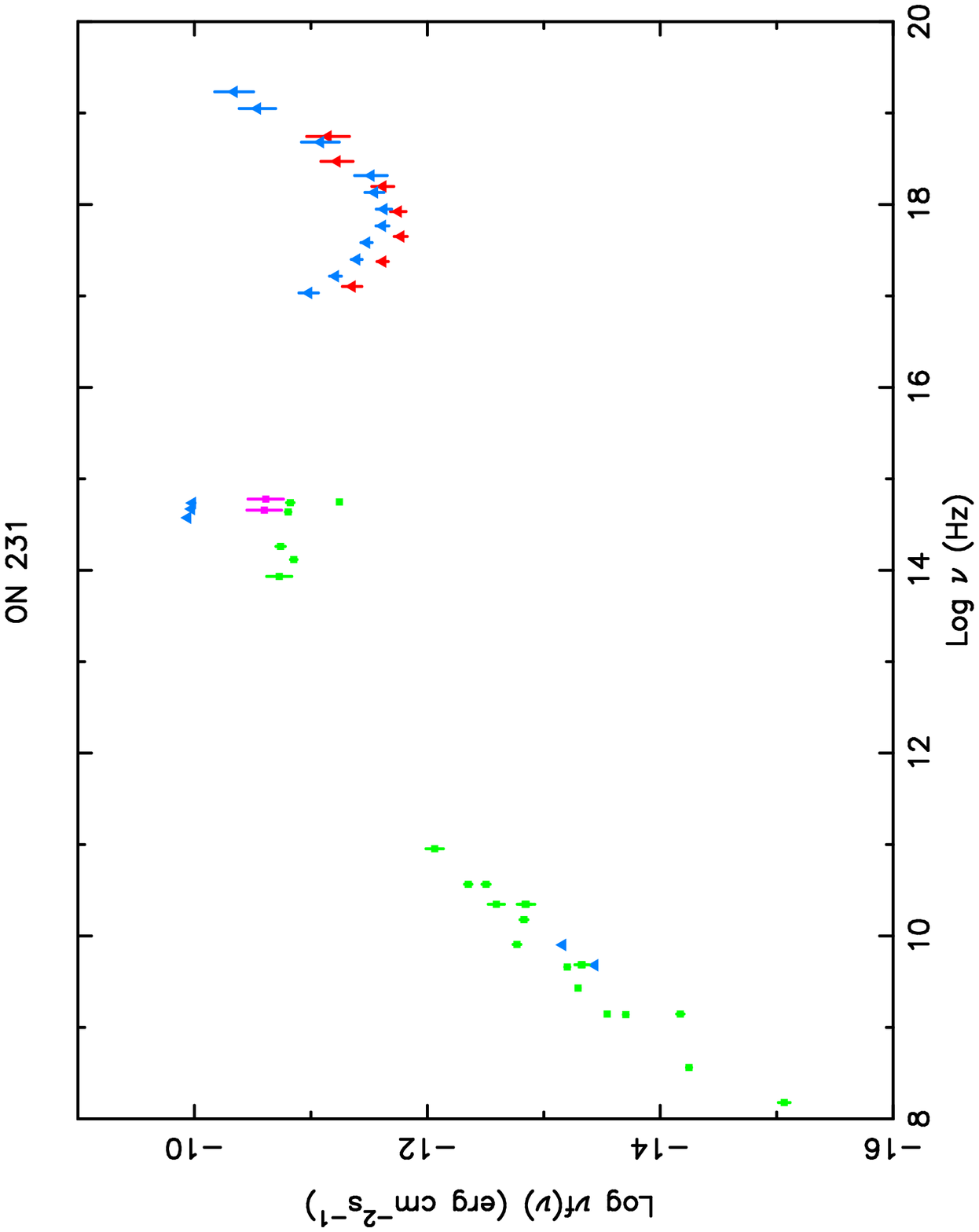} 
\includegraphics{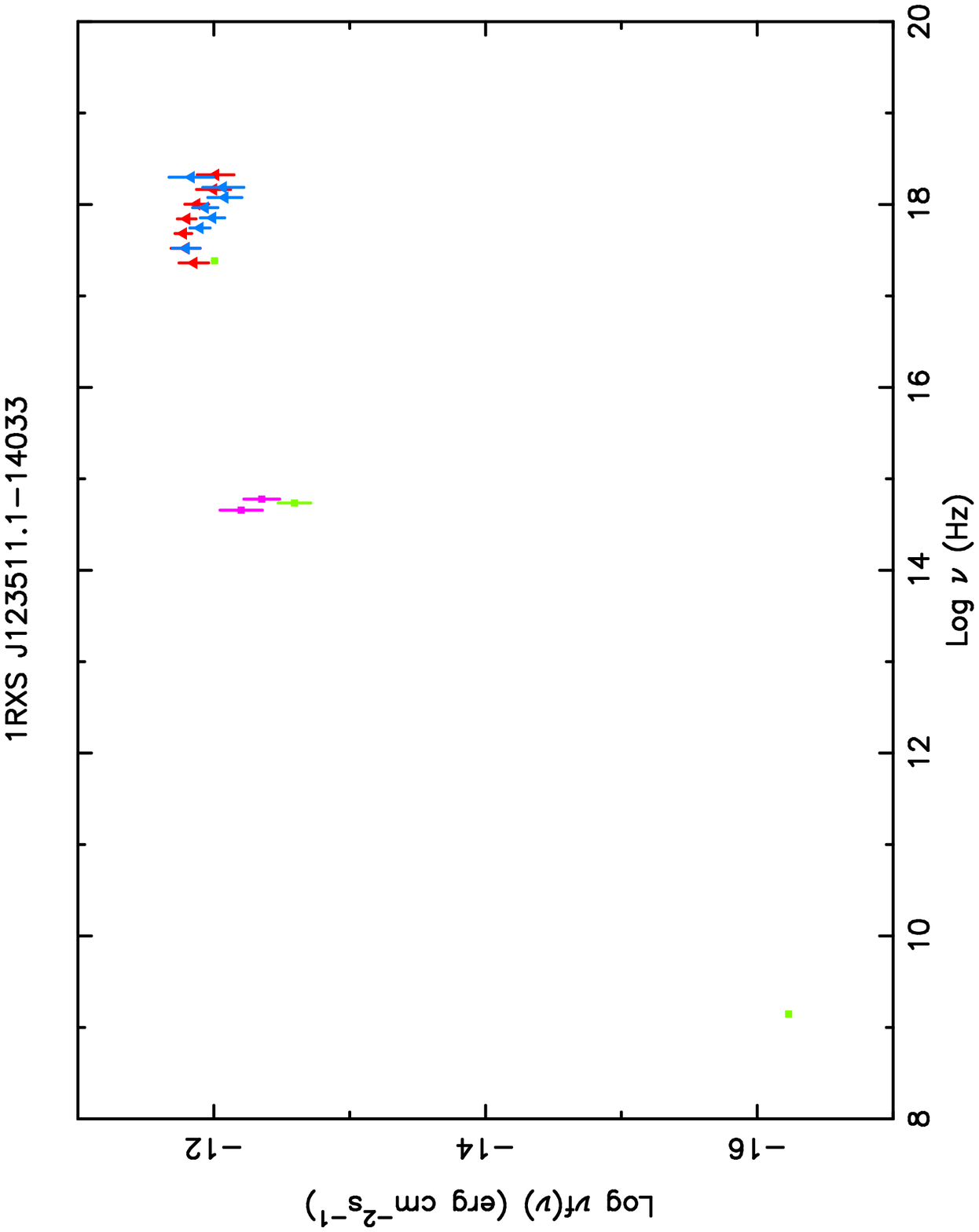} 
\vspace{19.0cm} 
\caption[t]{i- Spectral Energy Distribution of the BL Lacs EXO 1215.3+3022,
1H 1219+301, ON 231 and 1RXS J123511.1$-$14033} 
\label{fig1i} 
\end{figure} 
\clearpage 
\setcounter{figure}{1} 
\begin{figure}[t]
\centering
\includegraphics{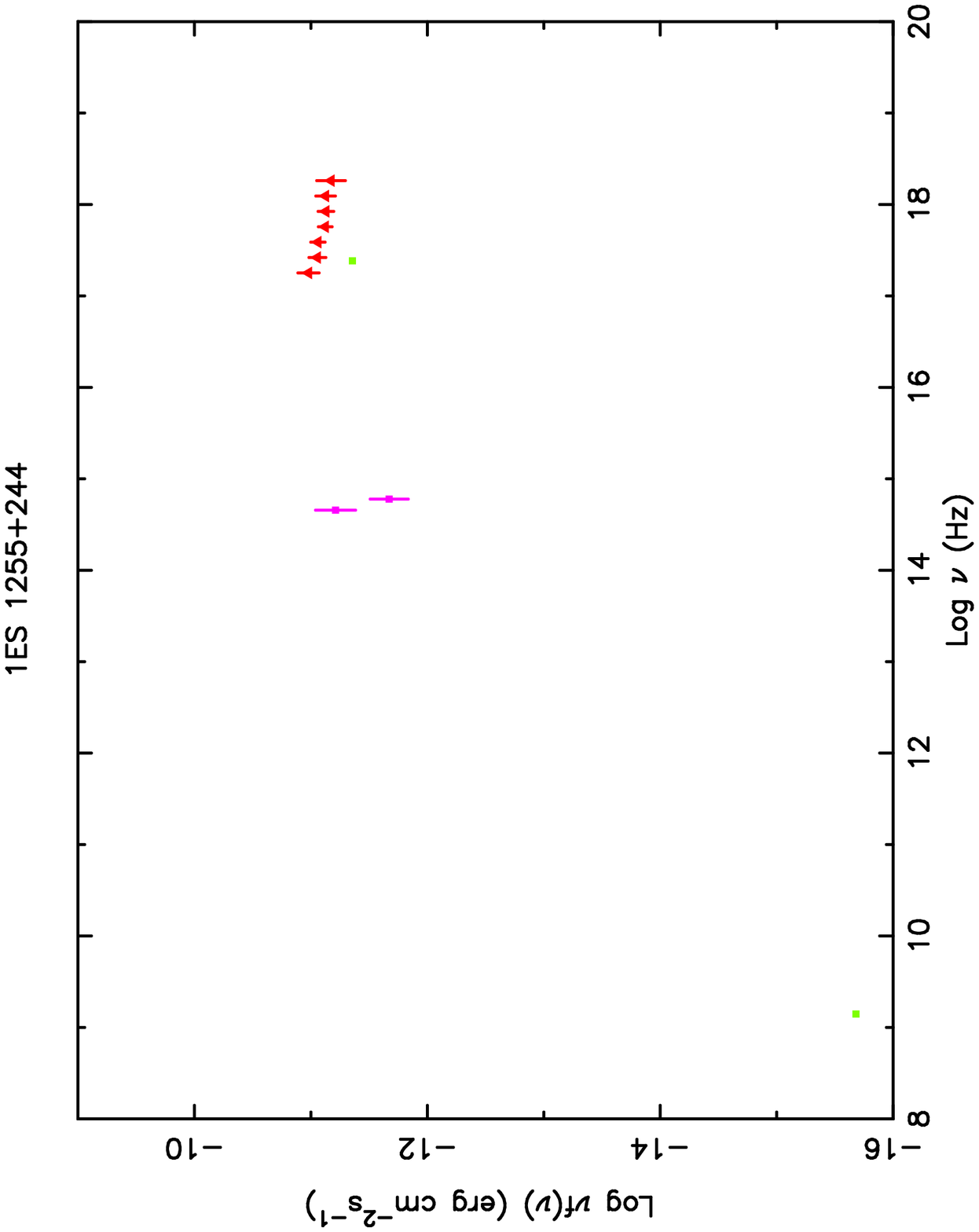} 
\includegraphics{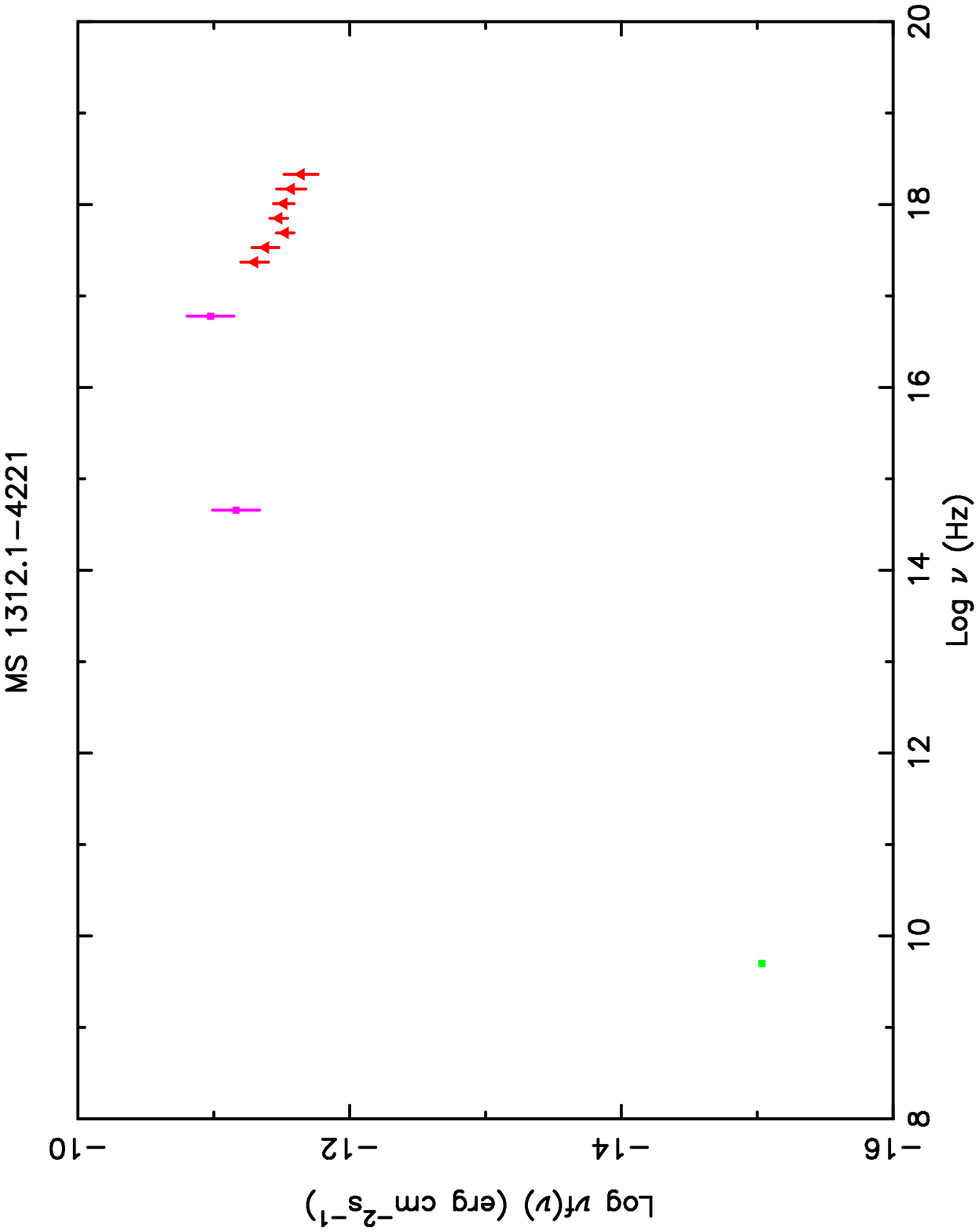} 
\includegraphics{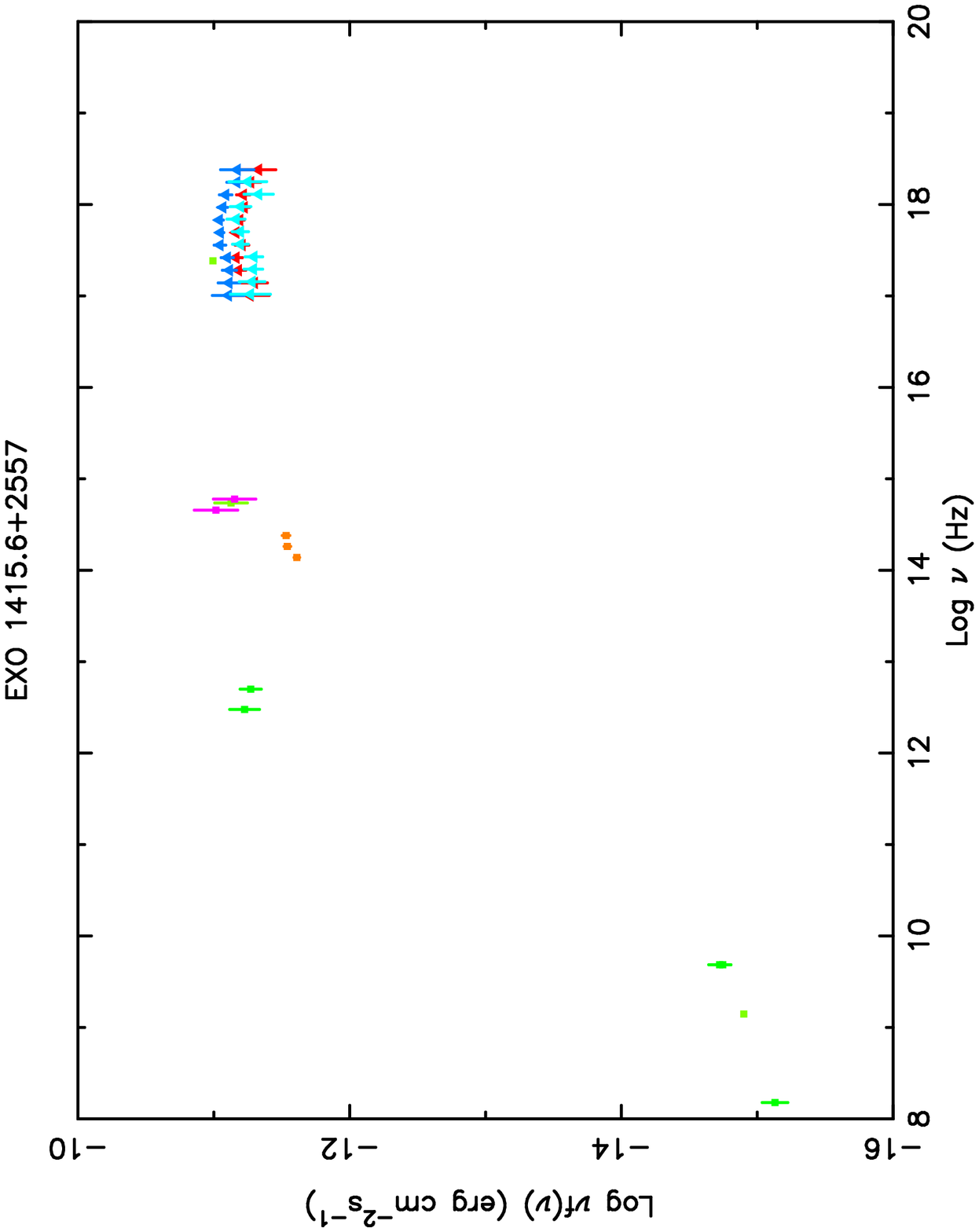} 
\includegraphics{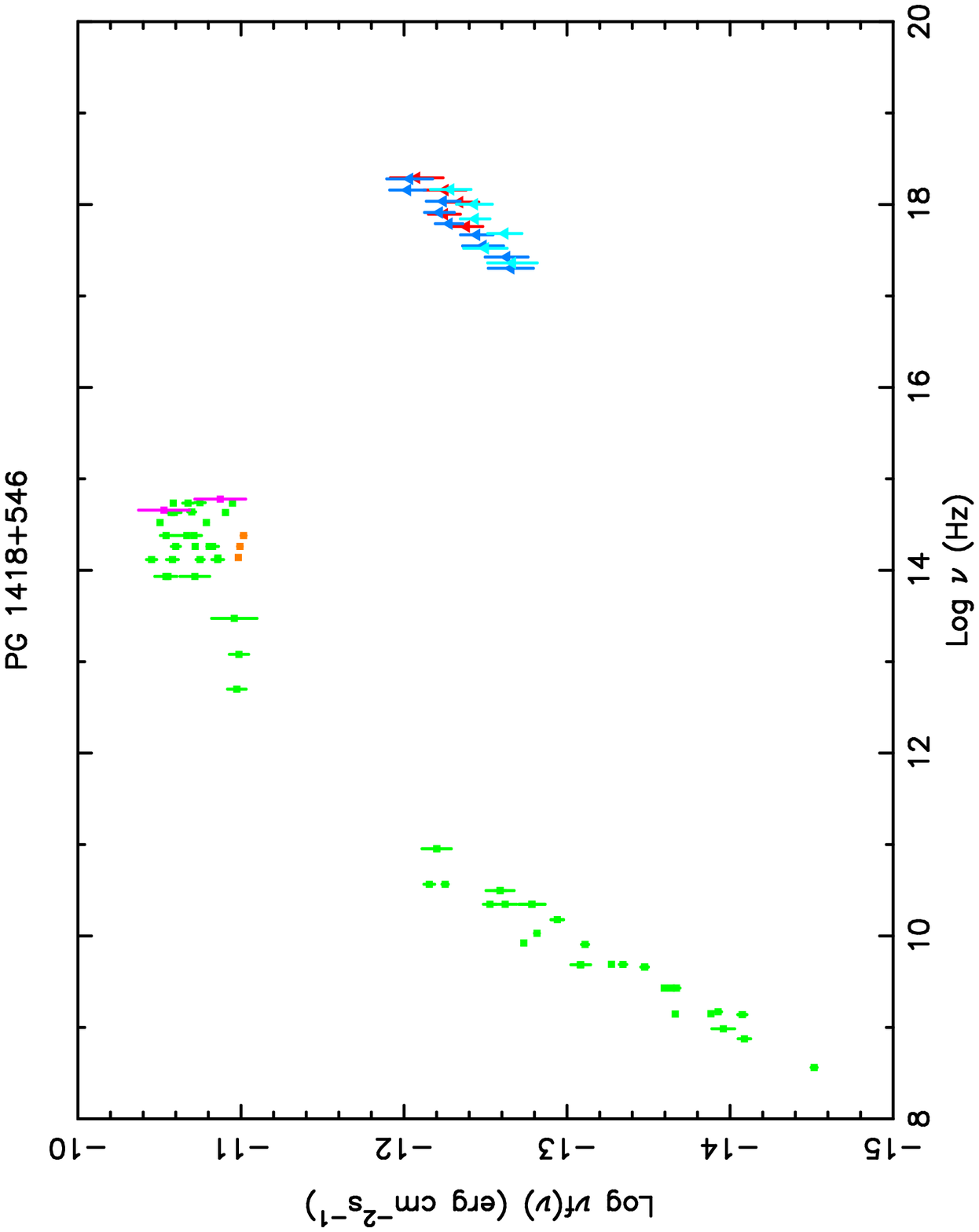} 
\vspace{19.0cm} 
\caption[t]{j- Spectral Energy Distribution of the BL Lacs 1ES 1255+244, 
MS 1312.1$-$4221, EXO~1415.6+2557 and PG~1418+546} 
\label{fig1j} 
\end{figure} 
\clearpage 
\setcounter{figure}{1} 
\begin{figure}[t]
\centering
\includegraphics{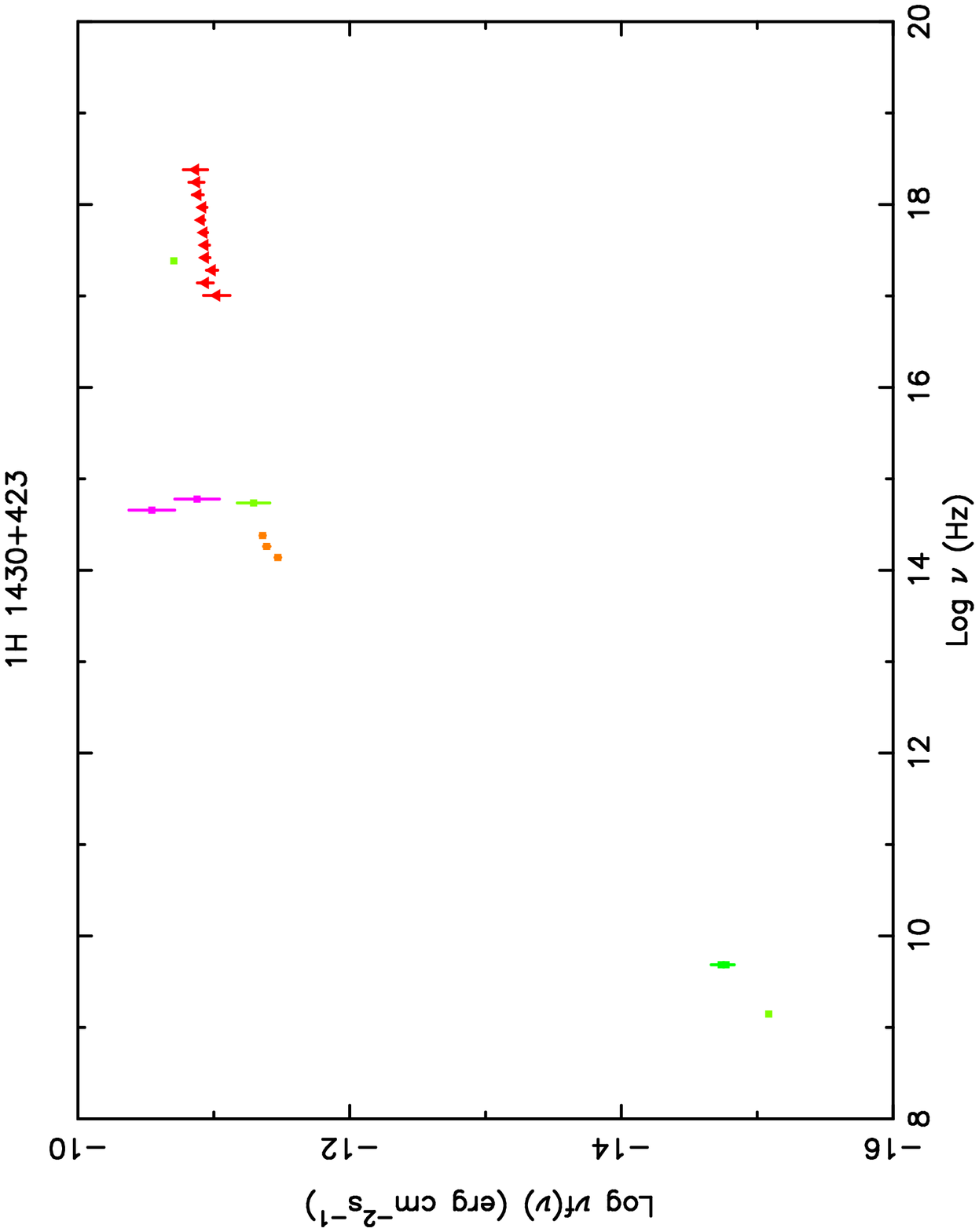} 
\includegraphics{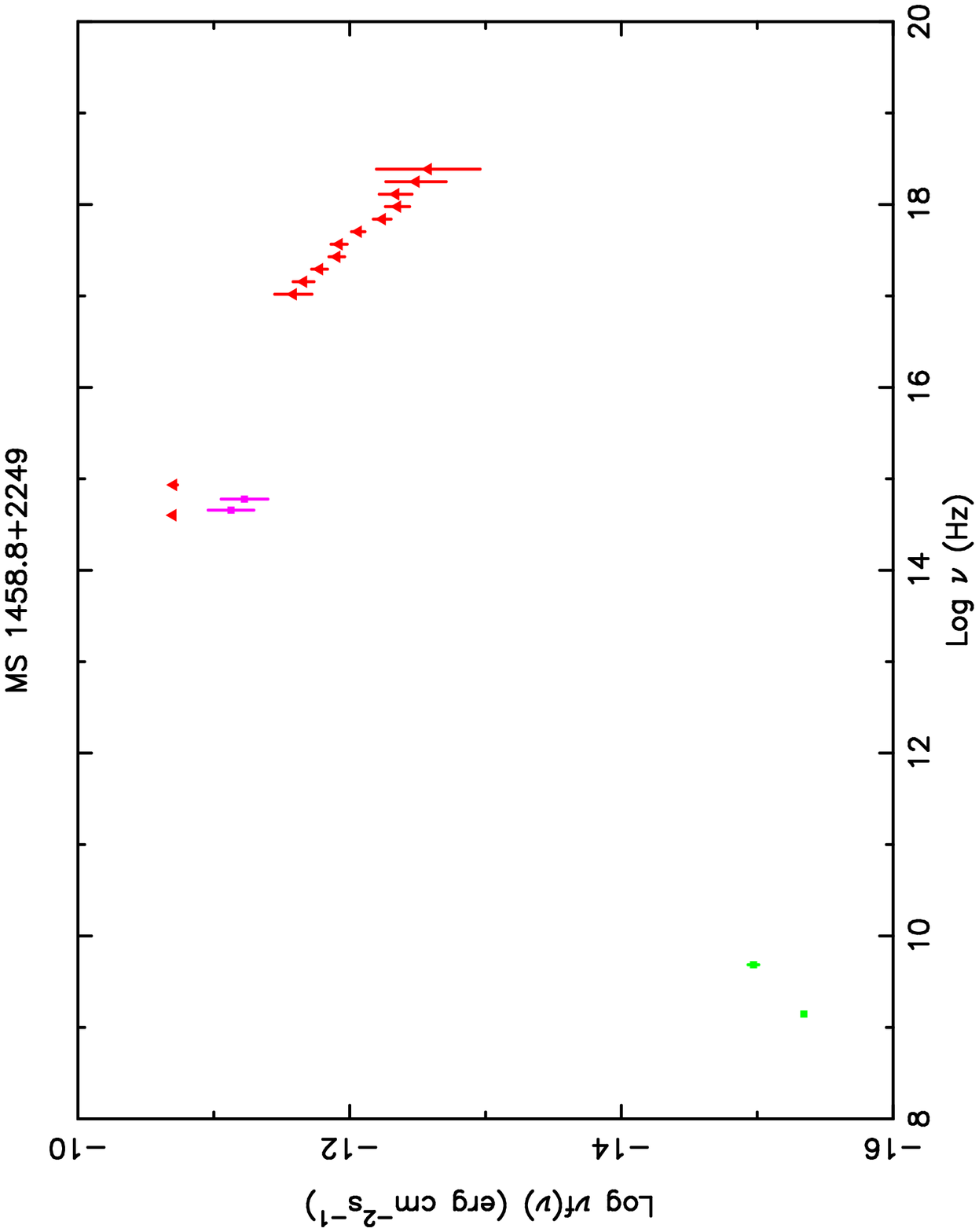} 
\includegraphics{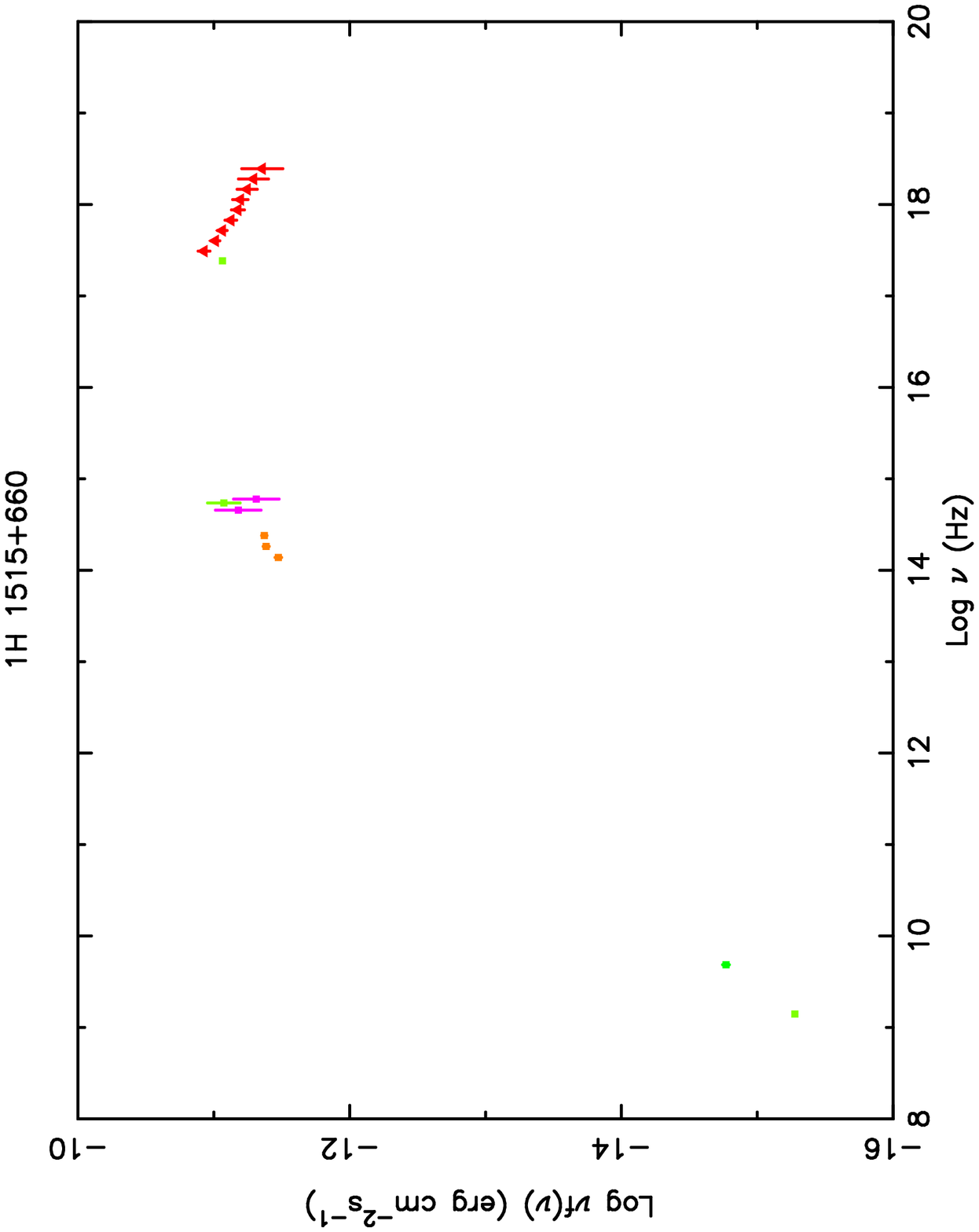} 
\includegraphics{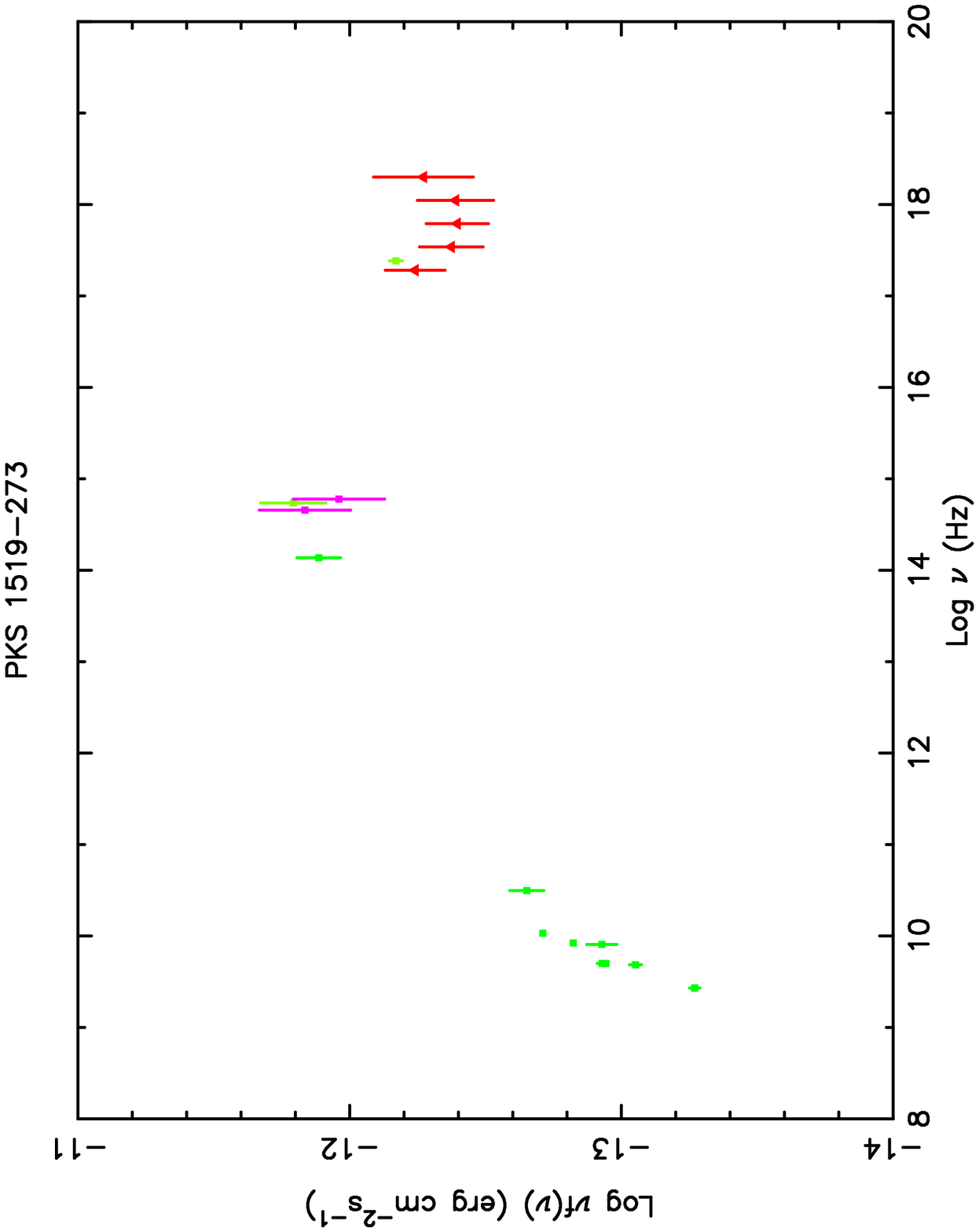} 
\vspace{19.0cm} 
\caption[t]{k- Spectral Energy Distribution of the BL Lacs 1H 1430+423,
MS 1458.8+2249, 1H 1515+660 and PKS 1519$-$273} 
\label{fig1k} 
\end{figure} 
\clearpage 
\setcounter{figure}{1} 
\begin{figure}[t]
\centering
\includegraphics{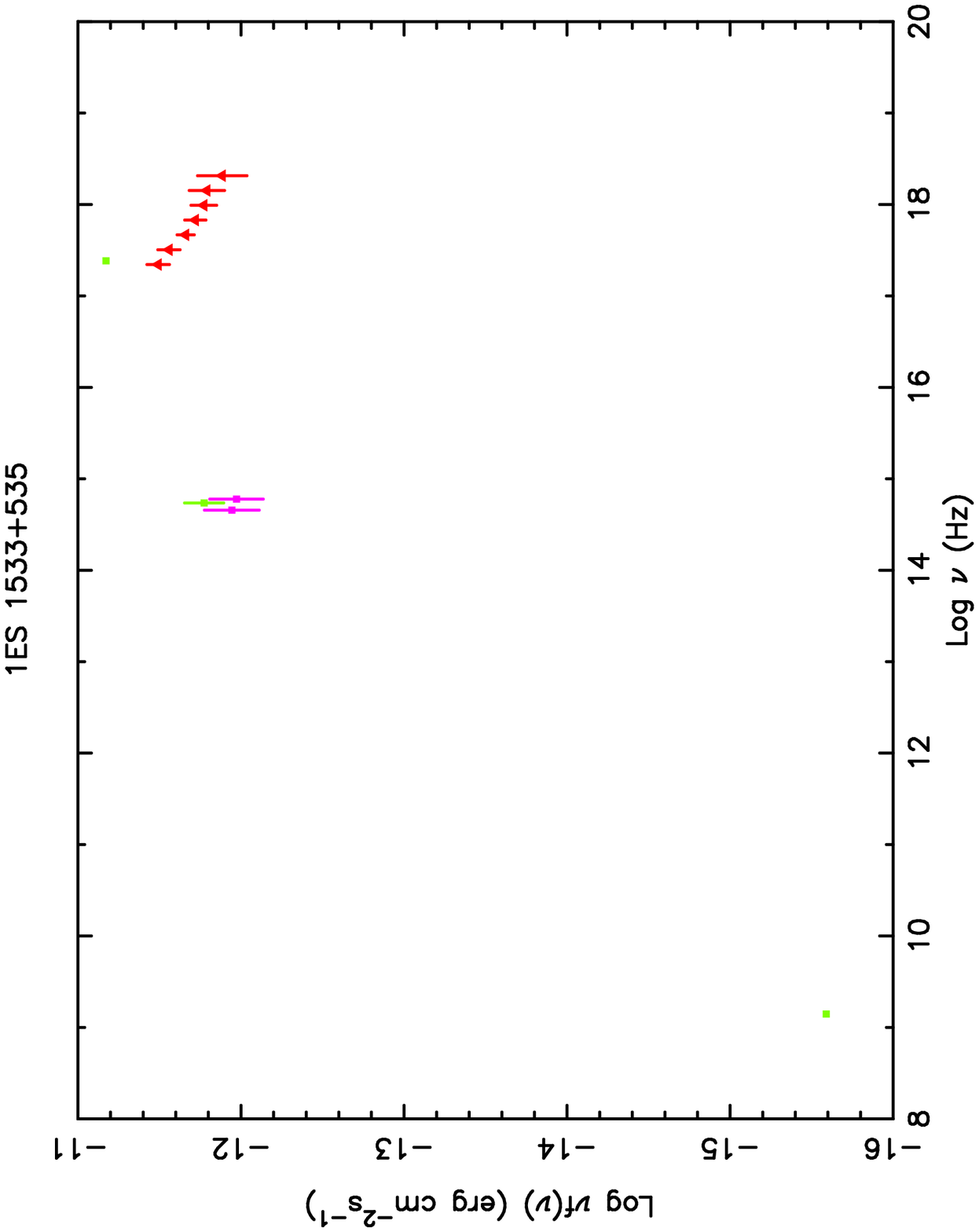} 
\includegraphics{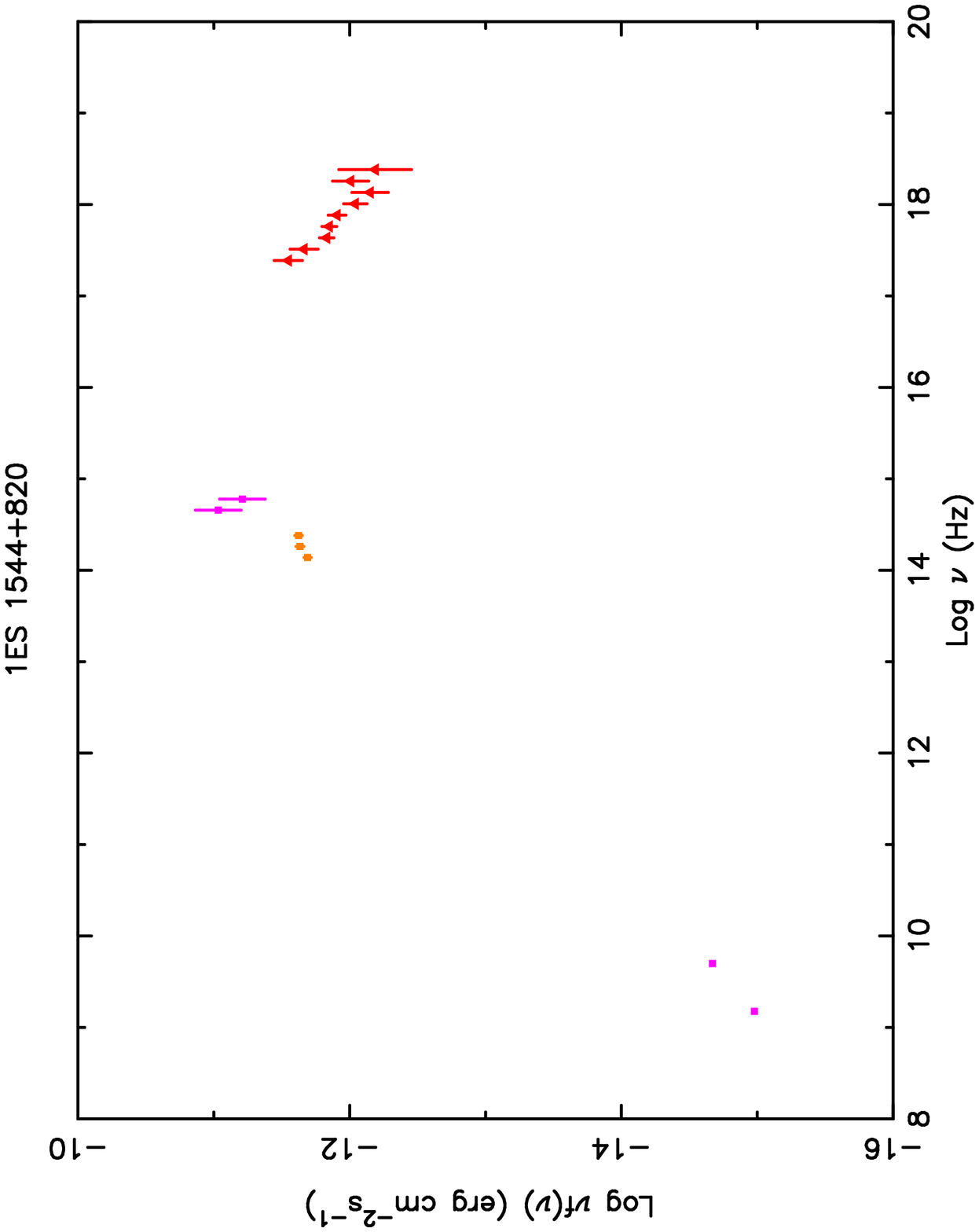} 
\includegraphics{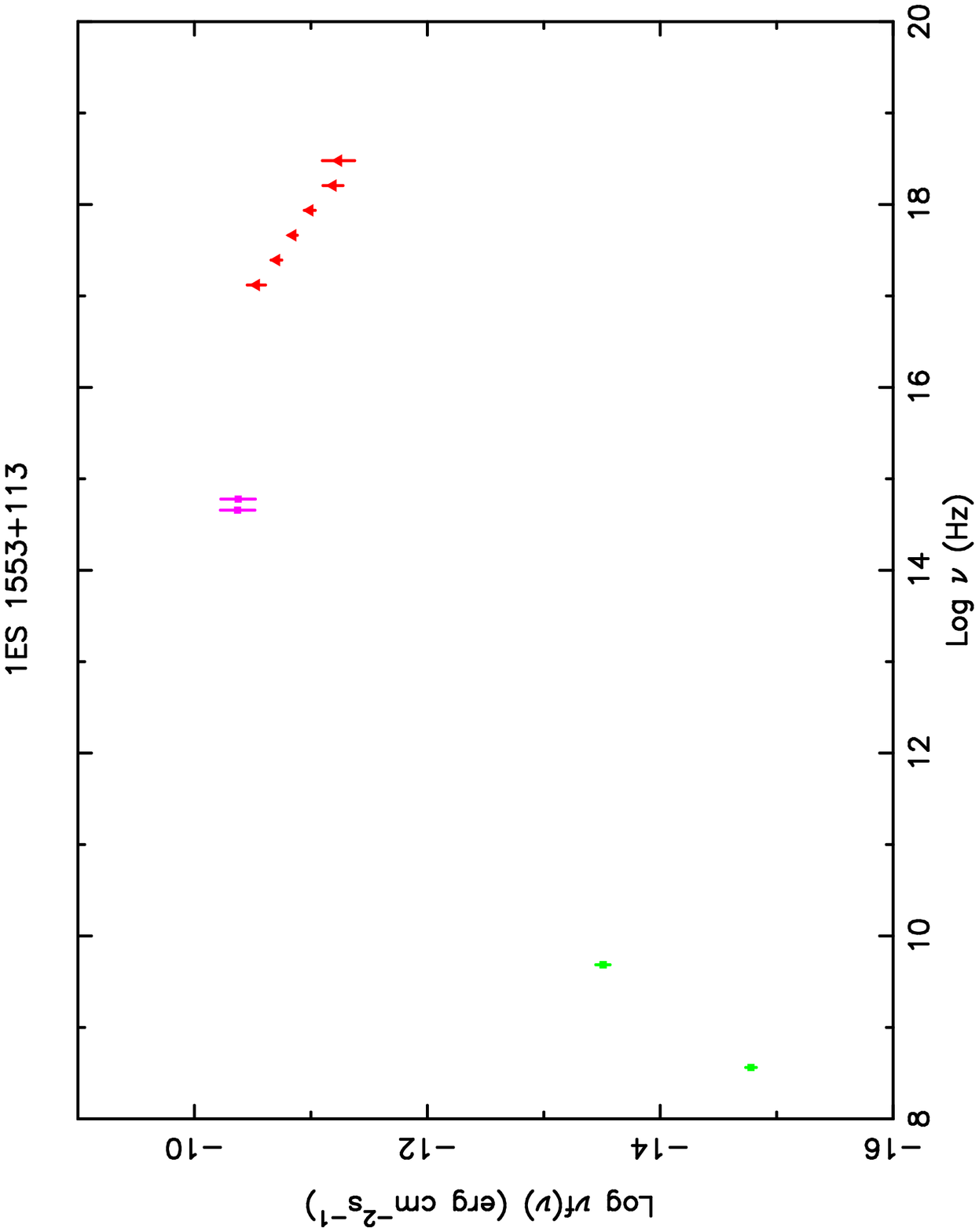} 
\includegraphics{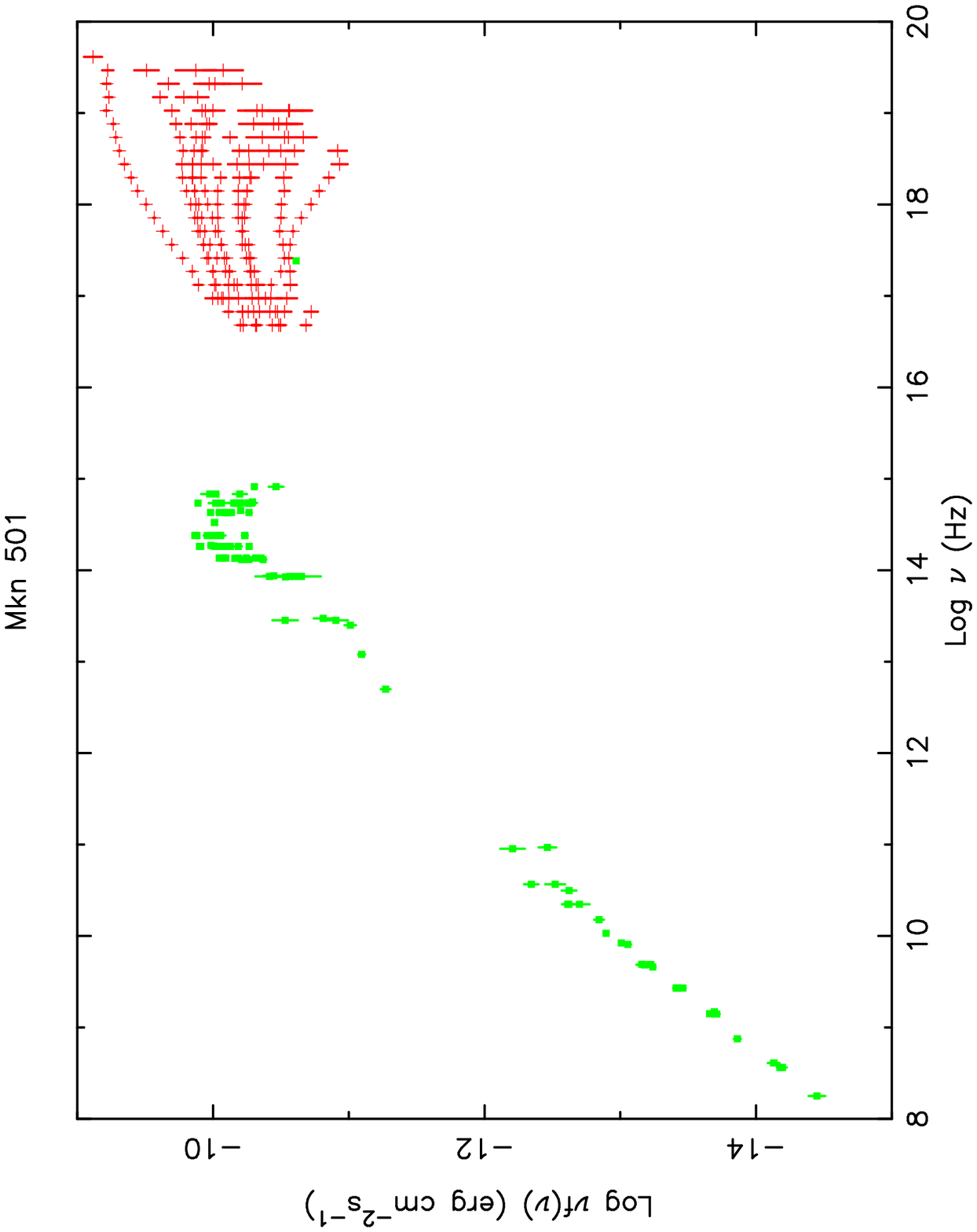} 
\vspace{19.0cm} 
\caption[t]{l- Spectral Energy Distribution of the BL Lacs 1ES 1533+535,
1ES 1544+820, 1ES 1553+113 and Mkn 501} 
\label{fig1l} 
\end{figure} 
\clearpage 
\setcounter{figure}{1} 
\begin{figure}[t]
\centering
\includegraphics{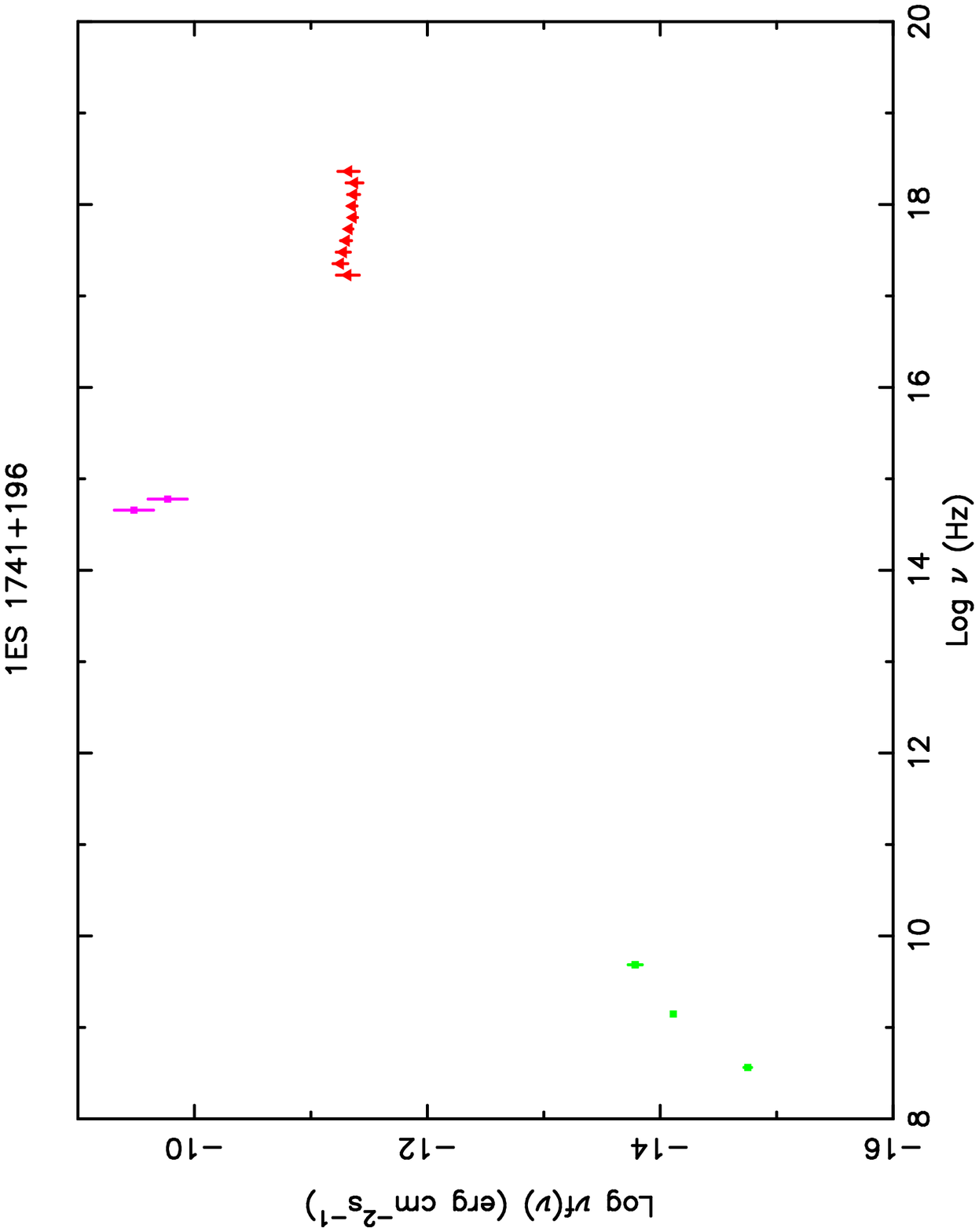} 
\includegraphics{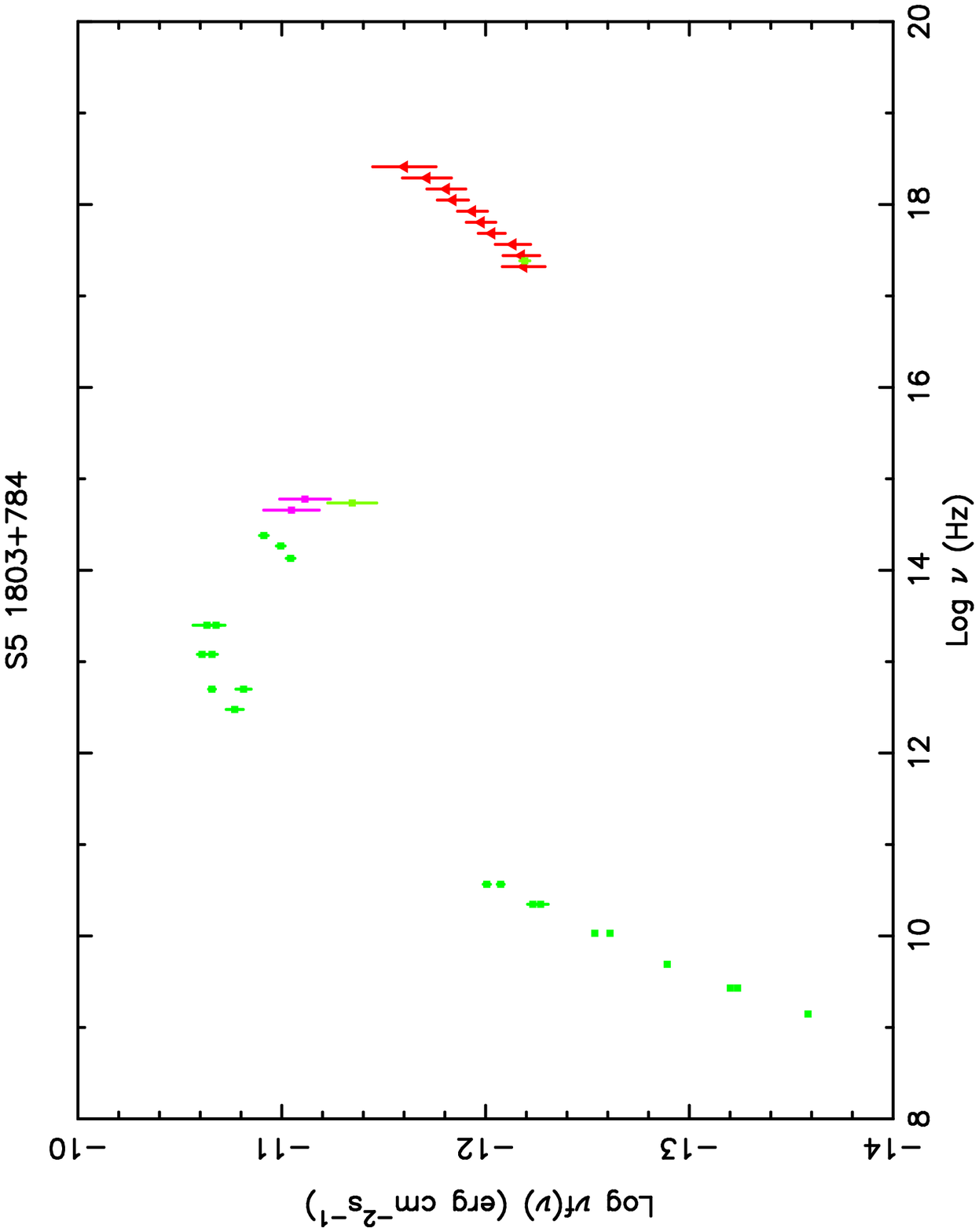} 
\includegraphics{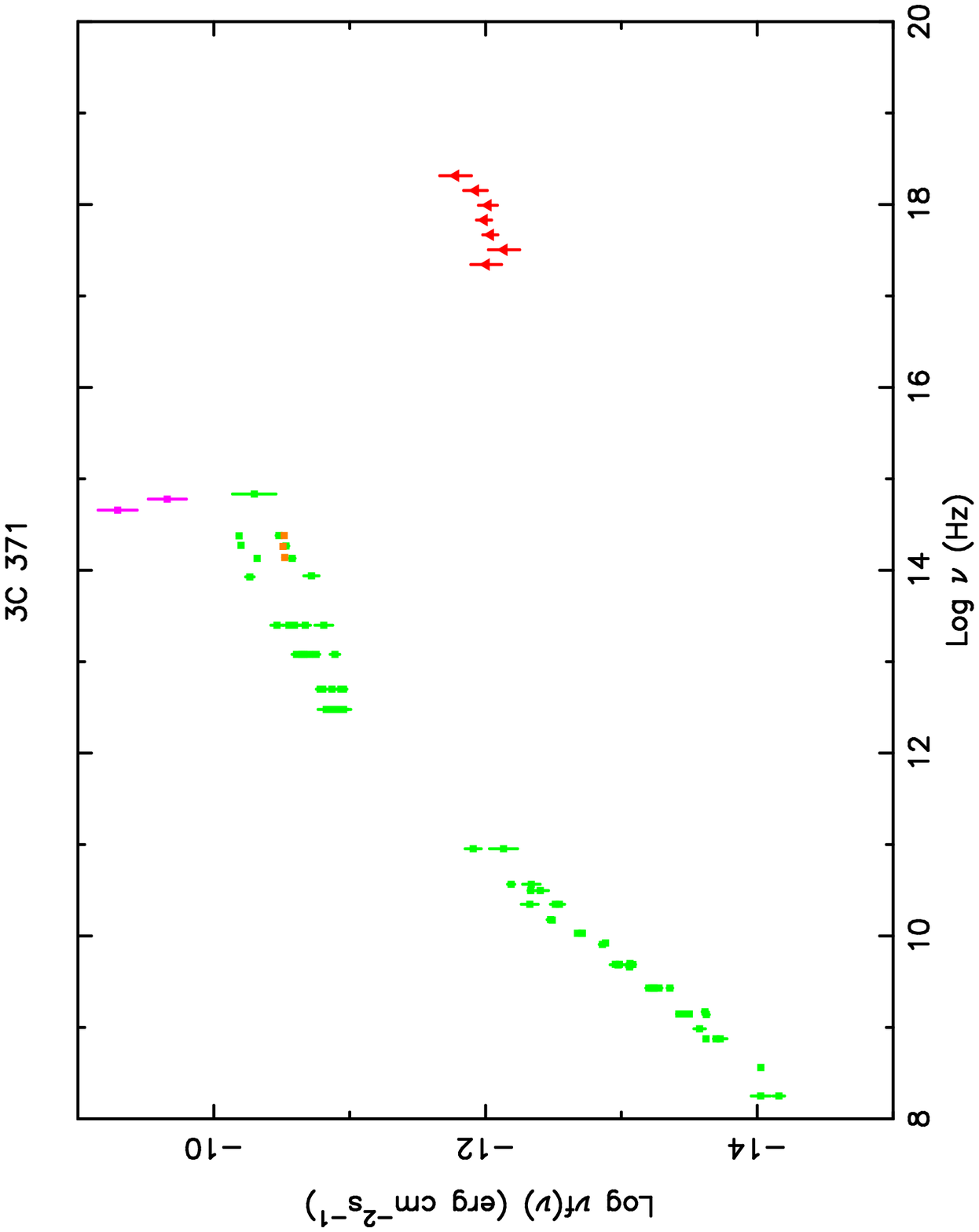} 
\includegraphics{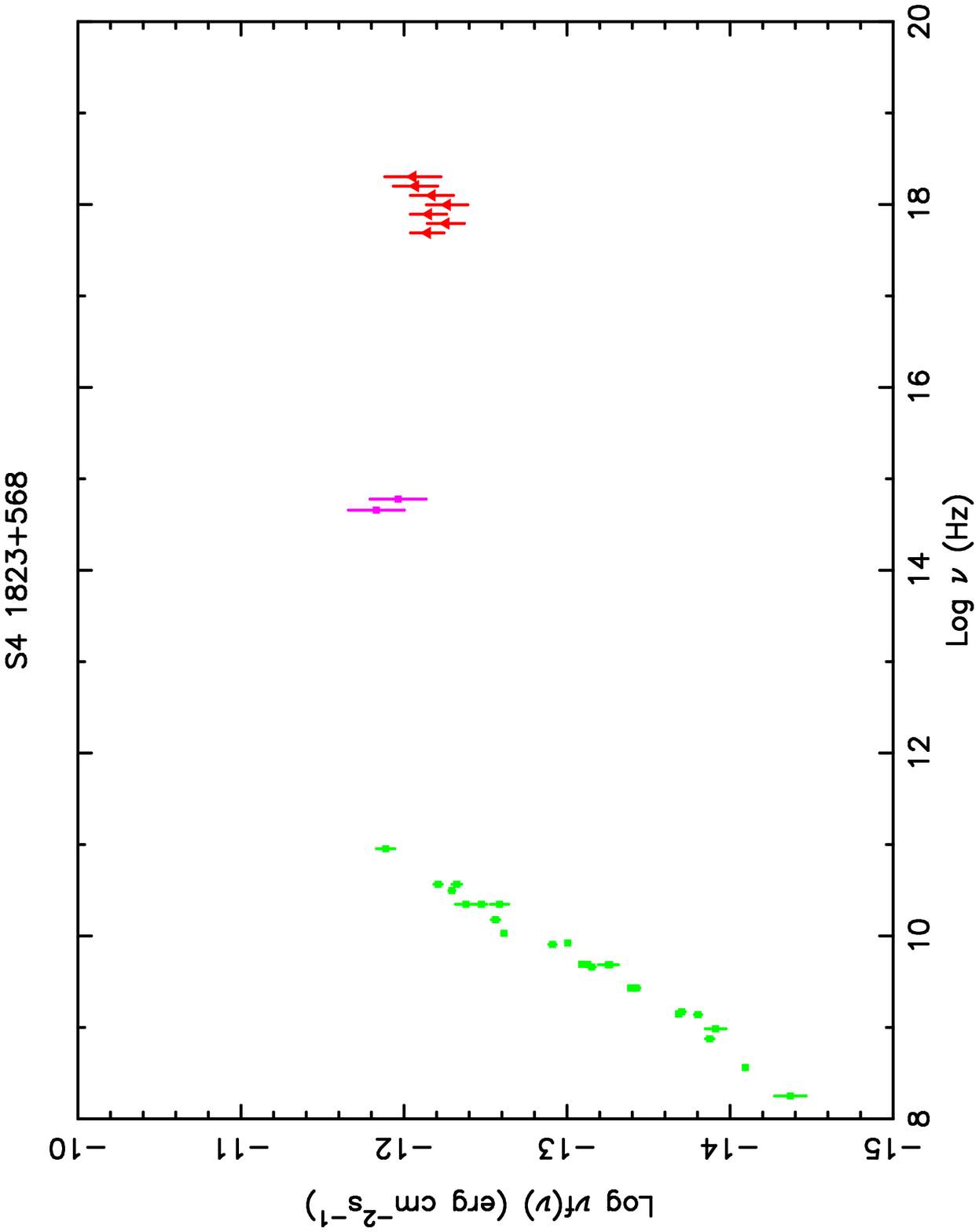} 
\vspace{19.0cm} 
\caption[t]{m- Spectral Energy Distribution of the BL Lacs 1ES 1741+196,
S5 1803+784, 3C 371 and S4 1823+568} 
\label{fig1m} 
\end{figure} 
\clearpage 
\setcounter{figure}{1} 
\begin{figure}[t]
\centering
\includegraphics{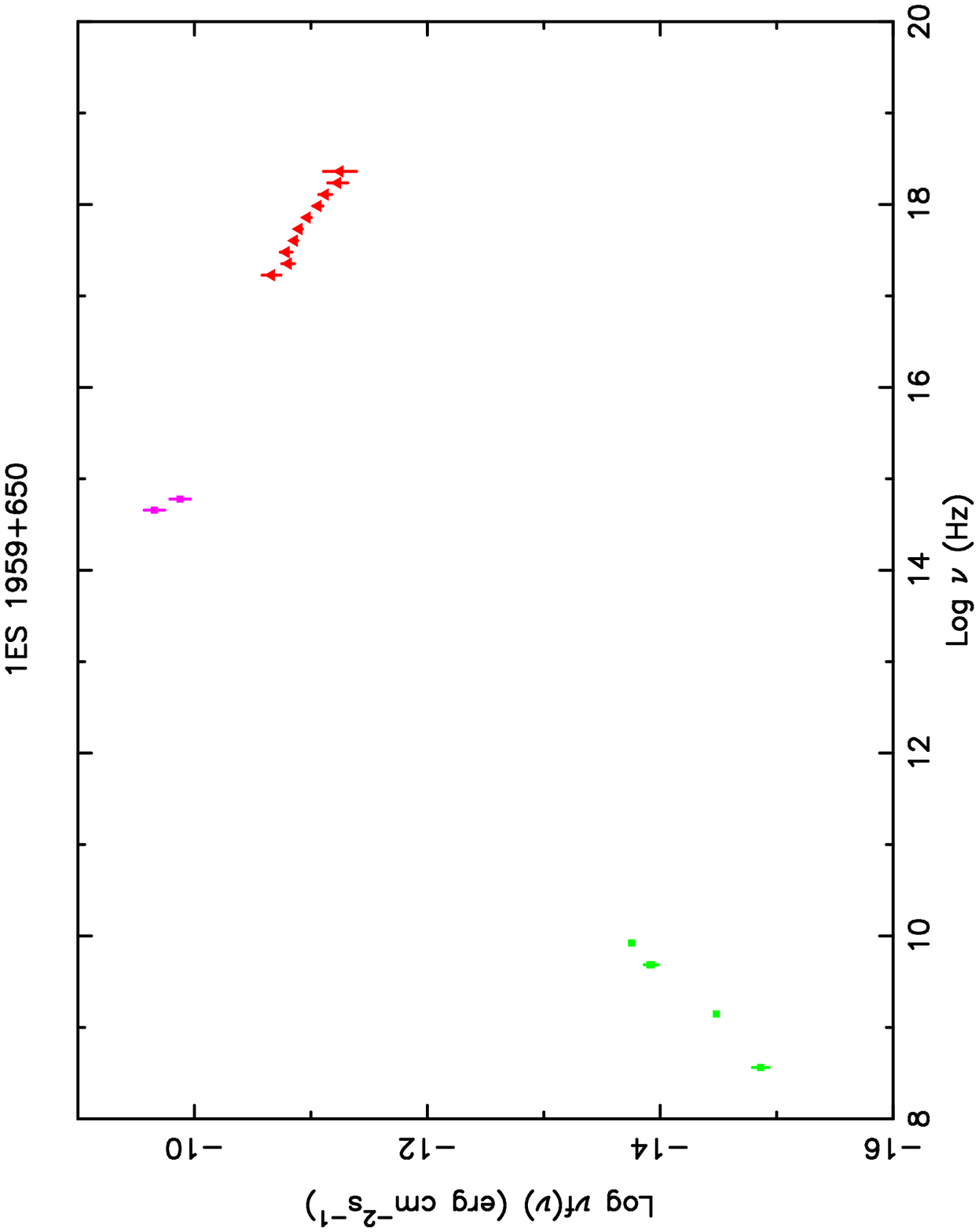} 
\includegraphics{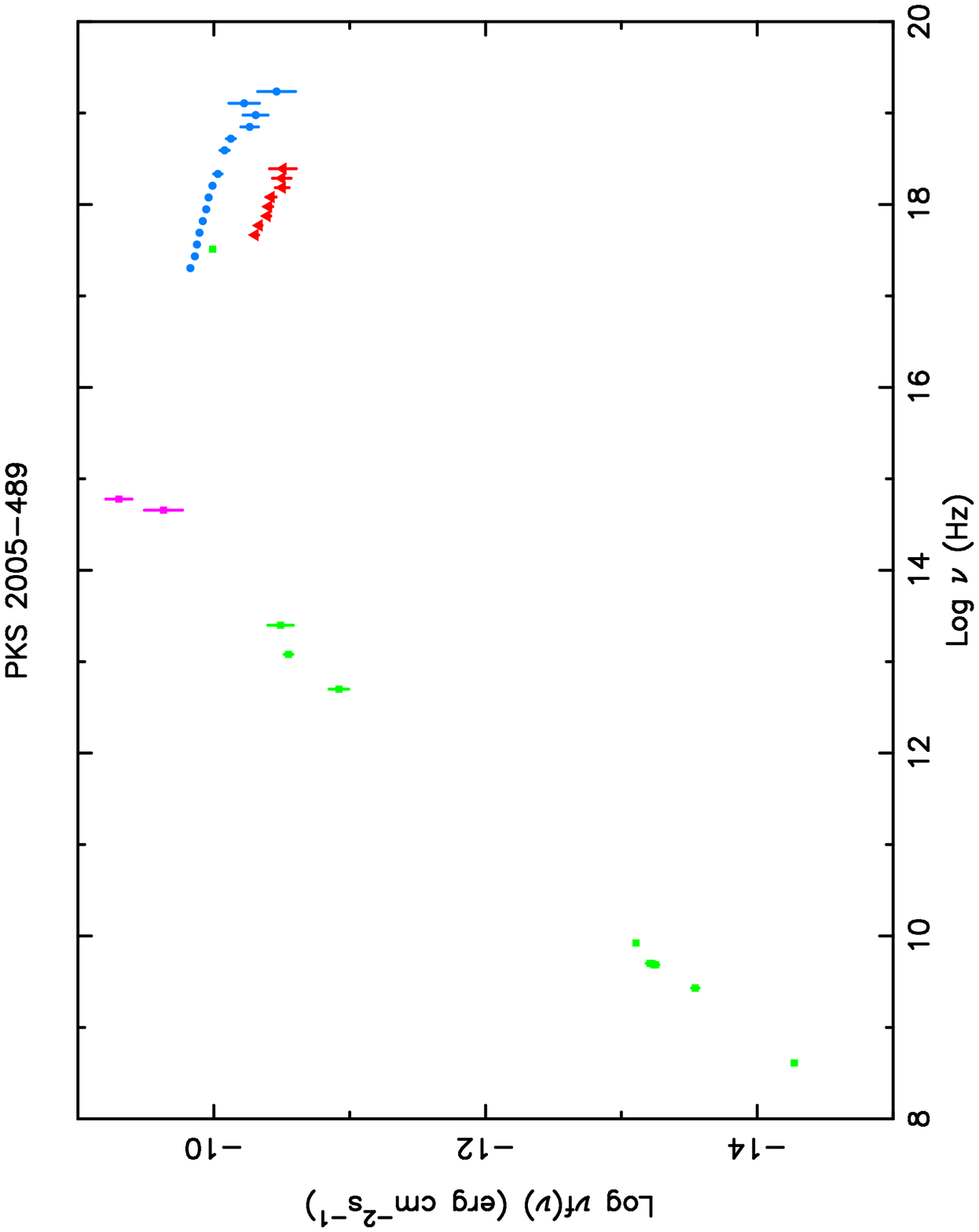} 
\includegraphics{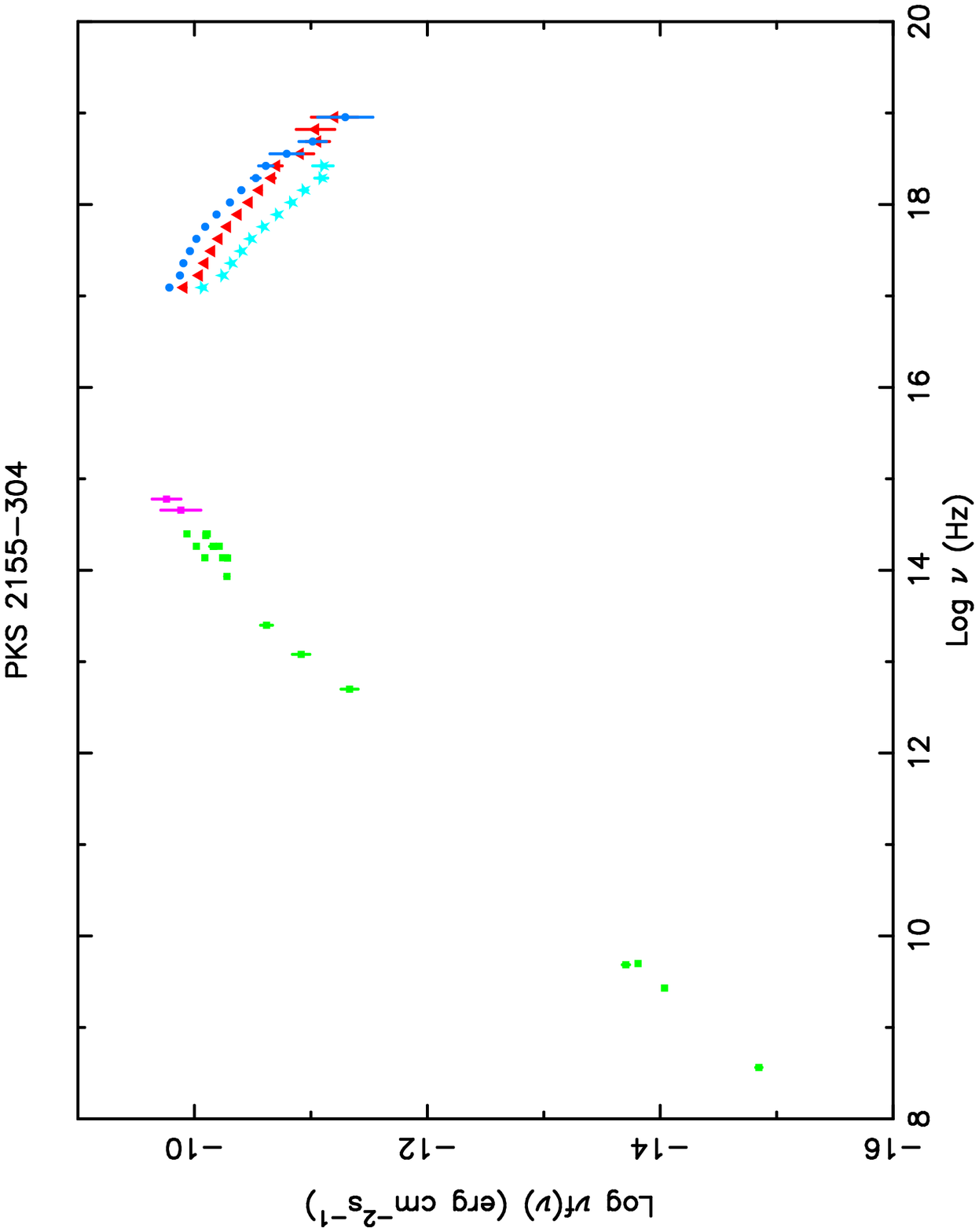} 
\includegraphics{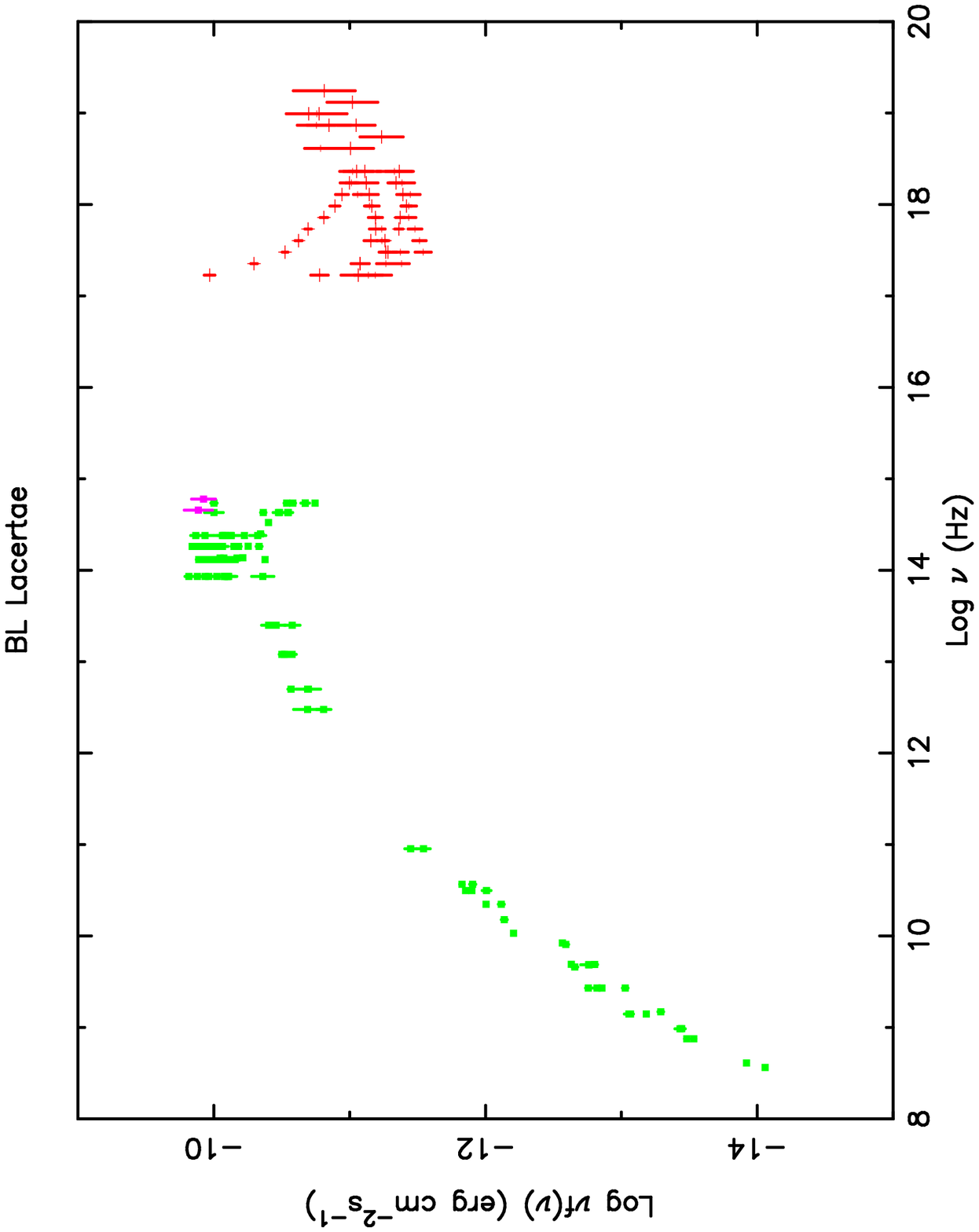} 
\vspace{19.0cm} 
\caption[t]{n- Spectral Energy Distribution of the BL Lacs 1ES 1959+650,
PKS 2005$-$489, PKS 2155$-$304 and BL Lacertae} 
\label{fig1n} 
\end{figure} 
\clearpage 
\setcounter{figure}{1} 
\begin{figure}[t]
\centering
\includegraphics{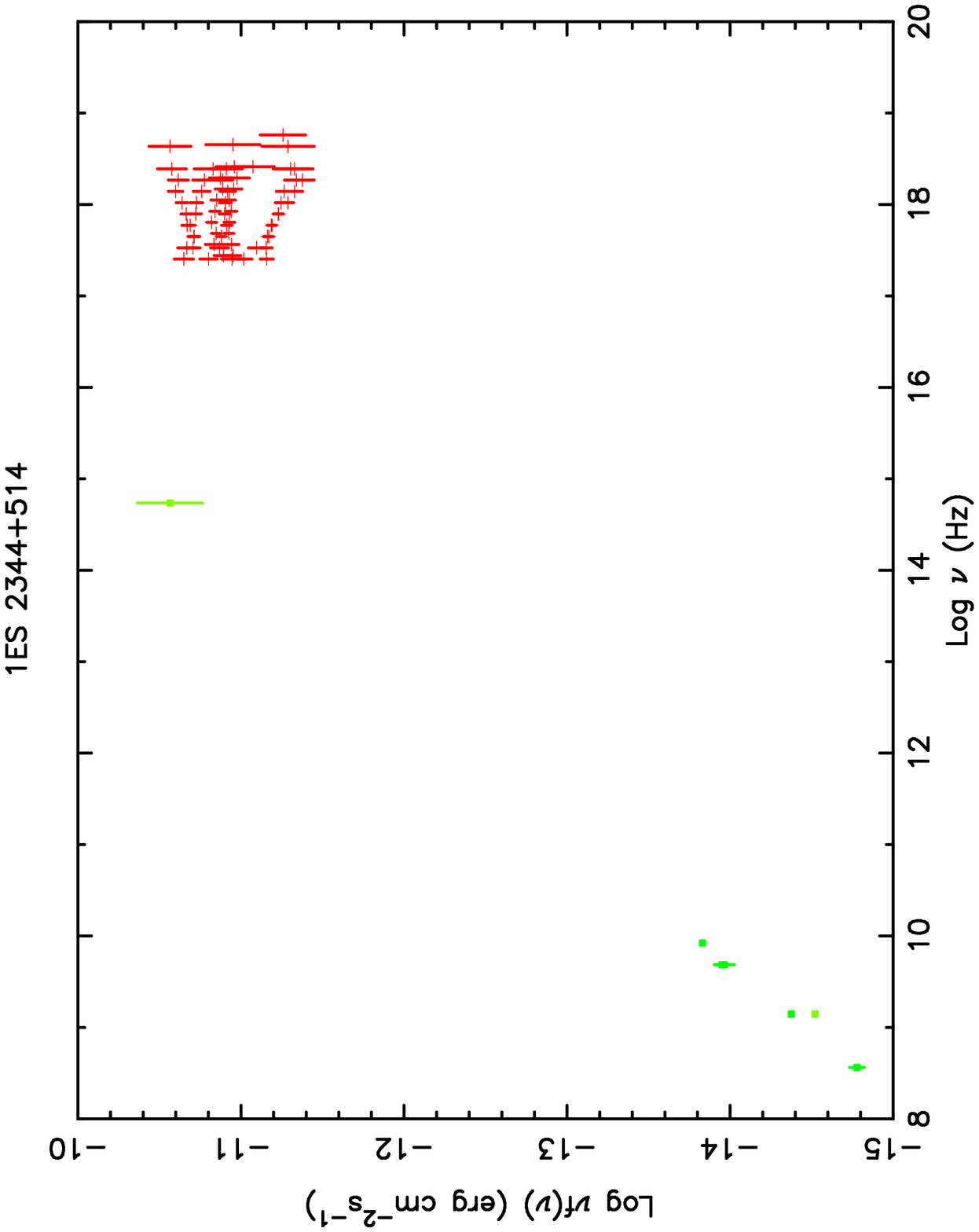} 
\includegraphics{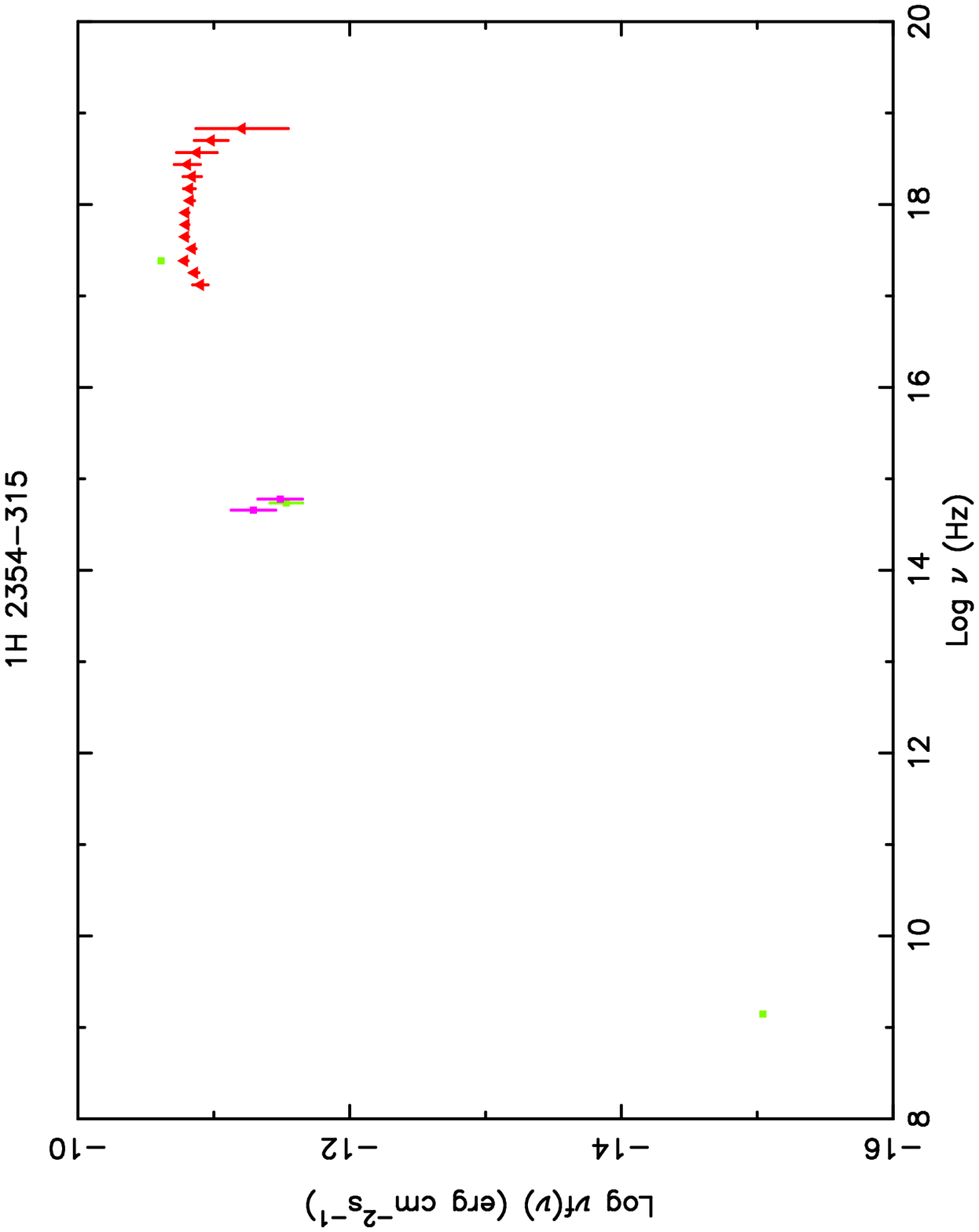} 
\vspace{19.0cm}
\caption[t]{o- Spectral Energy Distribution of the BL Lacs 1ES 2344+514 and 
1H 2354$-$315}
\label{fig1o}
\end{figure}
%---------------------------------------------------
\clearpage 
\setcounter{figure}{2} 
\begin{figure}[t]
\centering
\includegraphics{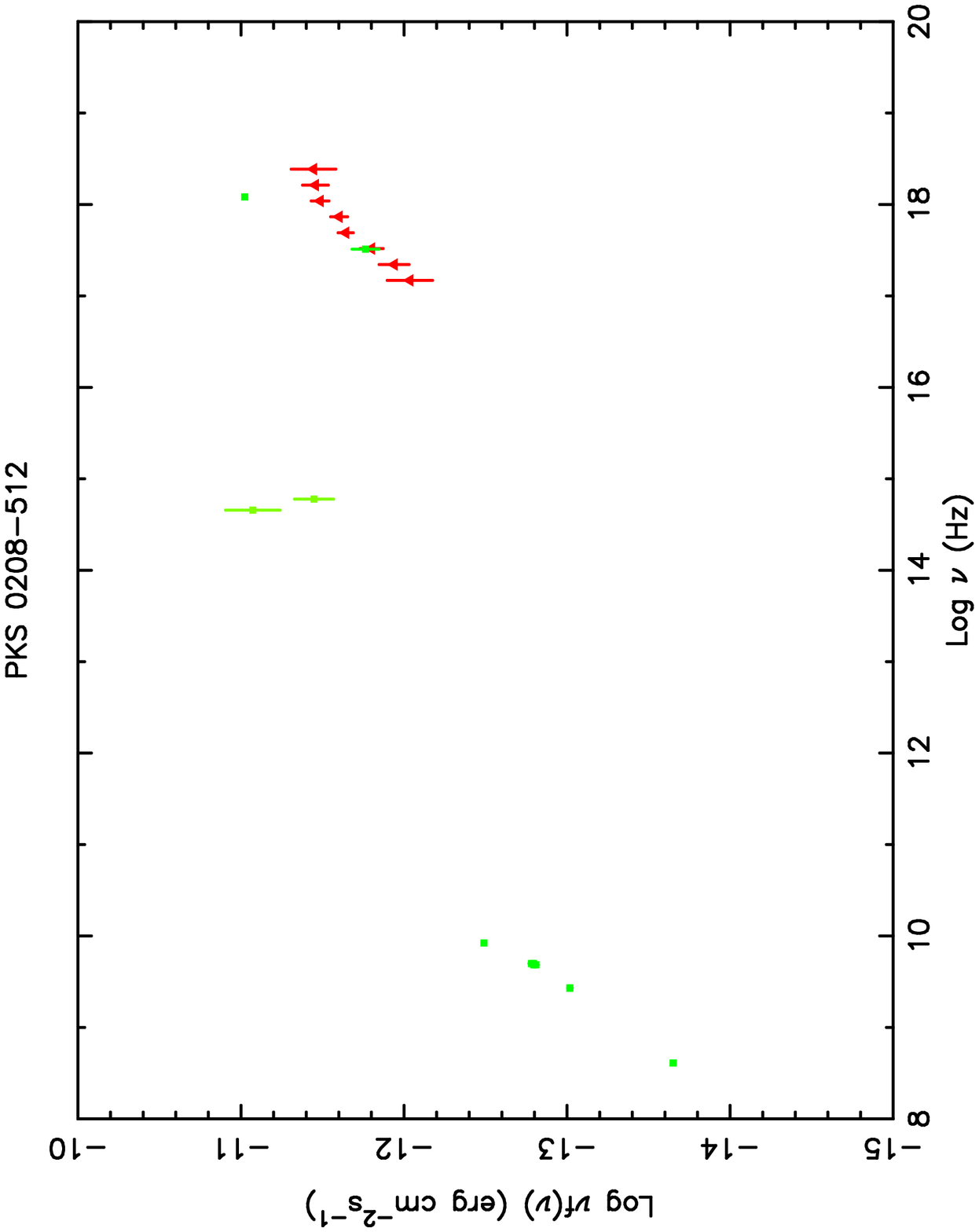} 
\includegraphics{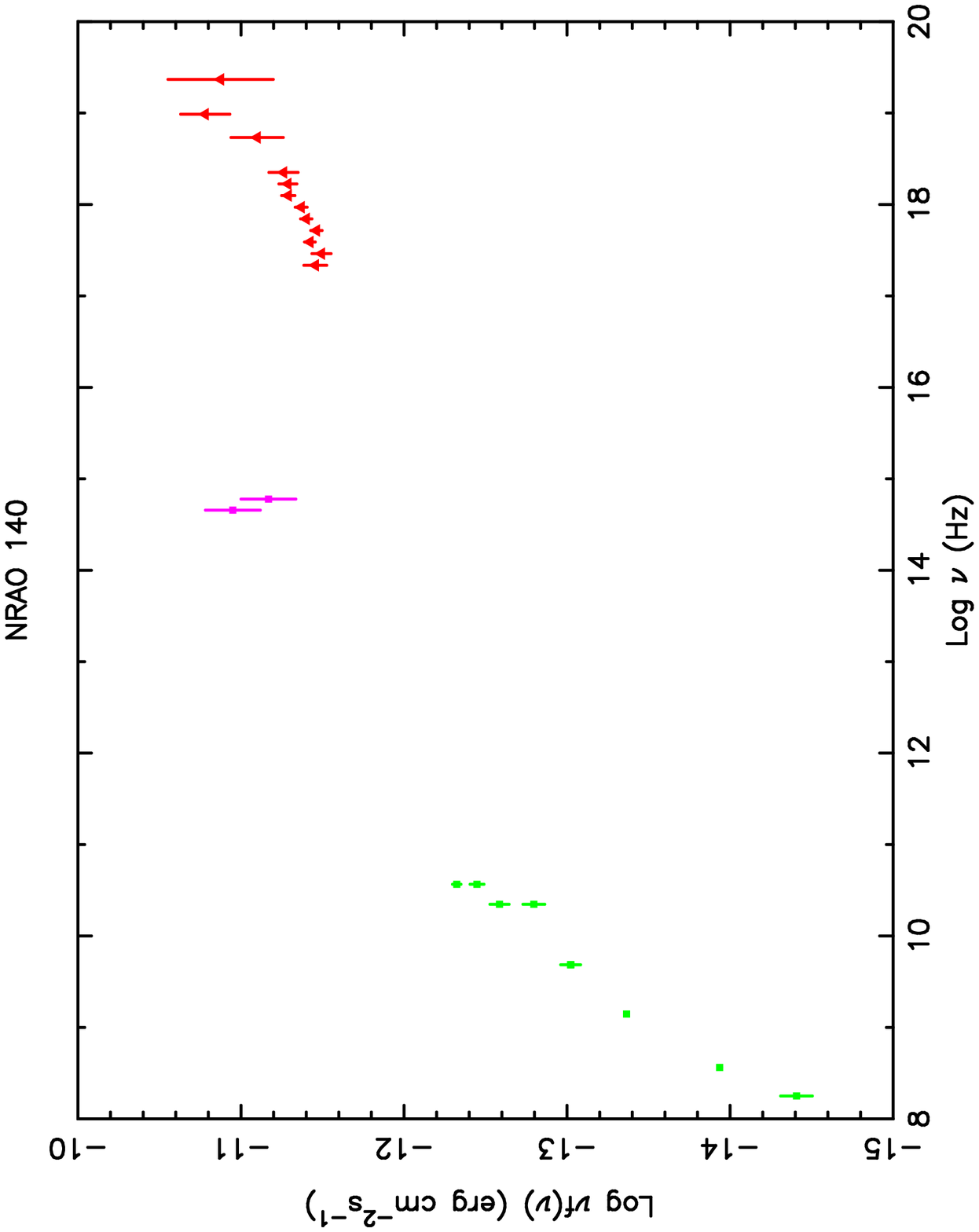} 
\includegraphics{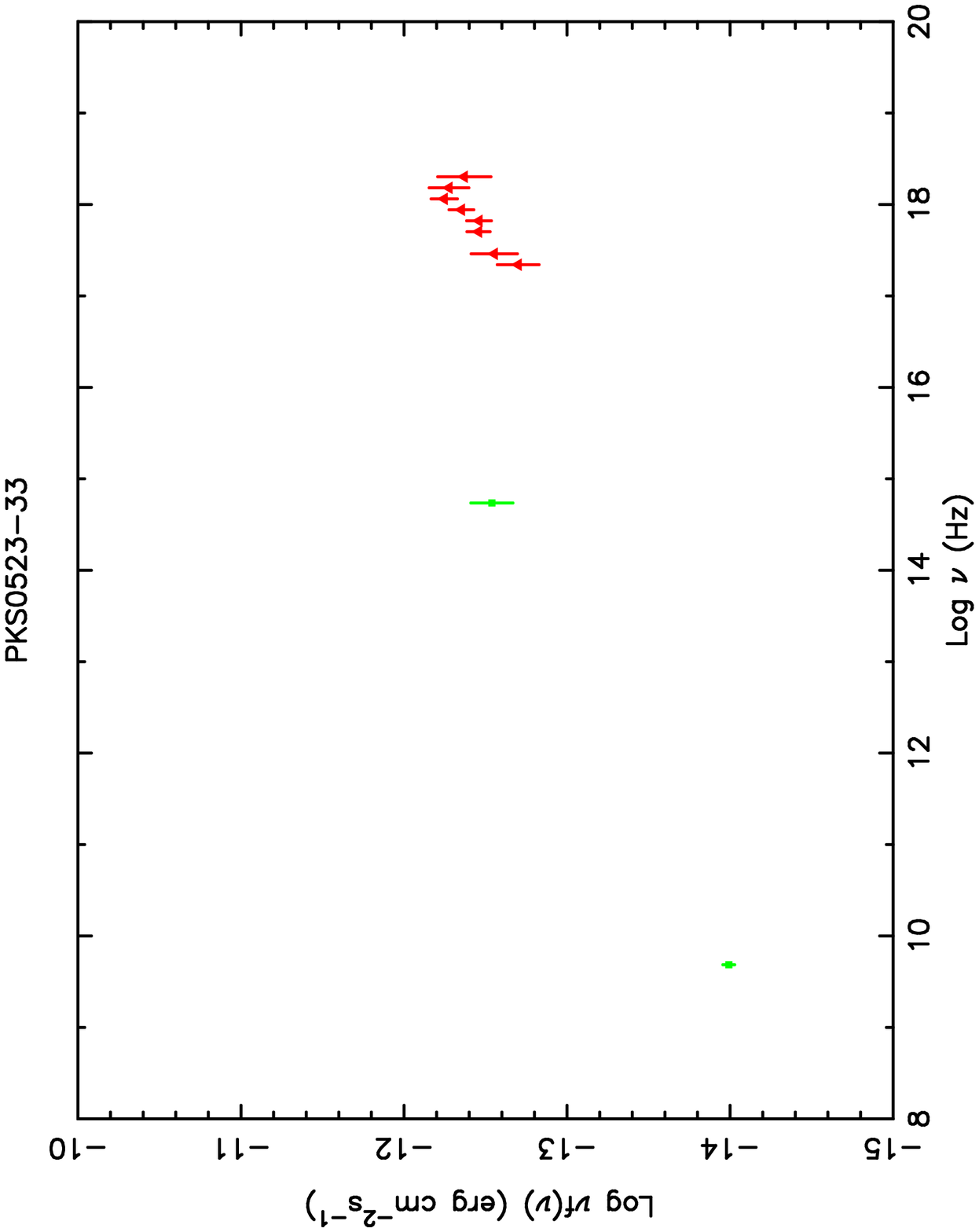} 
\includegraphics{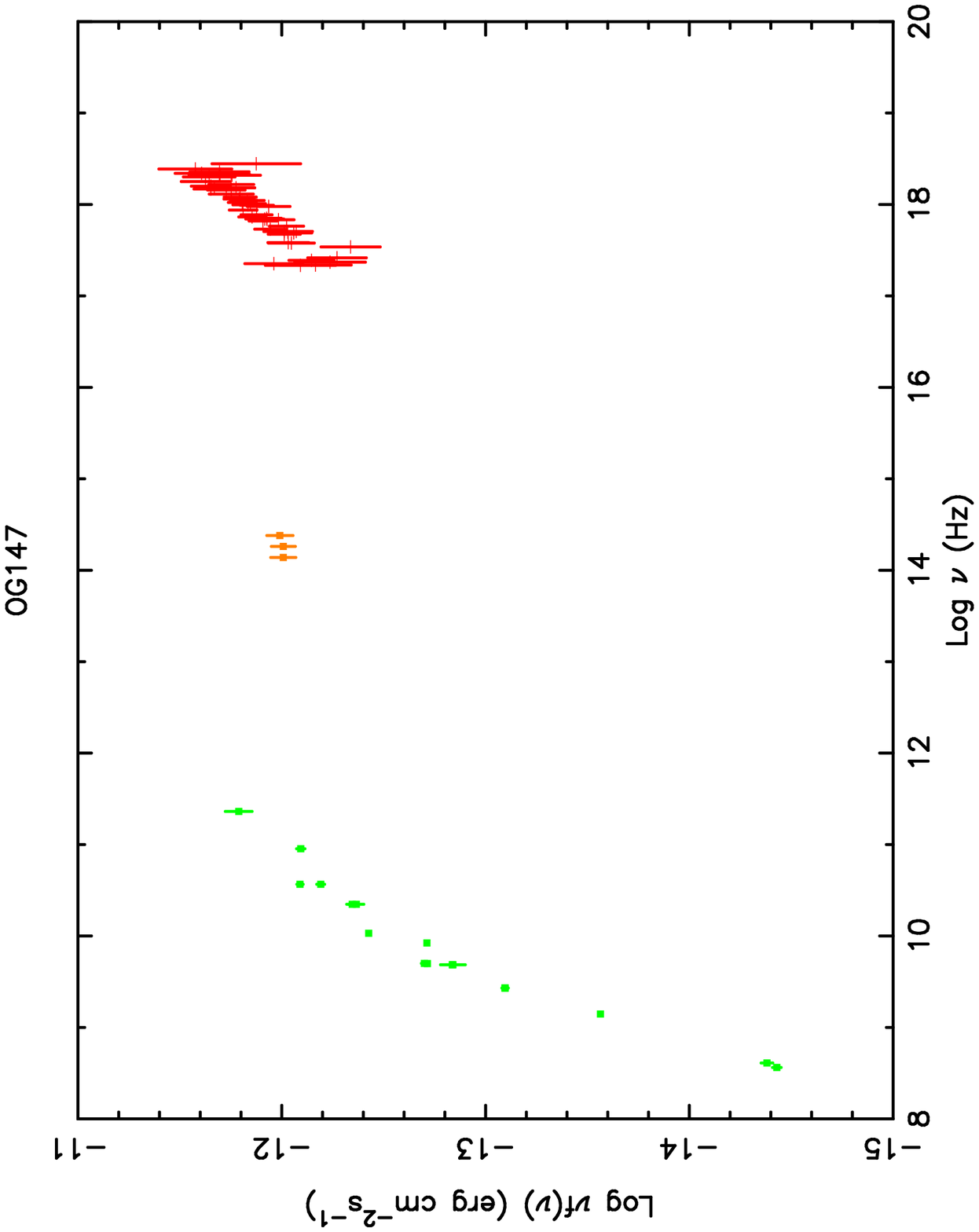} 
\vspace{19.0cm} 
\caption[t]{a- Spectral Energy Distribution of the FSRQs PKS 0208$-$512, NRAO 140,
PKS 0523$-$33 and OG 147} 
\label{fig2a} 
\end{figure} 
\clearpage 
\setcounter{figure}{2} 
\begin{figure}[t]
\centering
\includegraphics{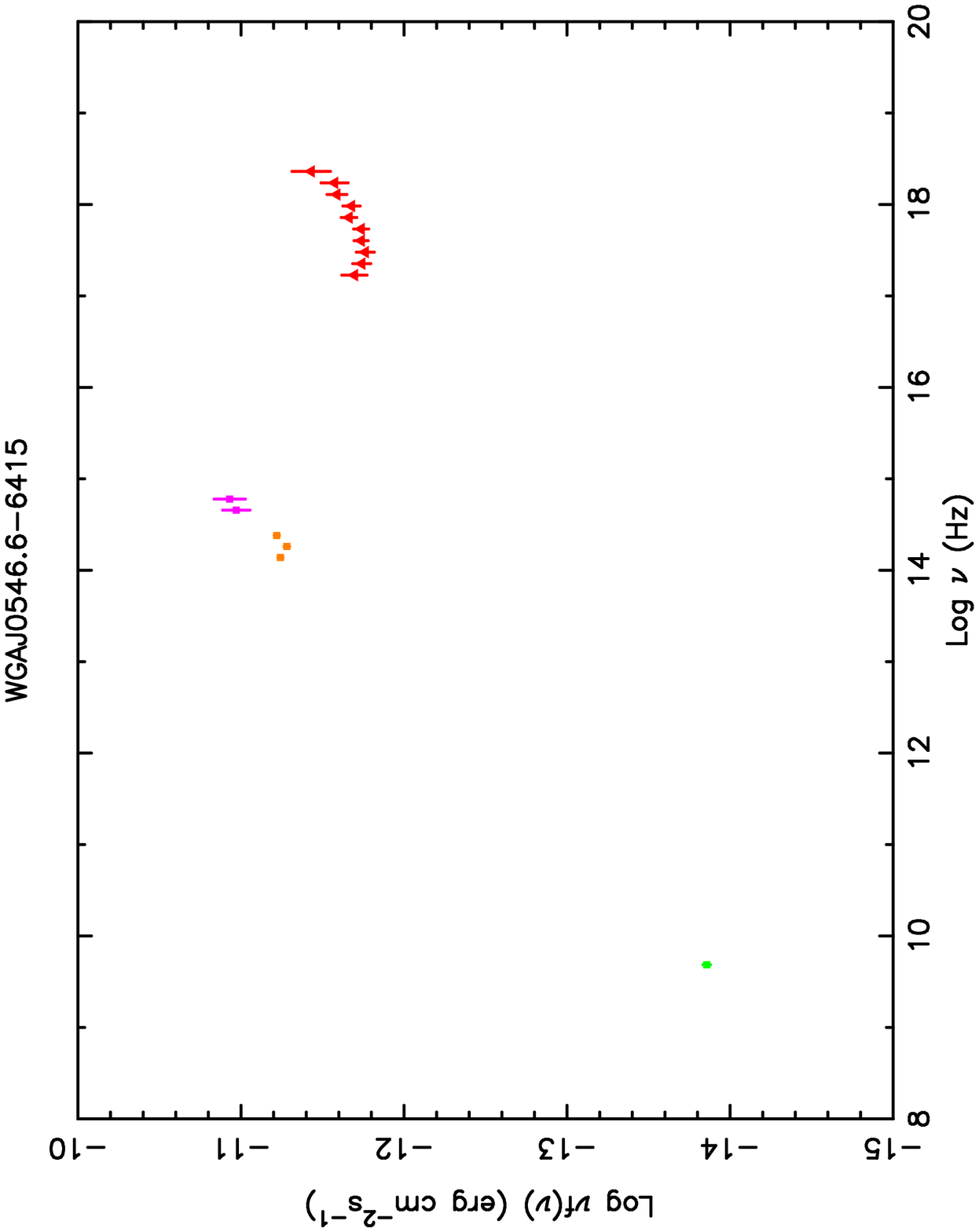} 
\includegraphics{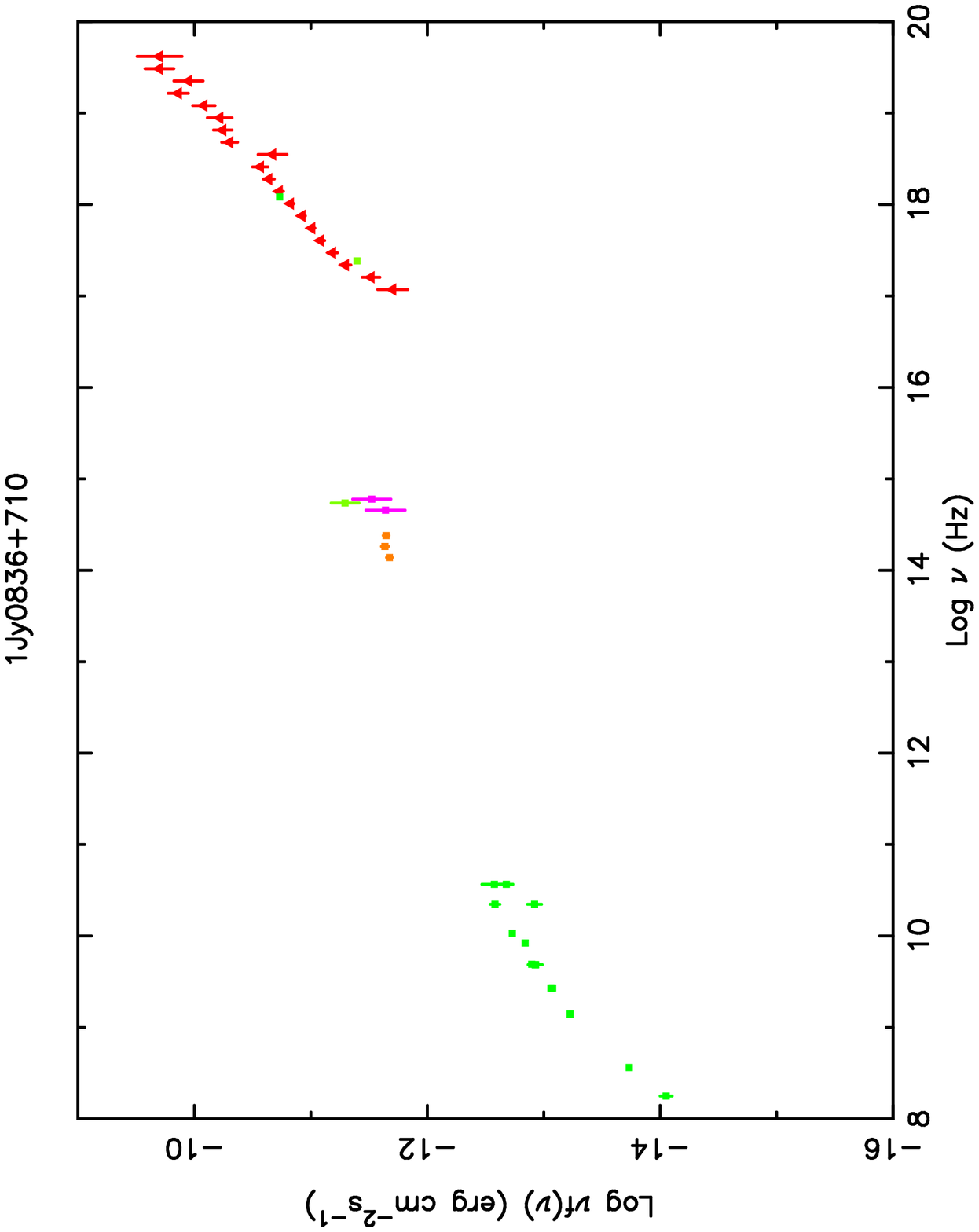} 
\includegraphics{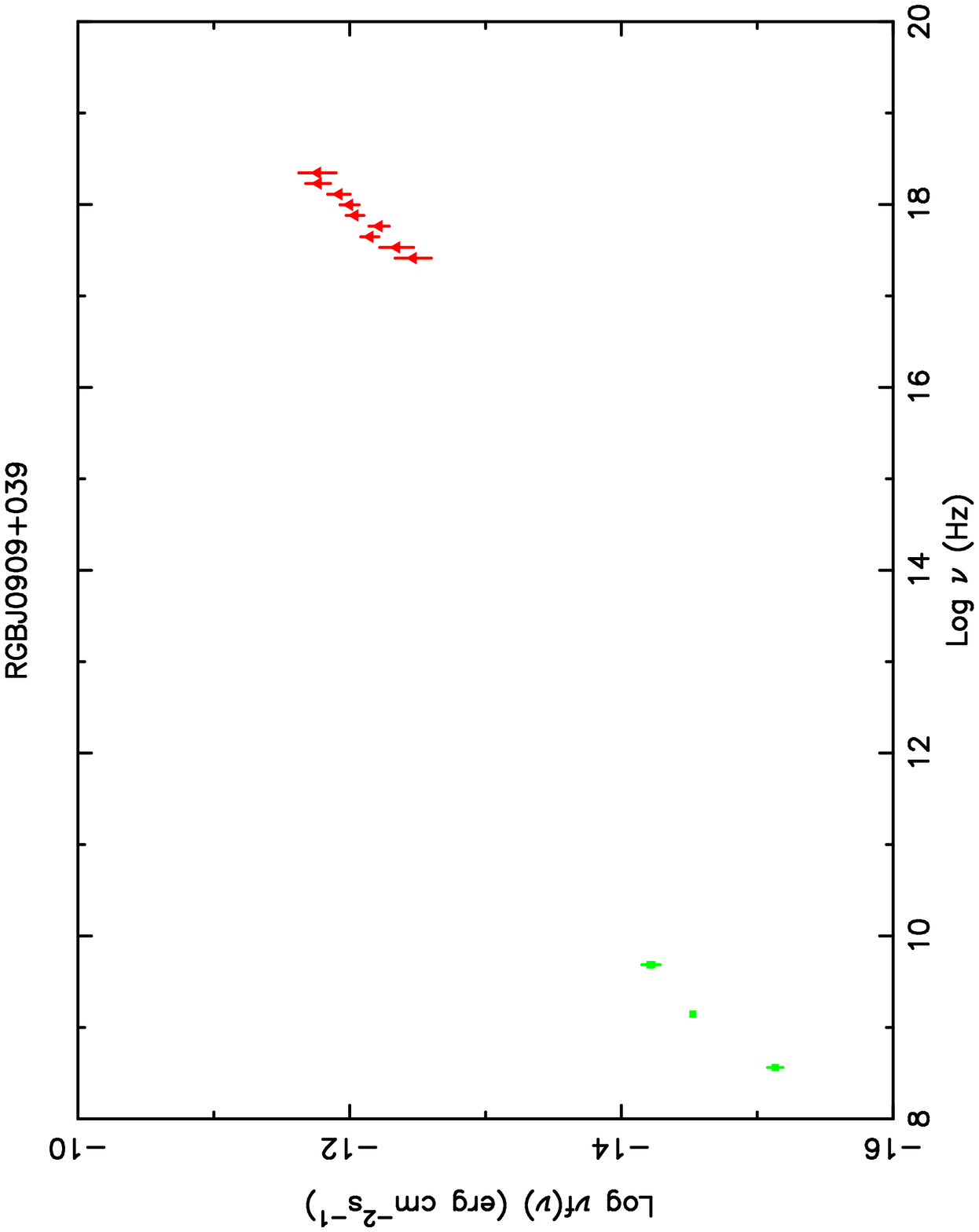} 
\includegraphics{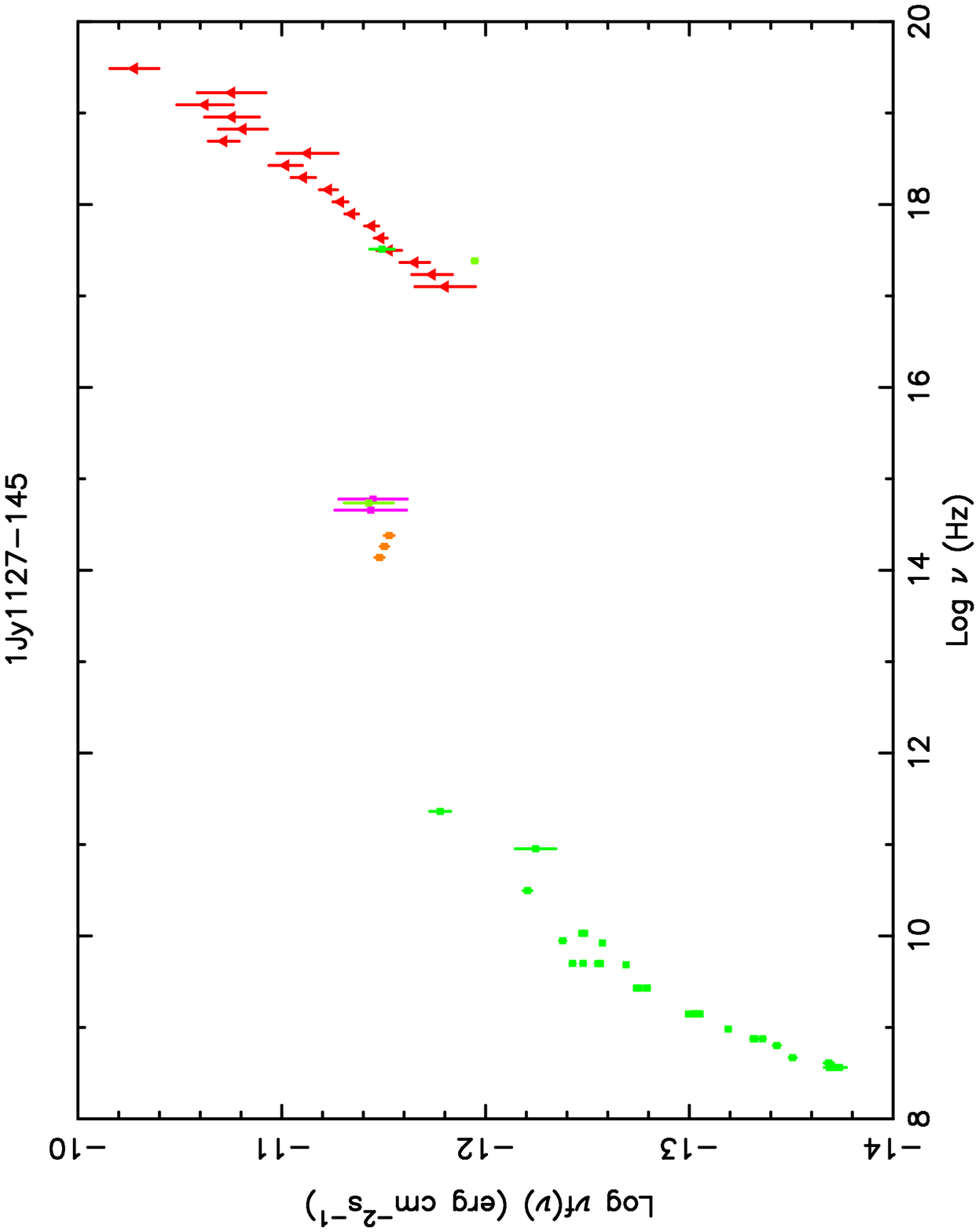} 
\vspace{19.0cm} 
\caption[t]{b- Spectral Energy Distribution of the FSRQs WGA J0546.6$-$6415,
1Jy 0836+710, RGB J0909+039 and 1Jy 1127$-$145} 
\label{fig2b} 
\end{figure} 
\clearpage 
\setcounter{figure}{2} 
\begin{figure}[t]
\centering
\includegraphics{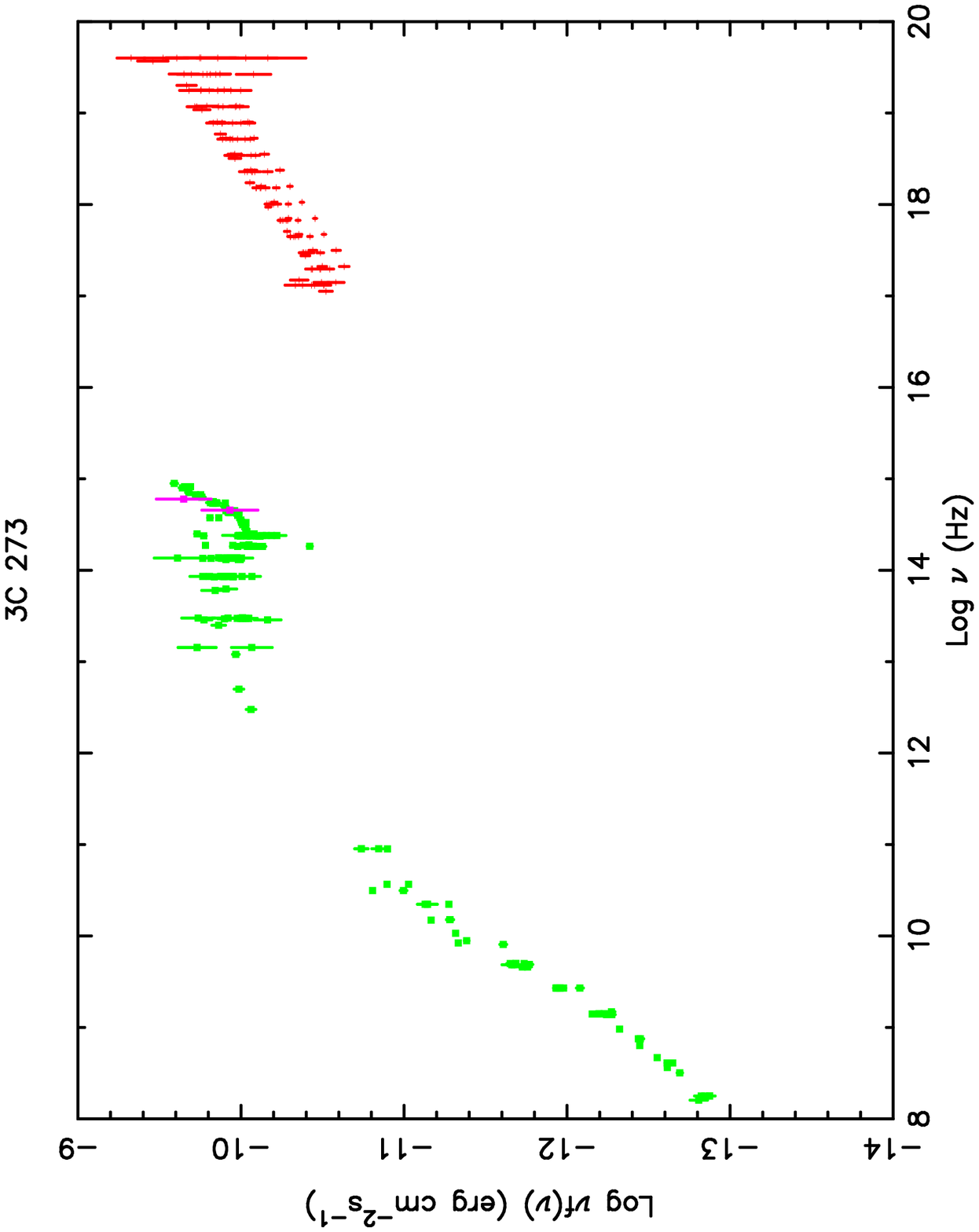} 
\includegraphics{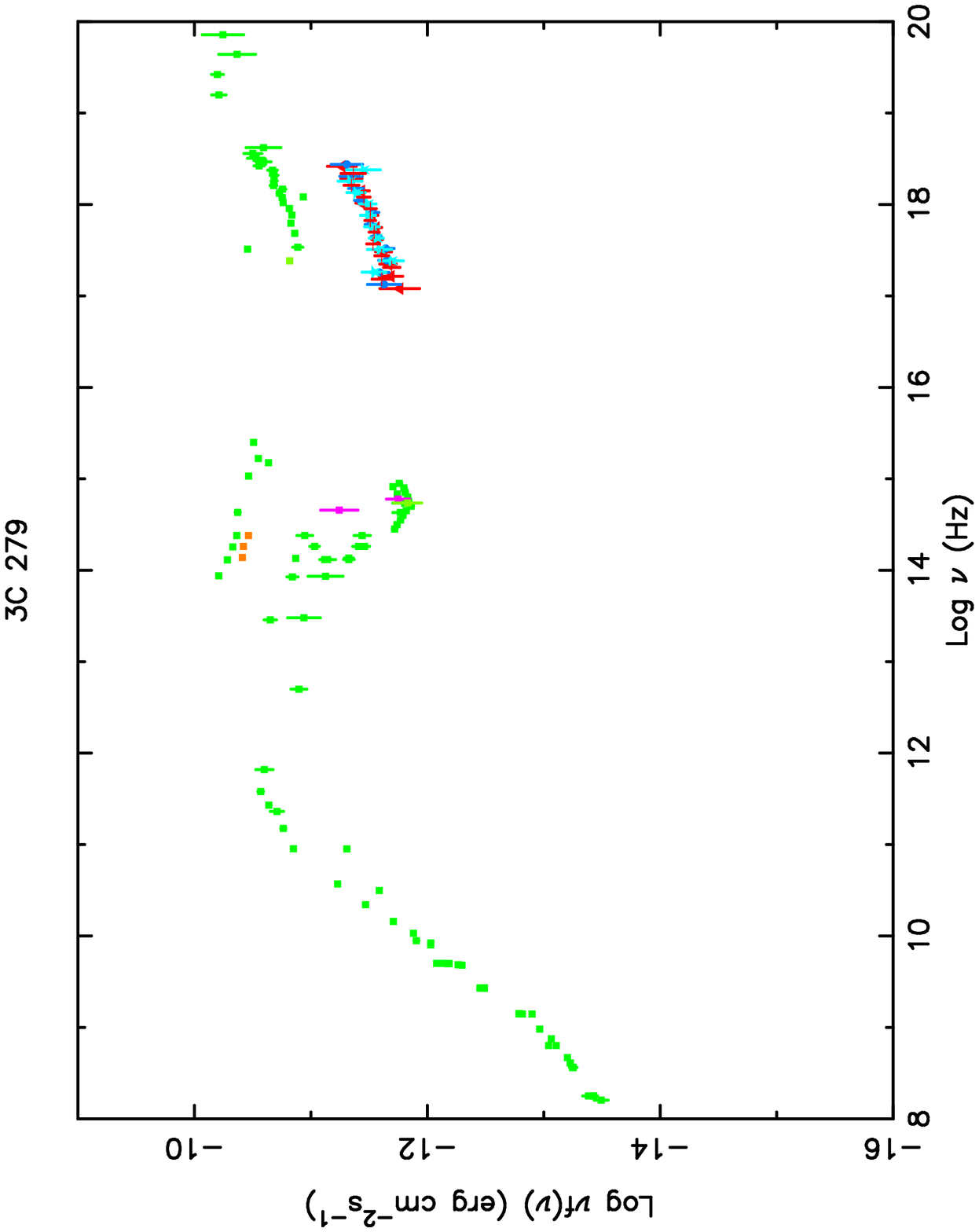} 
\includegraphics{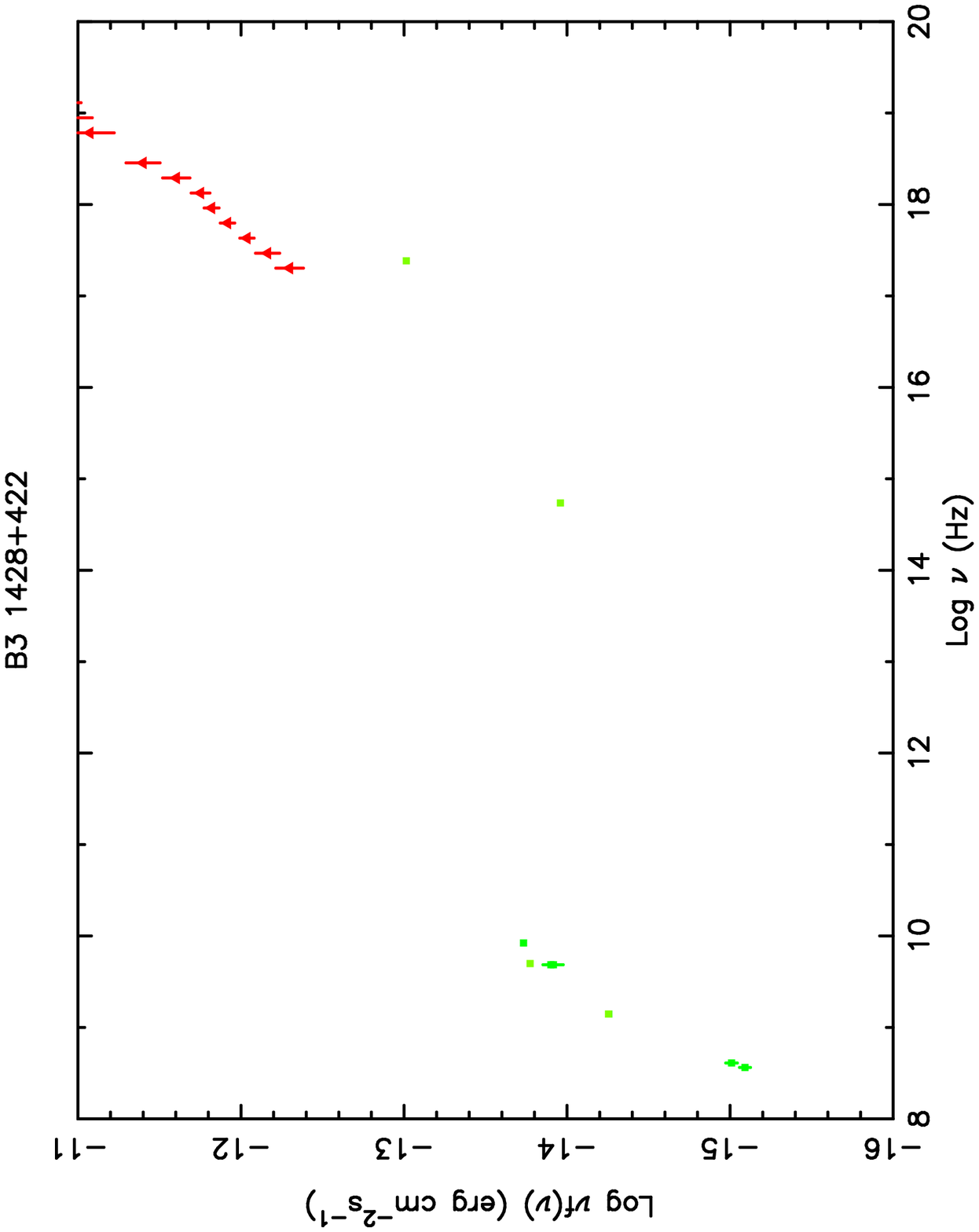} 
\includegraphics{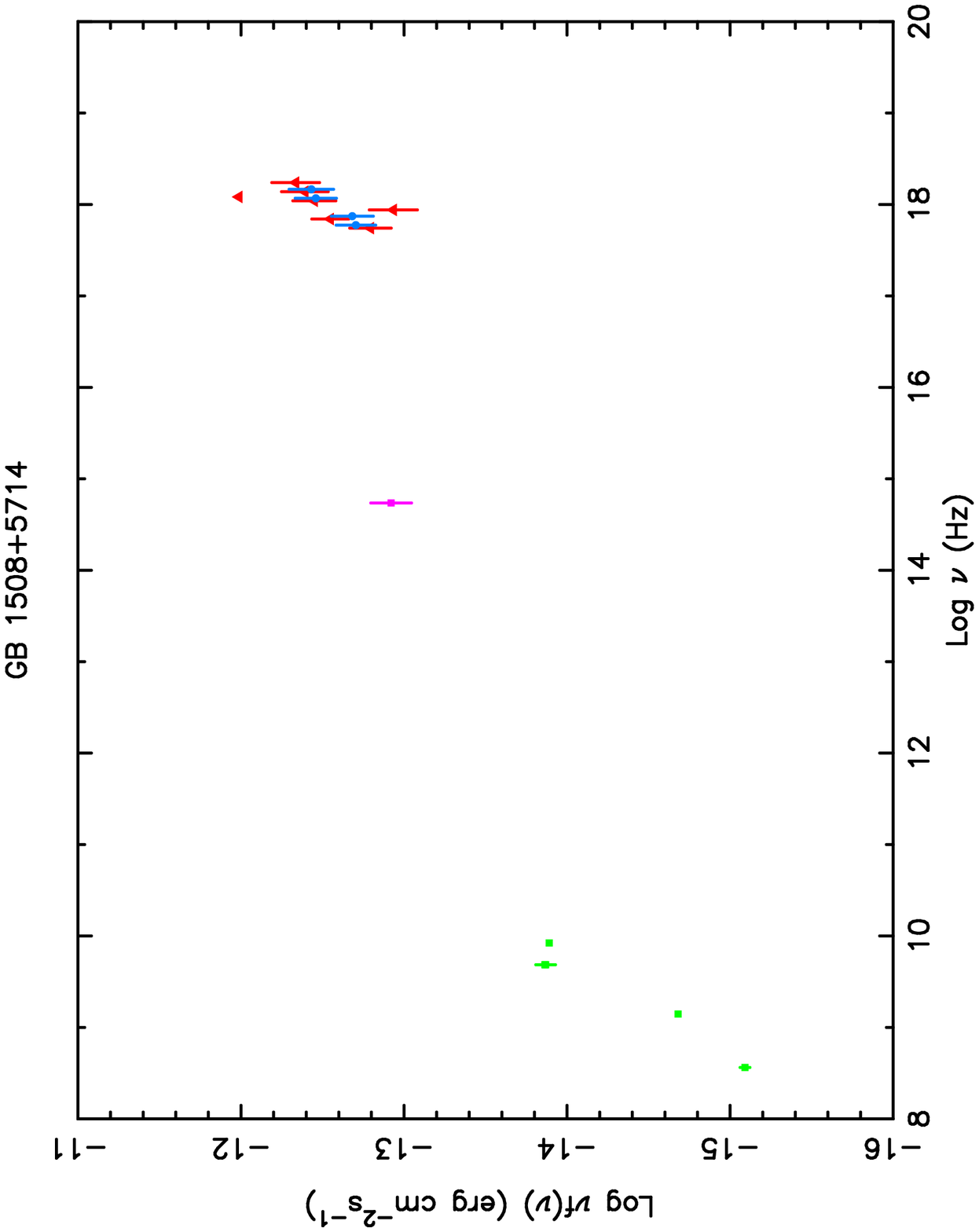} 
\vspace{19.0cm} 
\caption[t]{c- Spectral Energy Distribution of the FSRQs 3C 273, 3C 279,
B3 1428+422 and GB 1508+5714} 
\label{fig2c} 
\end{figure} 
\clearpage 
\setcounter{figure}{2} 
\begin{figure}[t]
\centering
\includegraphics{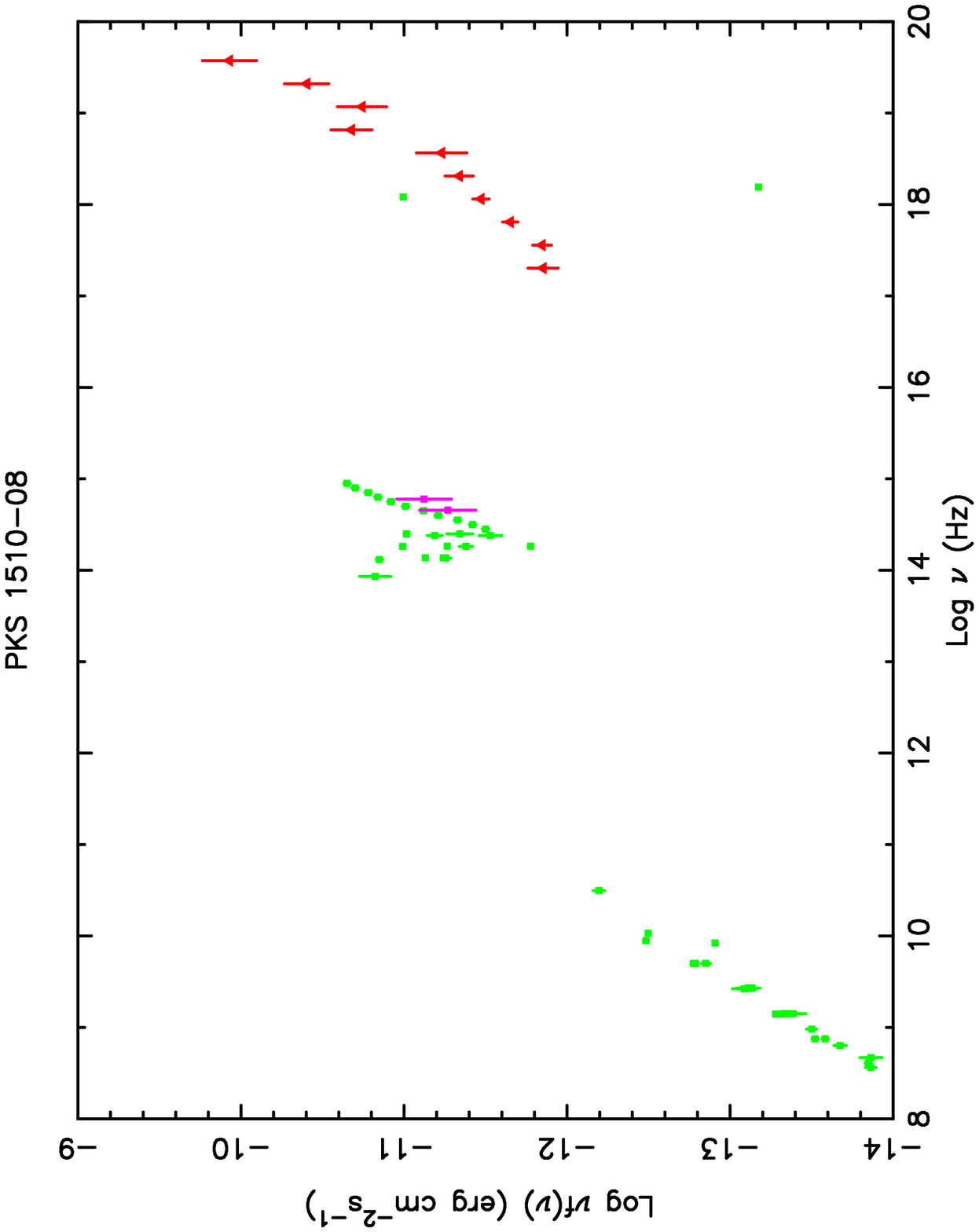} 
\includegraphics{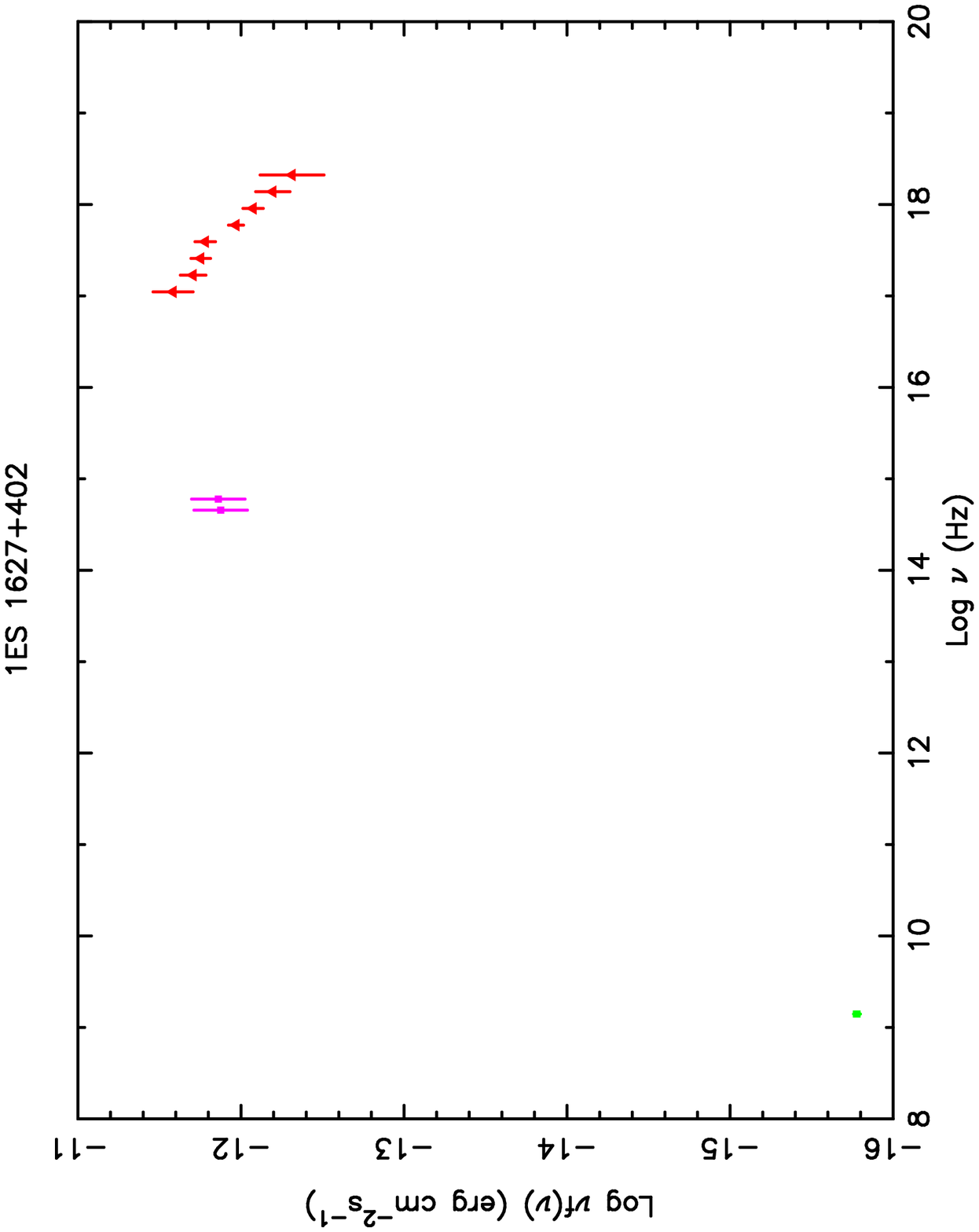} 
\includegraphics{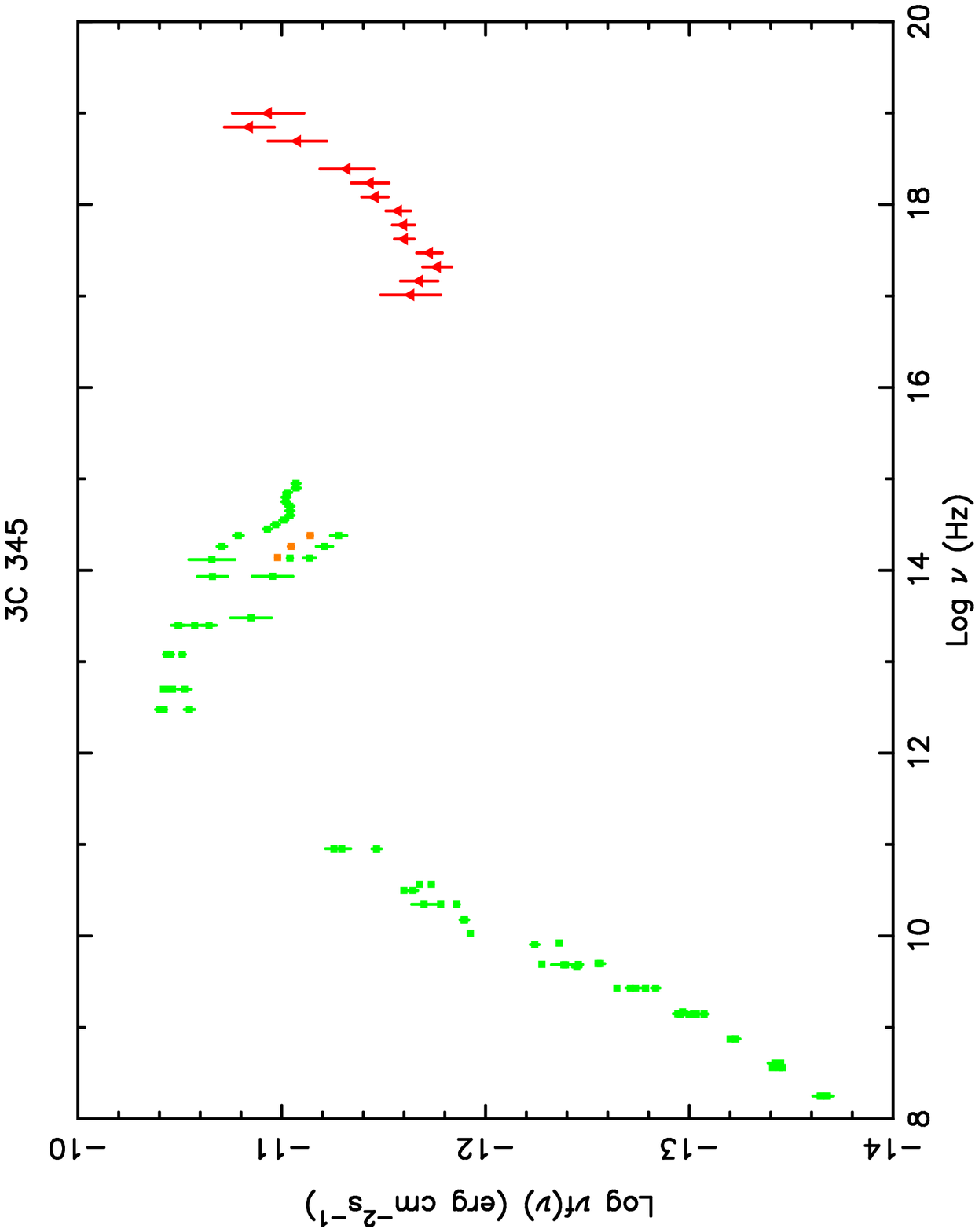} 
\includegraphics{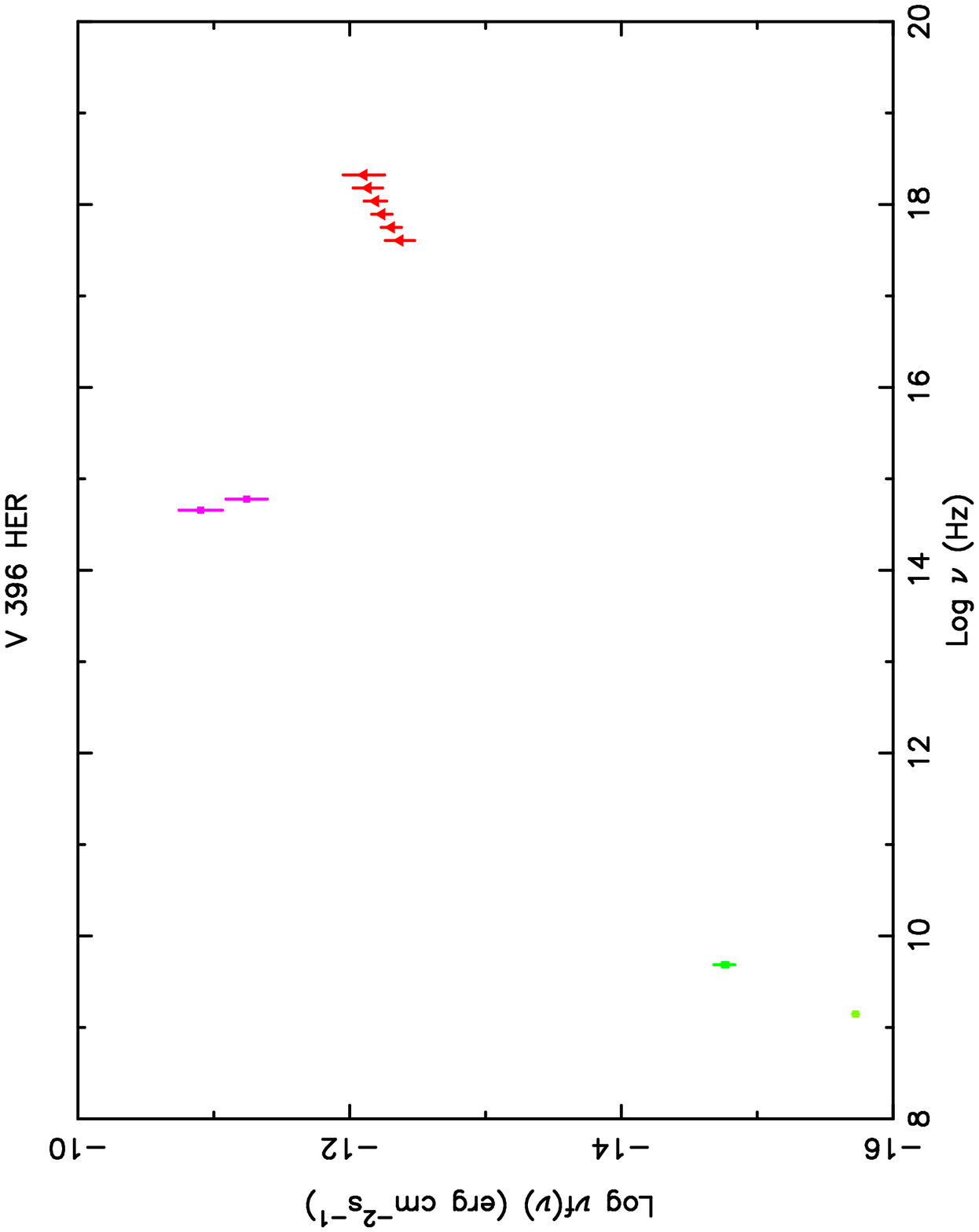} 
\vspace{19.0cm} 
\caption[t]{d- Spectral Energy Distribution of the FSRQs PKS 1510$-$08, 
1ES 1627+402, 3C 345 and V 396 HER} 
\label{fig2d} 
\end{figure} 
\clearpage 
\setcounter{figure}{2} 
\begin{figure}[t]
\centering
\includegraphics{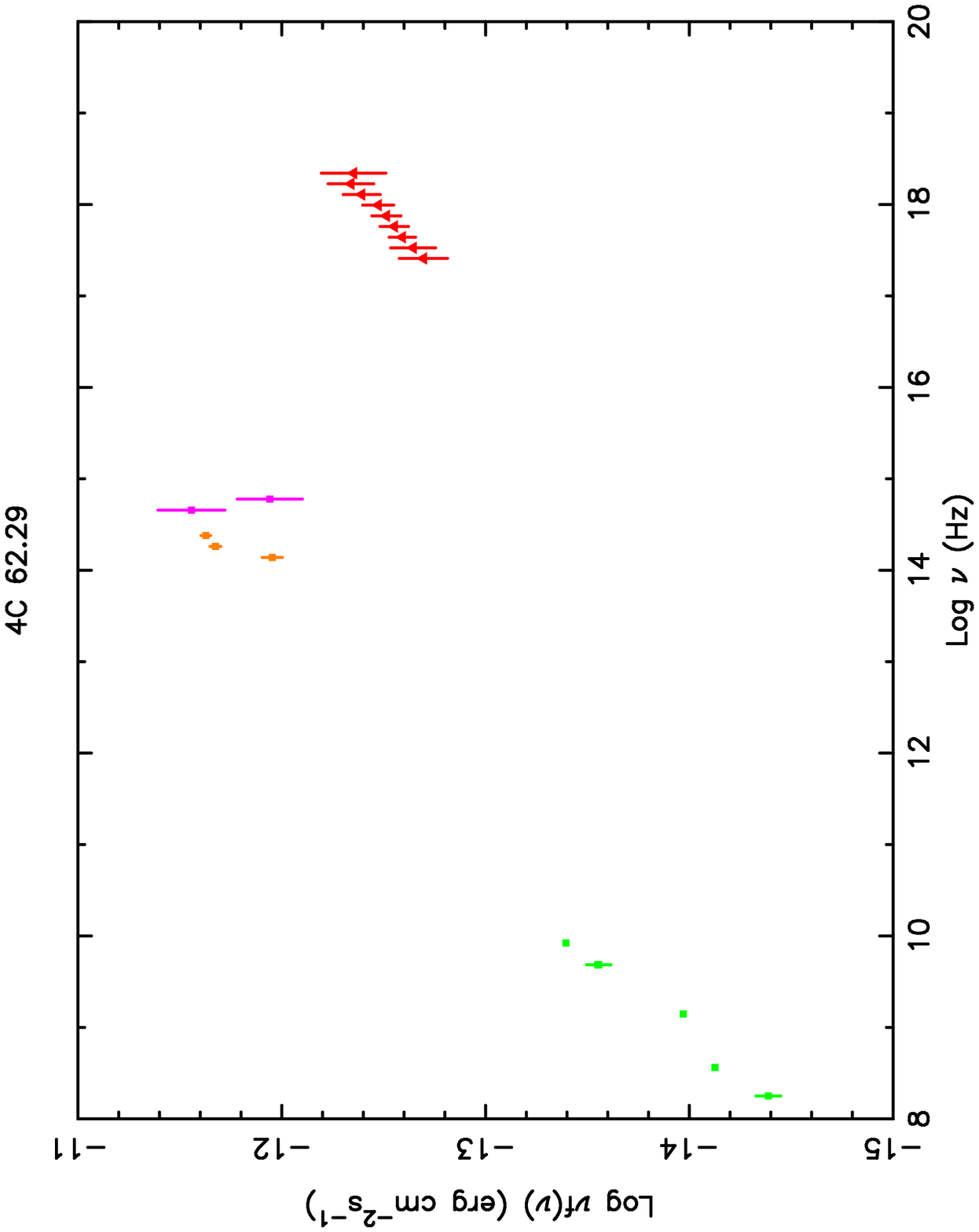} 
\includegraphics{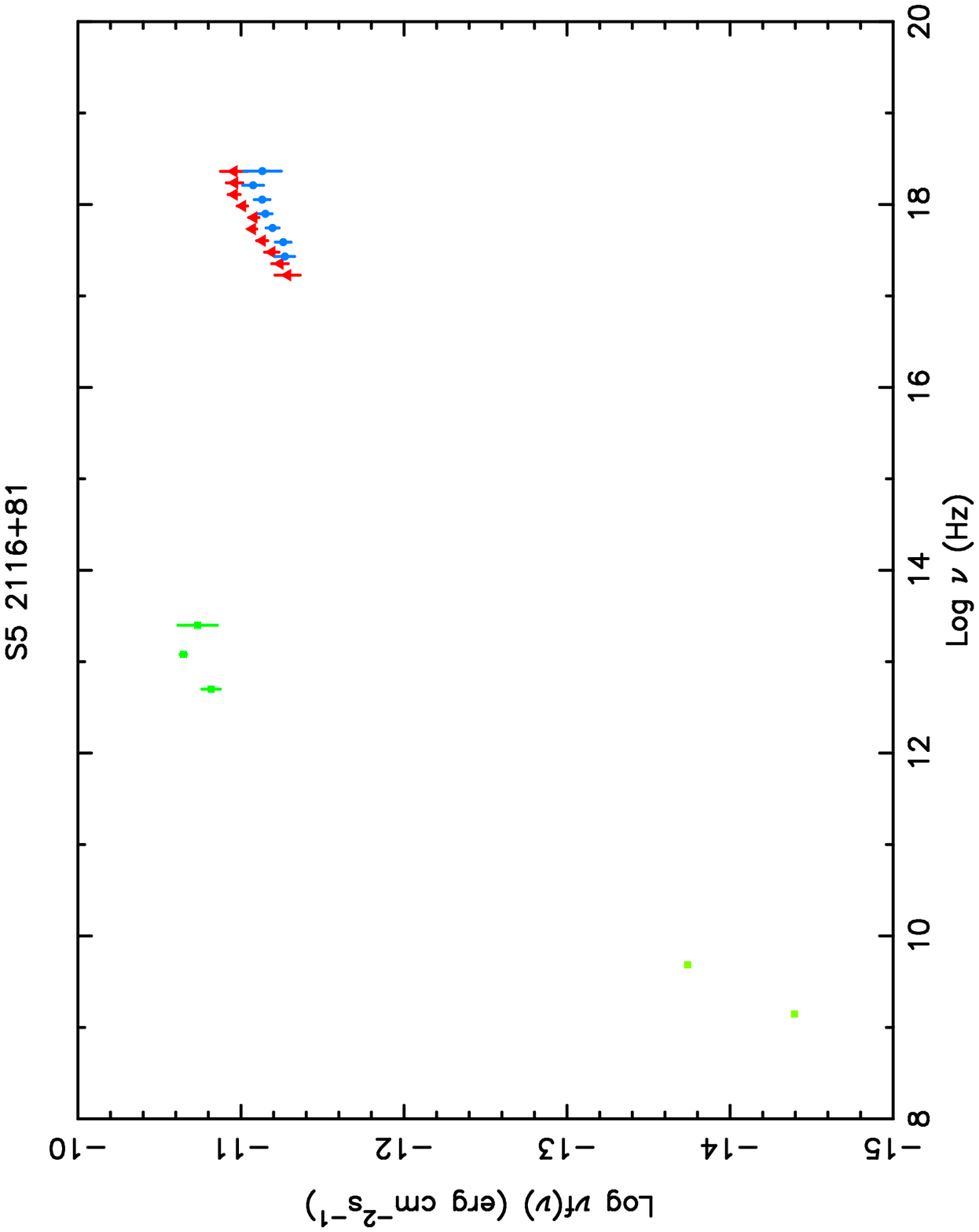} 
\includegraphics{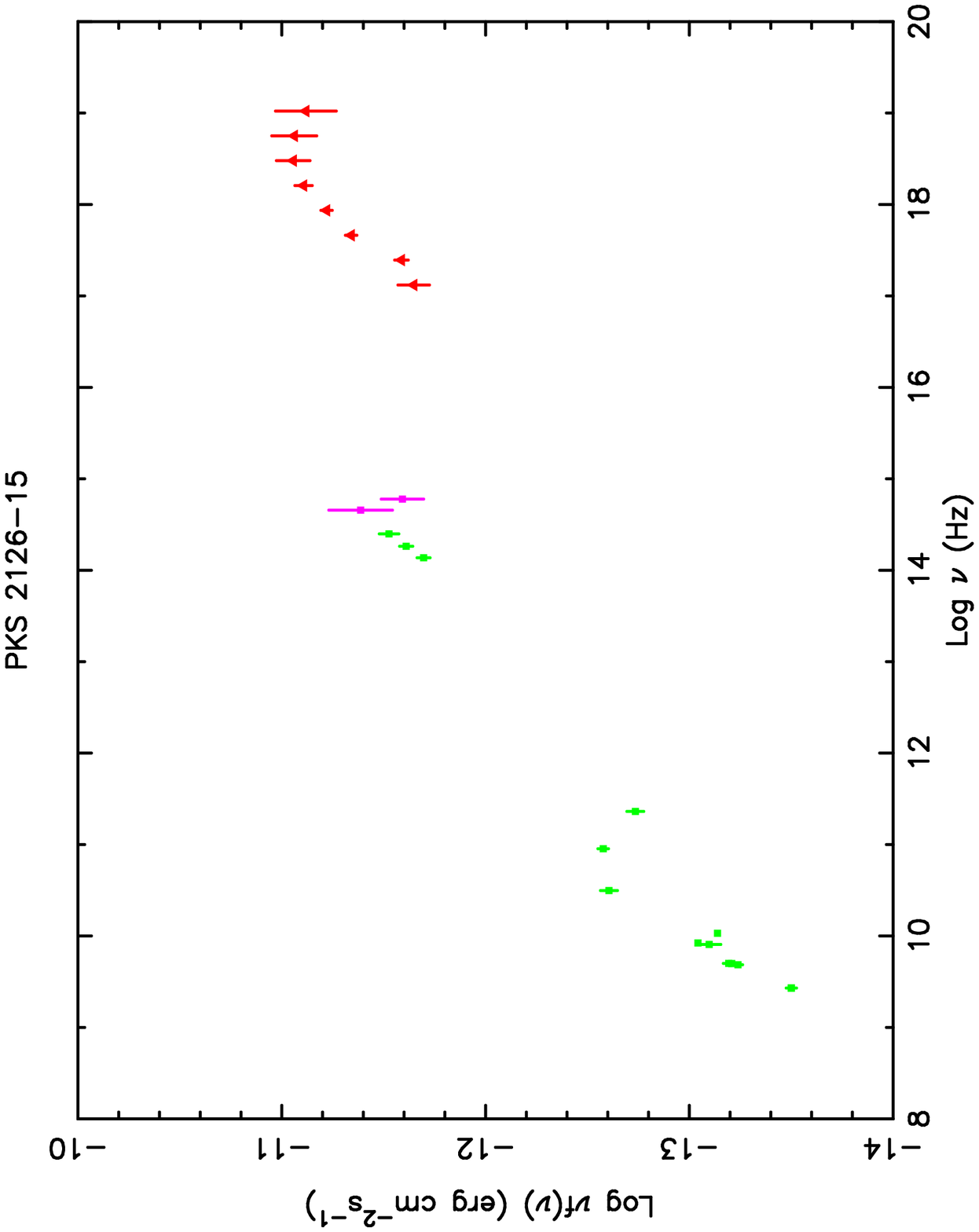} 
\includegraphics{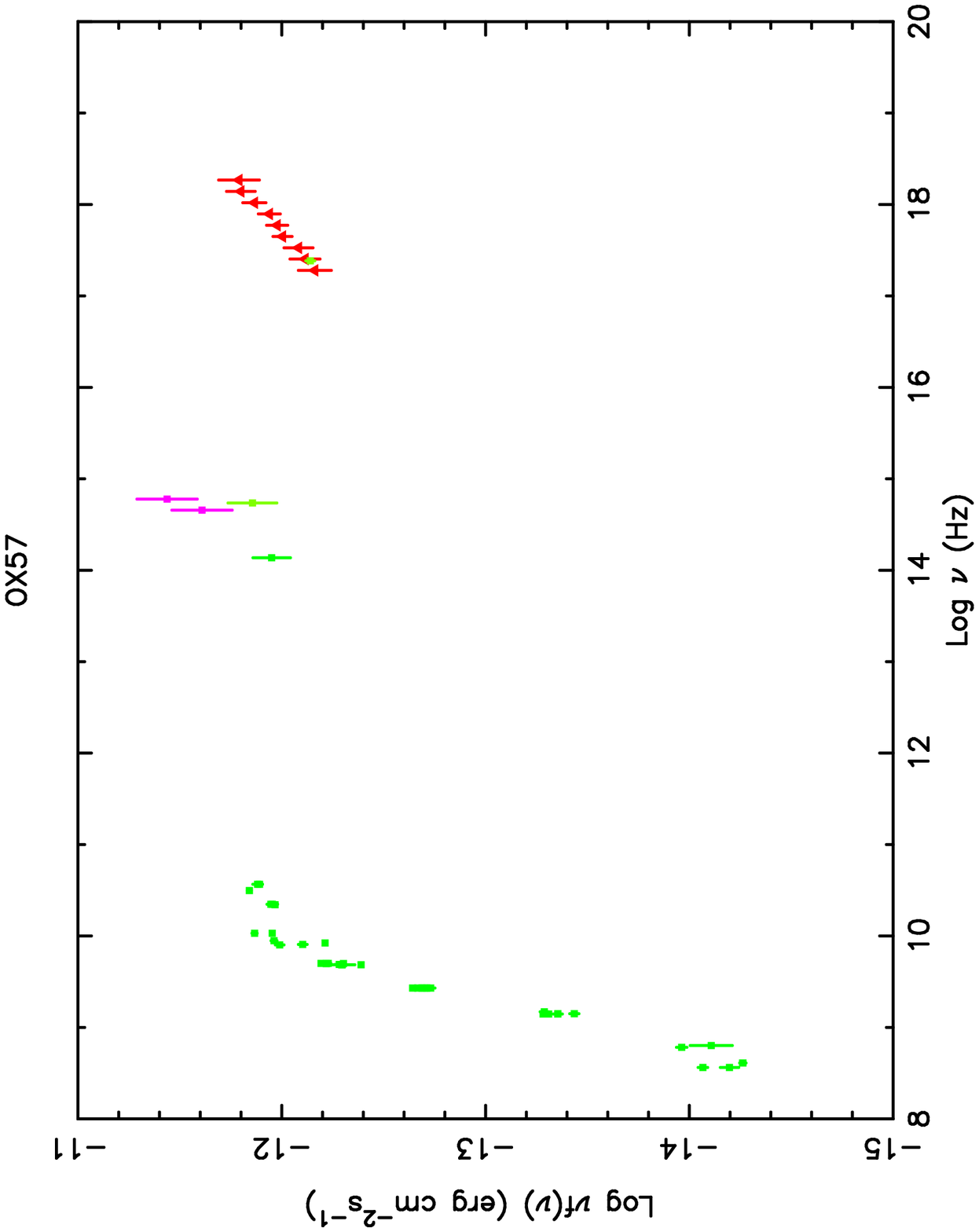} 
\vspace{19.0cm} 
\caption[t]{e- Spectral Energy Distribution of the FSRQs 4C 62.29, S5 2116+81,
PKS 2126$-$15 and OX 57} 
\label{fig2e} 
\end{figure} 
\clearpage 
\setcounter{figure}{2} 
\begin{figure}[t]
\centering
\includegraphics{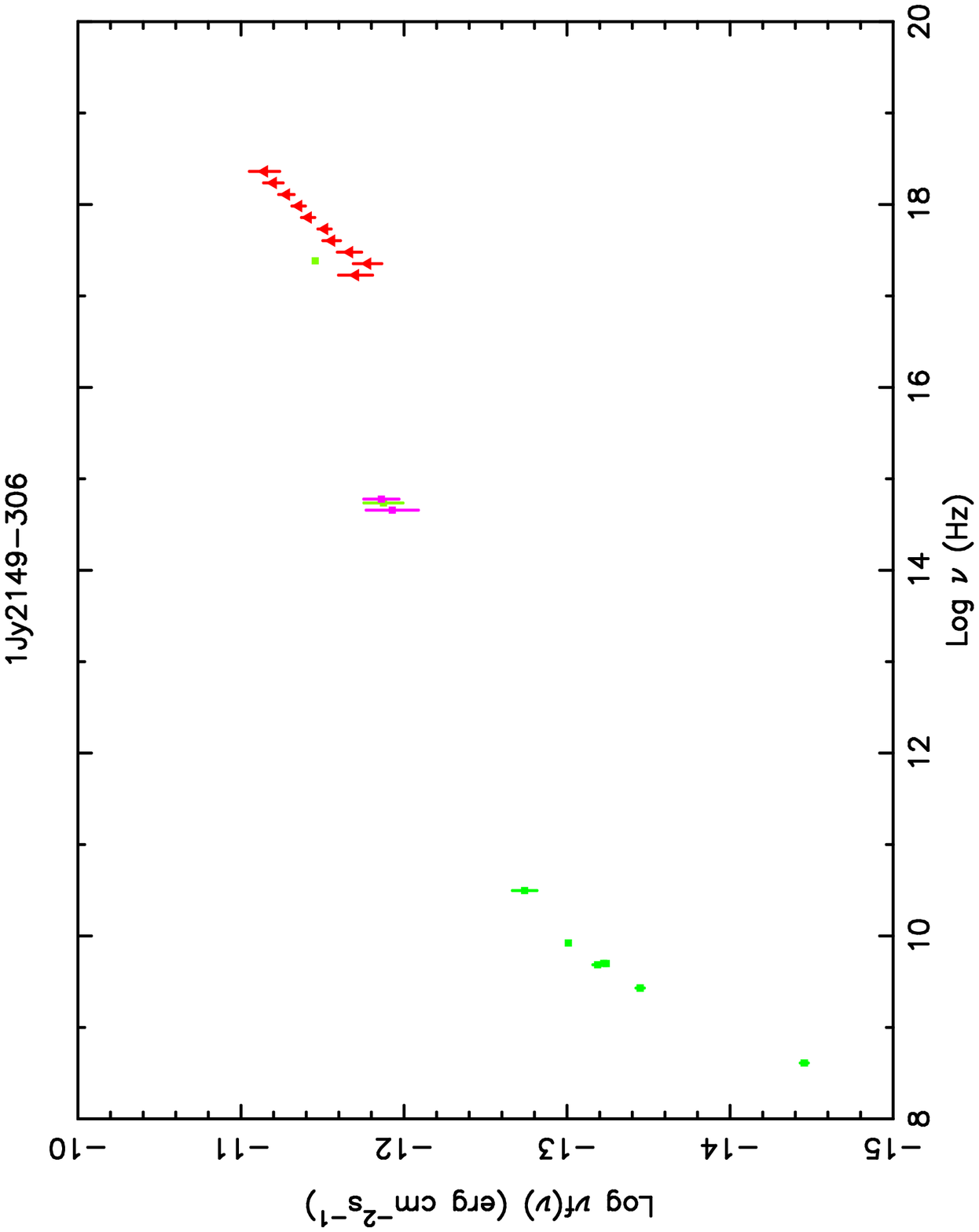} 
\includegraphics{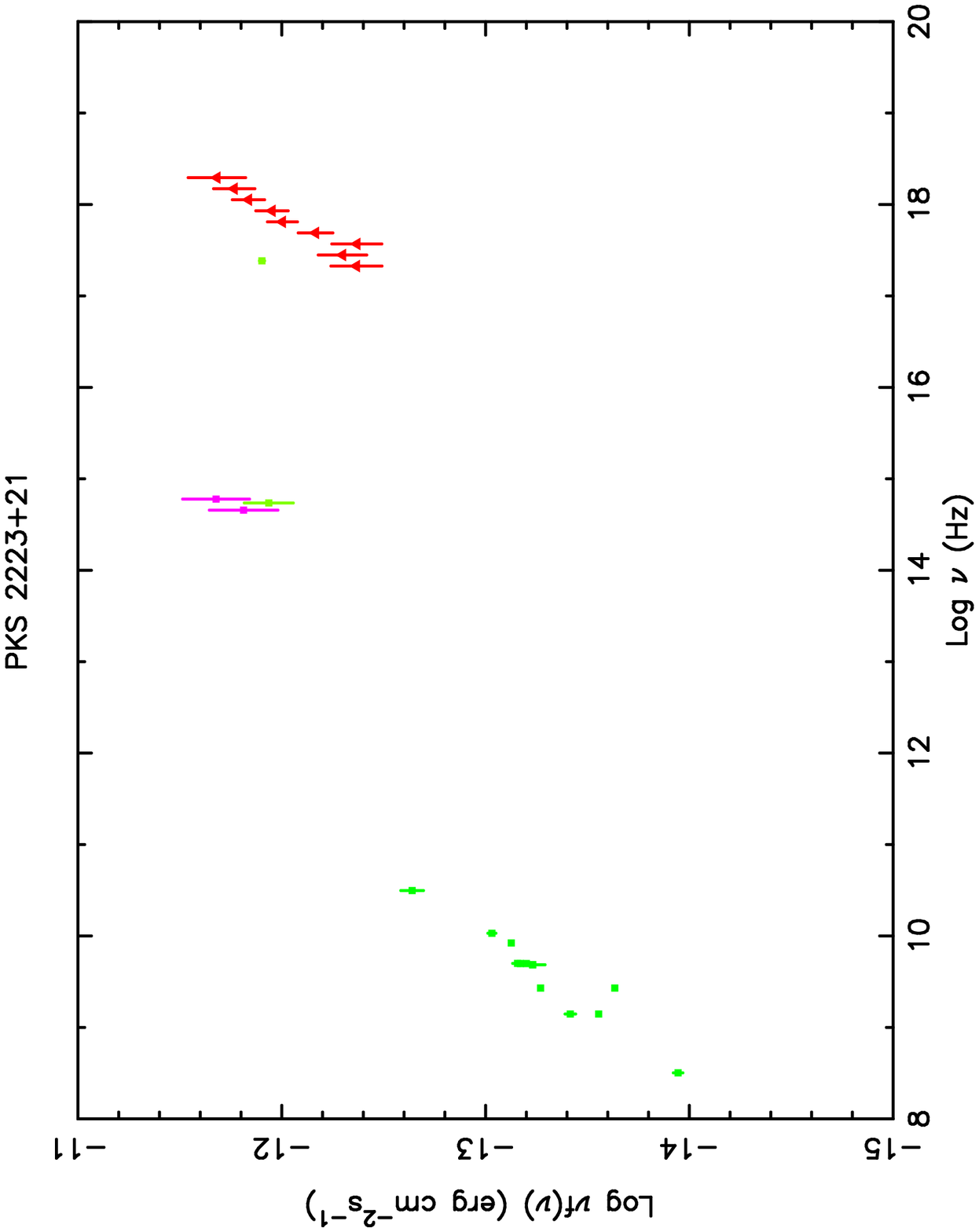} 
\includegraphics{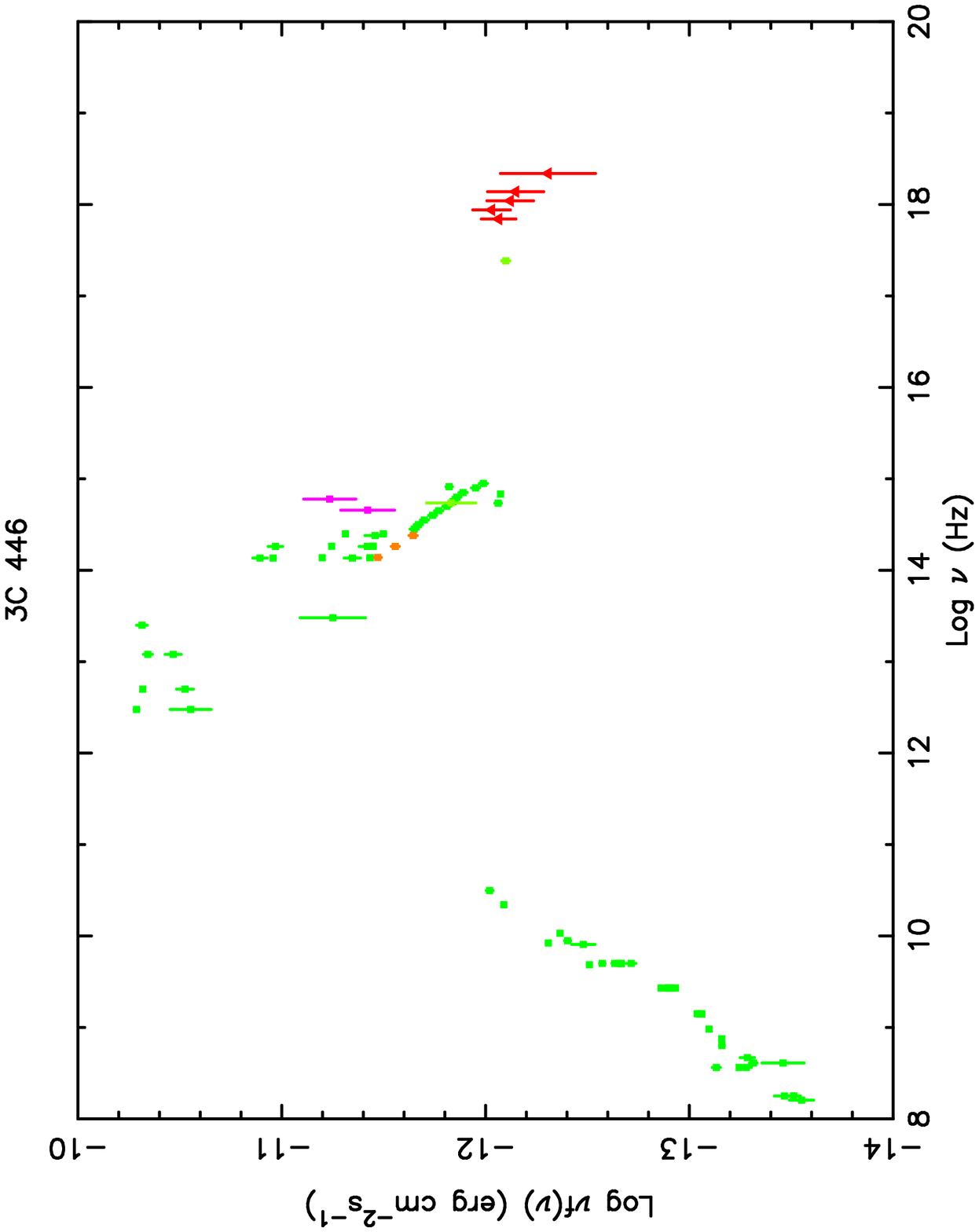} 
\includegraphics{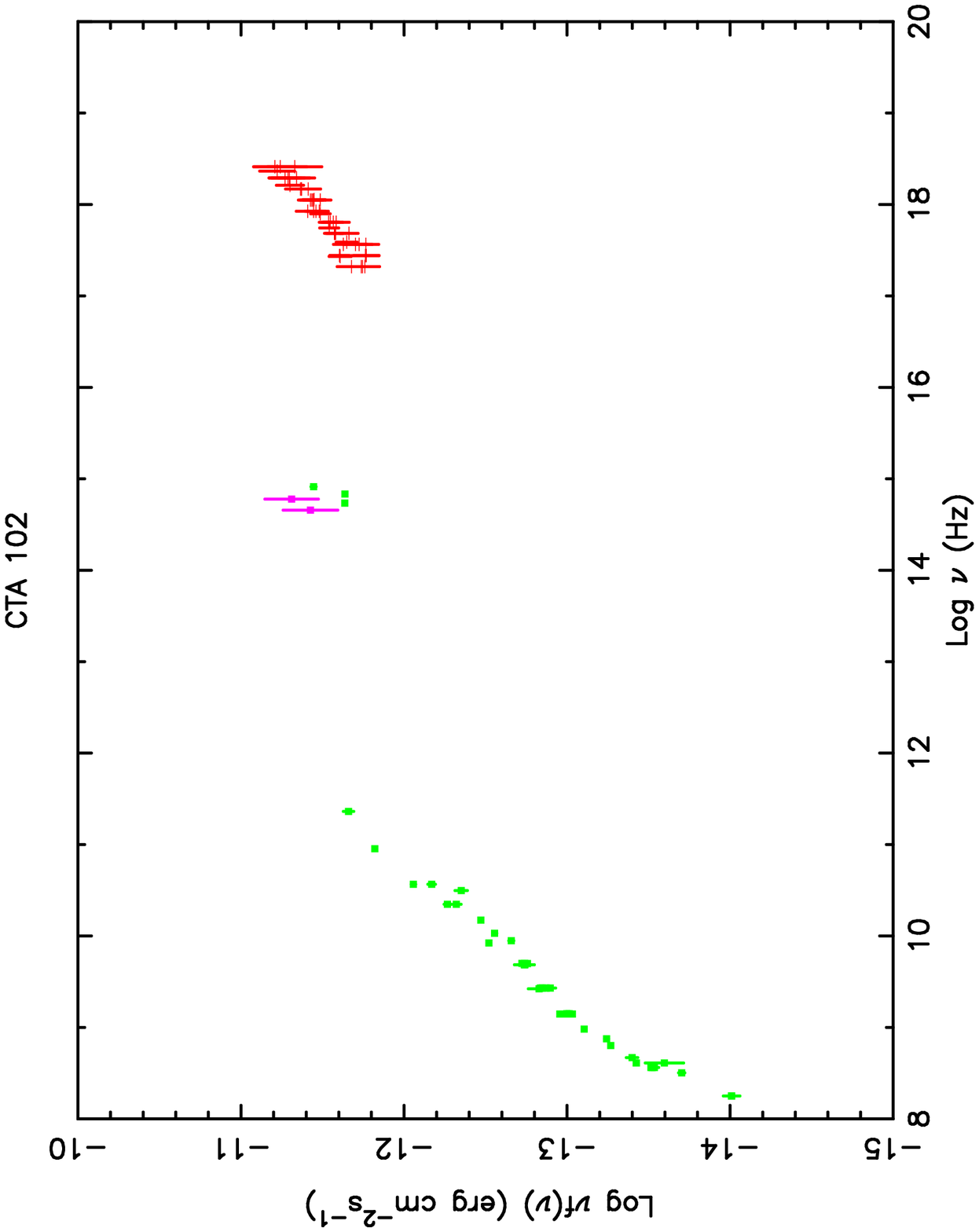} 
\vspace{19.0cm} 
\caption[t]{f- Spectral Energy Distribution of the FSRQs 1Jy 2149$-$306,
PKS 2223+21, 3C 446 and CTA 102} 
\label{fig2f} 
\end{figure} 
\clearpage 
\setcounter{figure}{2} 
\begin{figure}[t]
\centering
\includegraphics{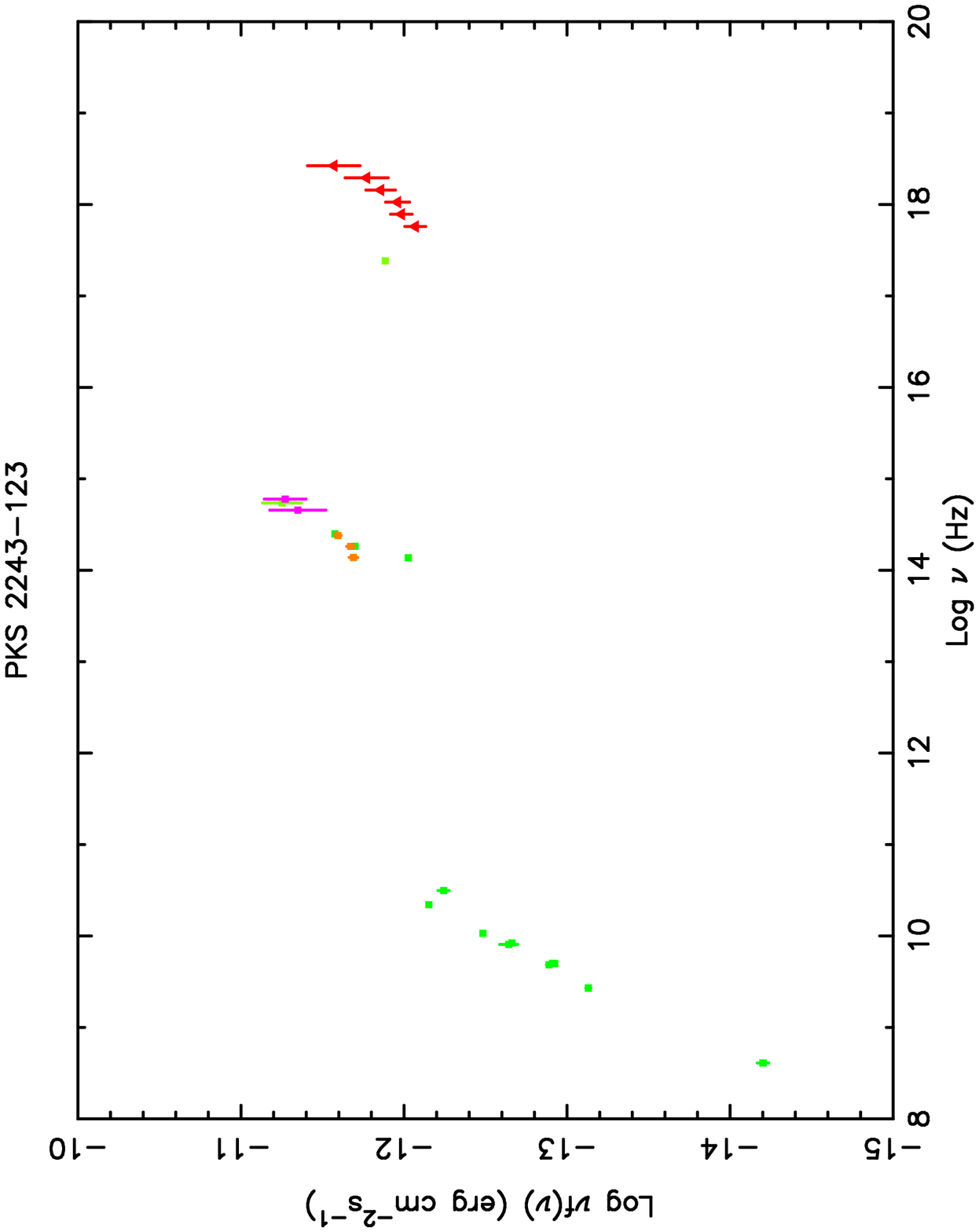} 
\includegraphics{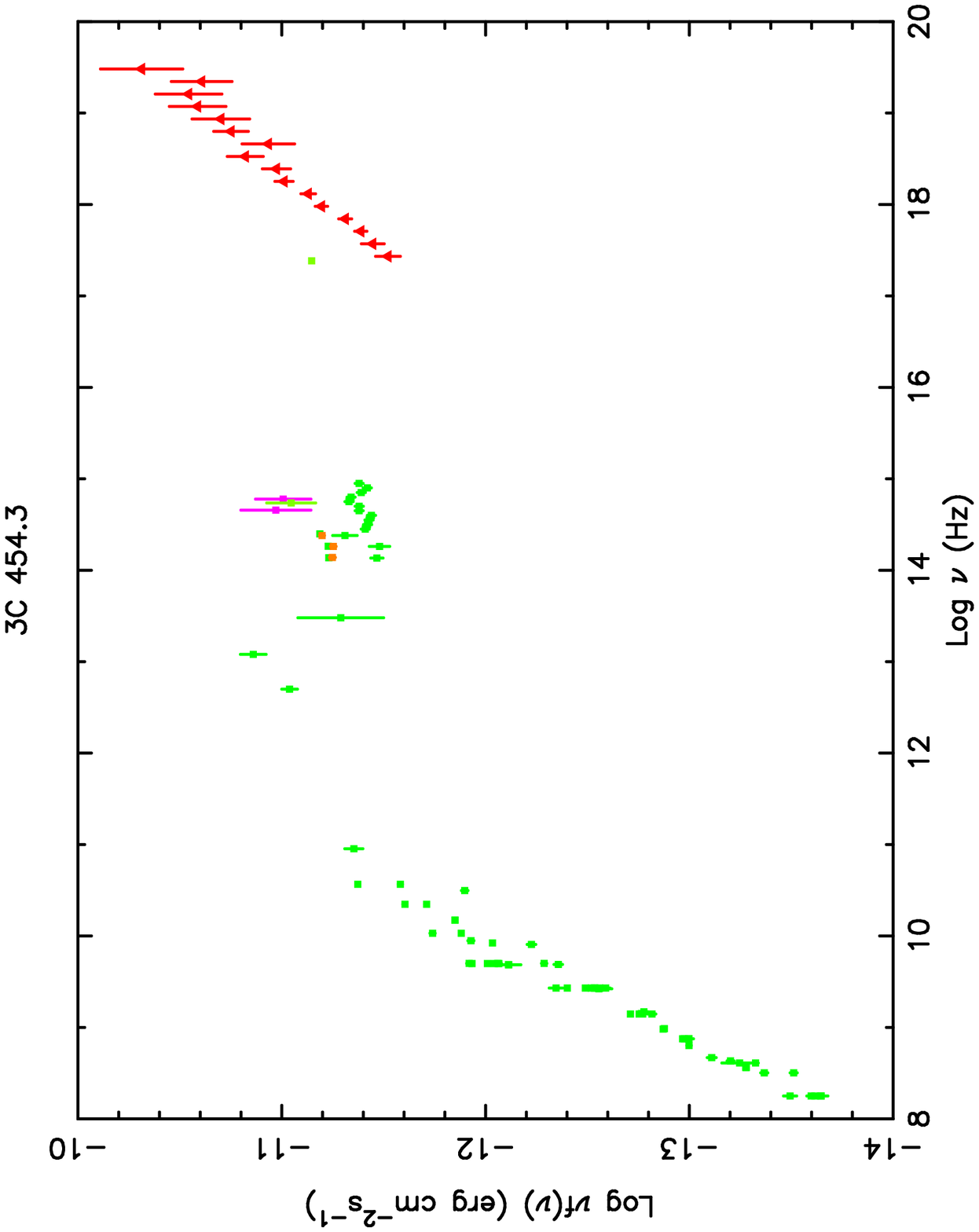} 
\vspace{19.0cm}
\caption[t]{g- Spectral Energy Distribution of the FSRQs PKS 2243$-$123 and 3C 454.3}
\label{fig2g}
\end{figure}
%---------------------------------------------------
\clearpage 

\section{Summary and Conclusion}

We have presented the X-ray spectrum and the Spectral Energy Distribution 
of a large sample of blazars observed by \sax with the aim of providing a 
single homogeneous reference for this type of \sax data. 
The collection of the results presented here together with all the data 
is also available as part 
of the \sax archive at the ASI Science Data Center (ASDC) at the following 
address \par 
\begin{center}
http://www.asdc.asi.it/blazars/
\end{center}

Given the heterogeneous nature of the sample we have not attempted to perform any 
deep statistical studies. In the following we summarize our work and give some 
remarks about possible interpretations. More detailed statistical studies or deeper 
interpretations are reported elsewhere or will be the subject of future 
publications.

\setcounter{figure}{3}
\begin{figure}[!ht]
\vspace*{-2.0cm}
\centering
\epsfysize=6.8cm\epsfbox{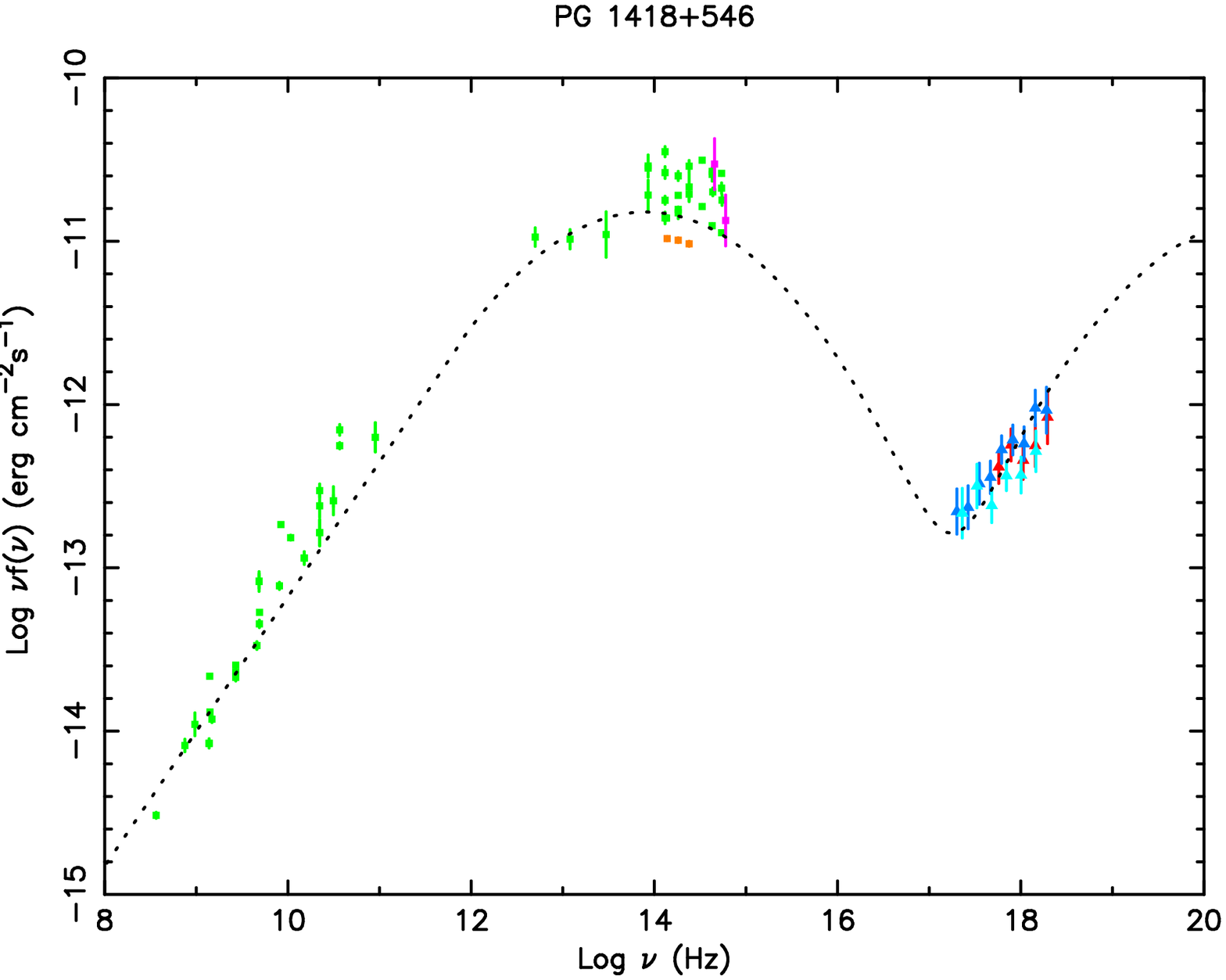}
\hspace{2.cm}\epsfysize=6.8cm\epsfbox{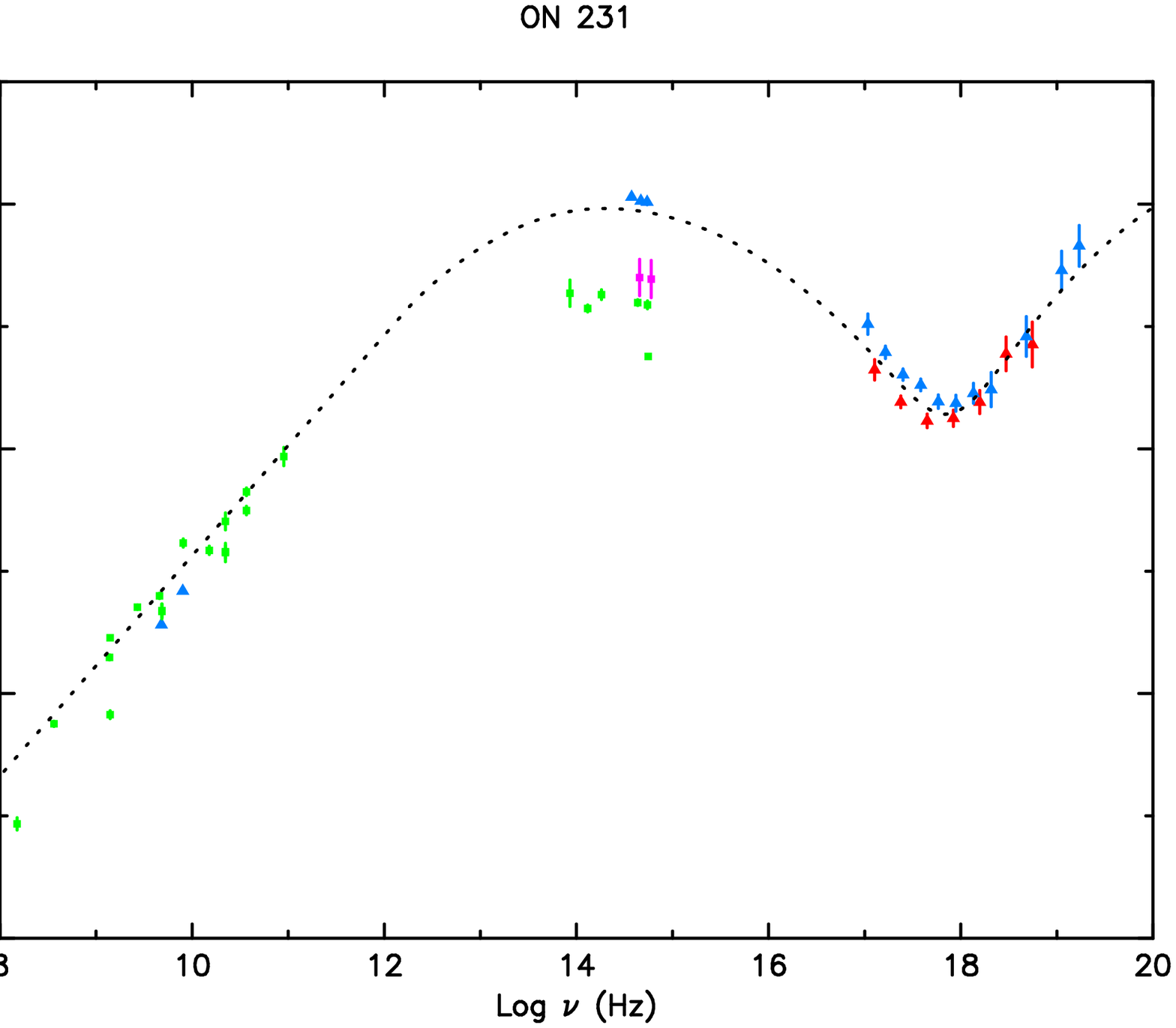} 
\caption[ht]{Spectral energy distributions of a typical LBL object  
(PG~1418+546, \nupeak $ \approx 0.4~$eV$~\approx 8\times 10^{13} $Hz) for which the X-ray emission
is dominated by the flat inverse Compton radiation and of an Intermediate BL Lac 
(ON 231, \nupeak $ \approx 1~$eV$~\approx 2\times 10^{14} $Hz) where the simultaneous optical 
and \sax observations (Tagliaferri et al. 2000) clearly show that the transition 
between the synchrotron and inverse Compton emission occurs in the soft X-ray band.}
\label{fig4}
\end{figure}

\setcounter{figure}{4}
\begin{figure}[!ht]
\vspace*{-2.5cm}
\centering
\epsfysize=6.8cm\epsfbox{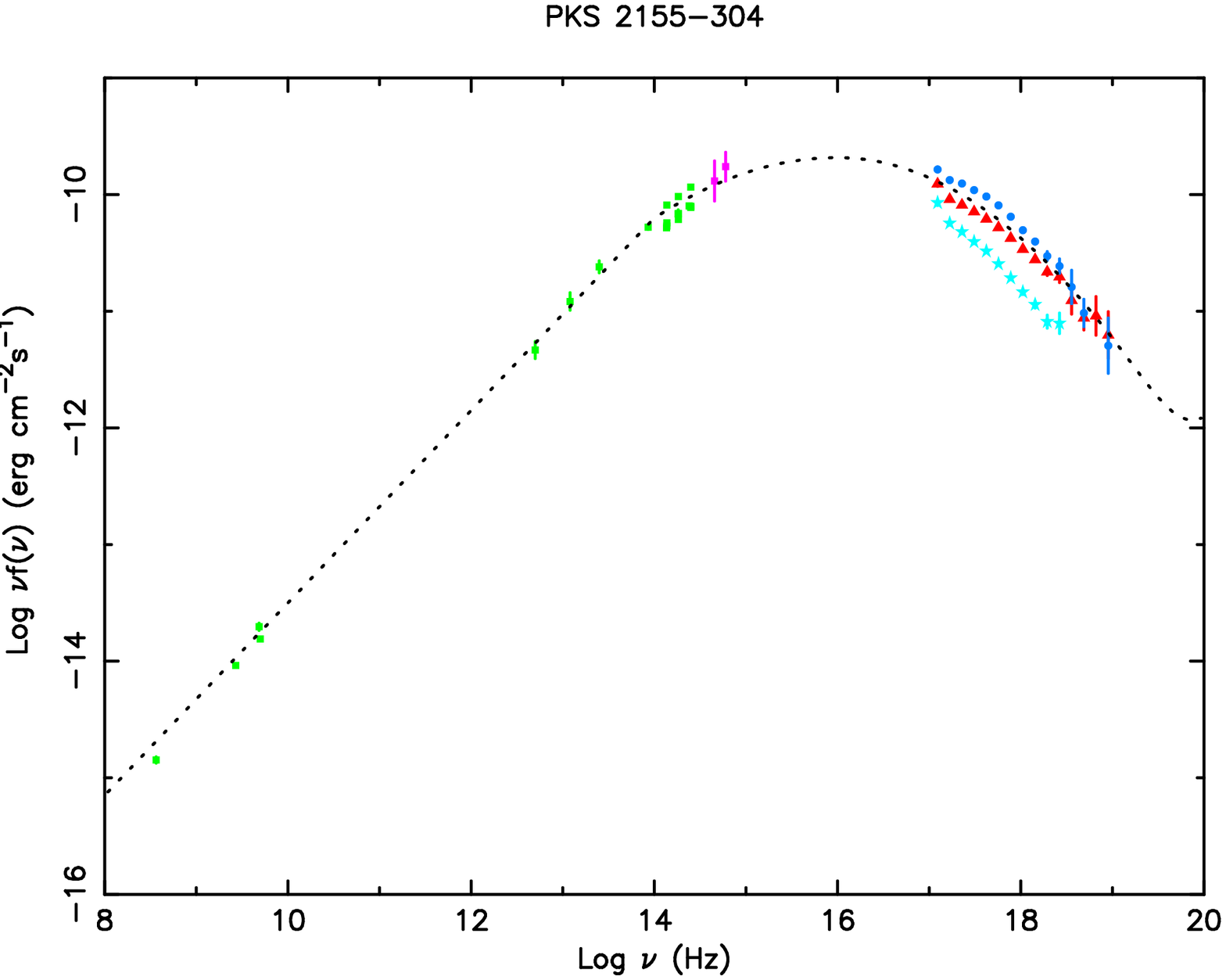}
\hspace{2.cm}\epsfysize=6.8cm\epsfbox{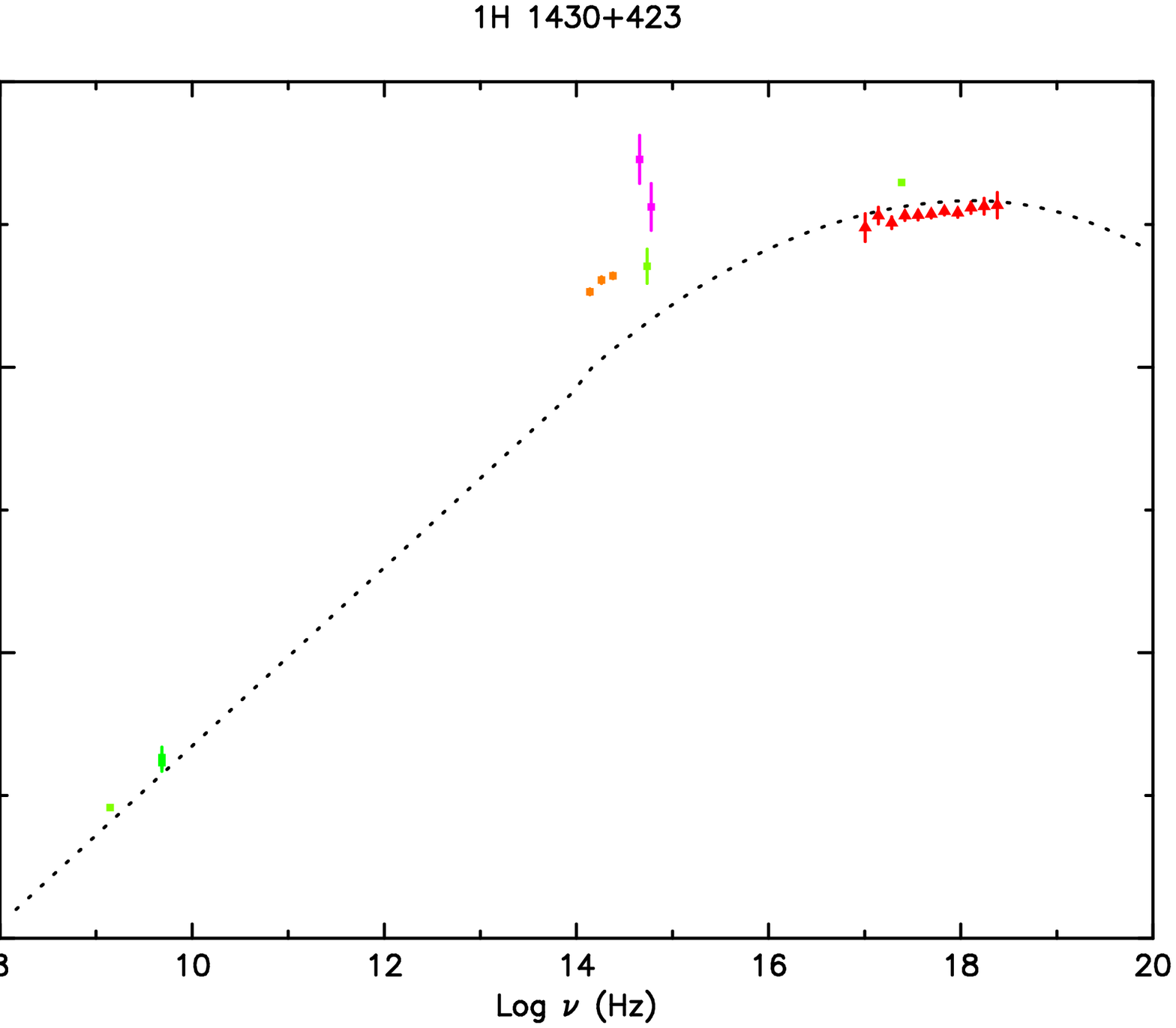} 
\caption[ht]{Spectral energy distributions of HBLs where the X-ray emission is completely dominated
by synchrotron radiation. In the case of PKS 2155$-$304 \nupeak is at $\approx$ 50 eV 
$\approx \times10^{16}$ Hz while for the extreme HBL 1H 1430+423 \nupeak is above 10 keV.}
\label{fig4}
\end{figure}

\setcounter{figure}{5}
\begin{figure}[!ht]
\vspace*{-1.5cm}
\centering
\epsfysize=12.0cm\epsfbox{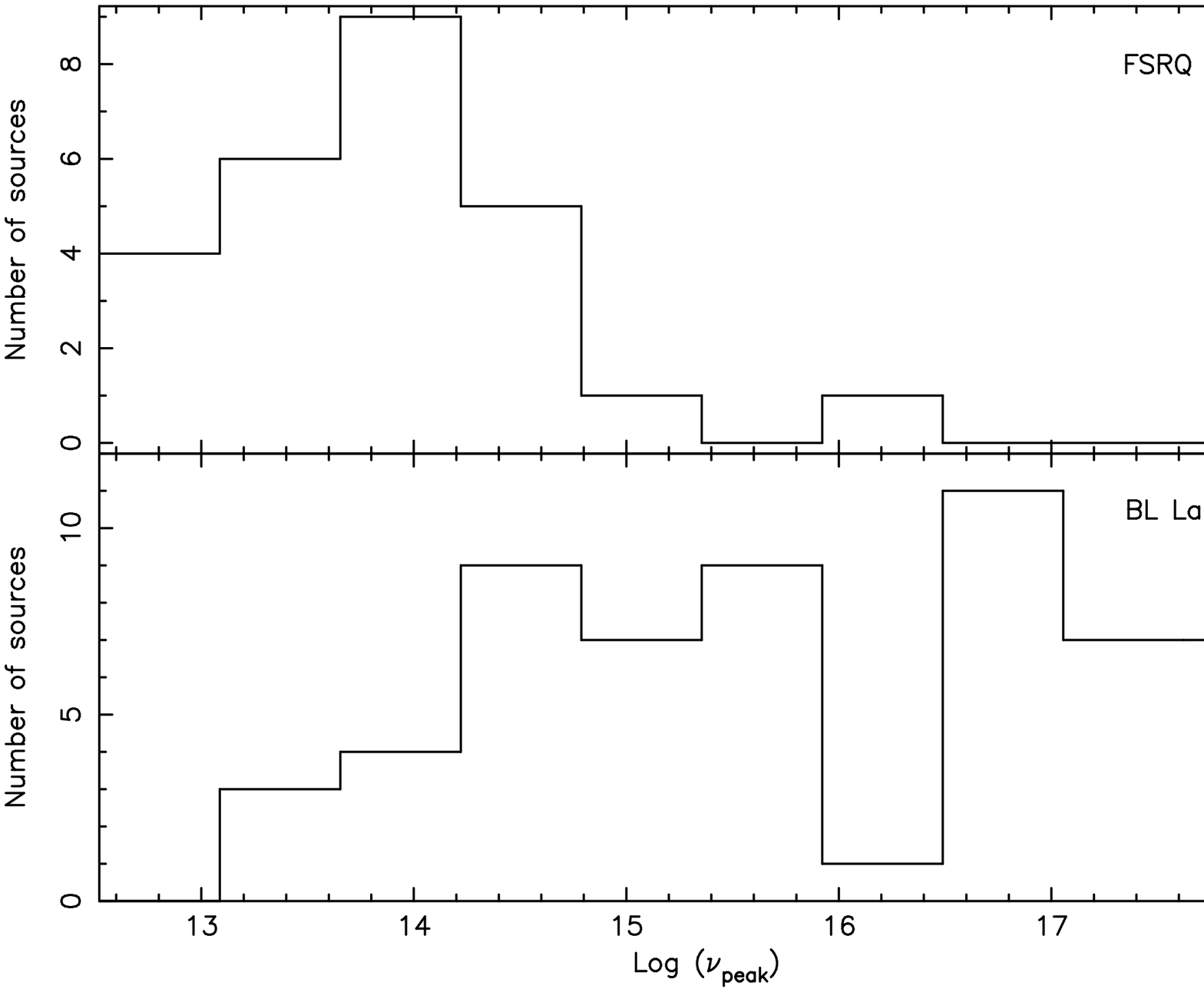}
\caption[h]{The distribution of the synchrotron peak frequencies in the FSRQ and in the 
BL Lac subsamples.
While the \nupeak values for FSRQs strongly cluster around $10^{14}$ Hz, a much wider distribution 
is present in the BL Lacs subsample. In the latter case the distribution is strongly biased 
towards high \nupeak values by the \sax time allocation process which favoured X-ray bright 
(and therefore high \nupeak) objects that could be detected by all NFI instruments.}
\label{fig5}
\end{figure}

%\setcounter{figure}{6}
%\begin{figure}[!ht]
%\vspace*{-1.0cm}
%\centering
%\epsfysize=12.0cm\epsfbox{arx_vs_nup.ps}
%\caption[h]{The Radio to X-ray spectral slope \arx of the BL Lacs (open circles) plotted 
%against \nupeak. GPS sources are not included in this plot since the simple SSC model 
%used to estimate \nupeak is not a good representation of all the SED data for these objects.
%}
%\label{fig5}
%\end{figure}

\newpage
%\par We note the following basic facts:

\bigskip
%\begin{enumerate}

\par The SED of the 84 blazars considered in this work confirms  
with large statistics the widely accepted scenario where blazar emission 
is smooth across several decades of energy and is characterized (in a  $\nu f(\nu)~vs~\nu$ 
representation) by two broad peaks which are usually interpreted 
as being due to synchrotron emission followed by inverse Compton radiation. 

\bigskip 

\par A wide range of X-ray spectral indices has been observed, ranging from very 
flat values in Compton dominated sources like PG~1418+546 to very steep spectral slopes in
objects where the tail of the synchrotron emission just reaches the X-ray 
band. A sharp transition from a very steep soft X-ray component to a much flatter 
hard X-ray spectrum, marking the transition between the synchrotron and Compton emission,
has been clearly detected in a number of objects. Examples  
are given in Figures 4 and 5 where, for four representative objects, we plot 
the observed SED together with the expected distribution from a SSC model peaking at 
appropriate \nupeak values.

\par When viewed only from the narrow soft X-ray band, as was done in the past, these 
SEDs clearly show that the local X-ray spectral index must be correlated to the X-ray to
radio flux ratio ($f_x/f_r$) as was first found by 
Padovani \& Giommi (1996) and by Lamer, Brunner \& Staubert (1996) in large samples 
of BL Lacs observed with ROSAT.
  
\bigskip 

\par The position of the synchrotron peak, estimated comparing the 
SEDs to SSC models such as those shown in Figures 4 and 5, spans at least six orders 
of magnitudes ranging from $\approx 0.1~$eV  in e.g. PKS~0048$-$097 
or S5 2116+81 to 10--100 keV in some extreme HBL BL Lacs like Mkn 501 and 1H 1430+423.

\par Very strong intensity and spectral variability can occur near 
the synchrotron (and inverse Compton) peak. The position of this peak can move 
to higher energy by up to two orders of magnitude (or perhaps more) during flares.

It is not clear what is the maximum \nupeak that can be reached and whether Ultra High
energy synchrotron peaked BL Lacs (UHBL) exist. A few potential UHBL sources 
may be present in the Sedentary survey (Giommi, Menna \& Padovani 1999, Perri 
et al. 2002), which by definition only includes extreme 
HBL objects, especially those few that are located within the error circle 
of unidentified EGRET sources. If these candidates turn out to be the real counterpart 
of the EGRET gamma-ray sources their \nupeak would be so high that their synchrotron 
radiation would reach the gamma-ray band. 
One such object, 1RXS~J123511.1$-$14033 (see Figure 2i), was observed by \sax  
on three occasions but always with short integration times giving inconclusive 
results (Giommi et al. 2002, in preparation).

\par
The observed distributions of \nupeak values (rest frame), obtained by fitting SSC models 
to the multi-frequency data shown in Figure 2a-2o and 3a-3g, are plotted in Figure 6 for
the FSRQ (top panel) and the BL Lac (bottom panel) subsamples. 
The two distributions are certainly affected by selection effects, including that induced by  
the \sax time allocation process which, by necessity, favoured high \nupeak/X-ray strong sources 
which could be detected by the high energy instruments. Although this bias is clearly present 
in the BL Lac subsample where a large fraction of the sources are X-ray bright HBL 
objects, in the case of FSRQs most of the objects have low \nupeak . This is because 
FSRQs are in general more luminous than BL Lacs and especially because FSRQs with 
\nupeak $ > 10^{16}$ Hz are very rare. To date 
the only FSRQ (RGB J1629+4008 = 1ES 1627+402, see Figure 3d) whose synchrotron emission 
reaches the X-ray band was found by Padovani et al. 2002b.

\bigskip 

\par A logarithmic parabola model, which can describe the spectral curvature 
of blazars in a very wide energy band with only three parameters (see Landau 
et al. 1986), fits better than other models (e.g. broken power law) the spectrum 
of HBL objects whose X-ray emission is still due to synchrotron radiation. 
The average amount of spectral curvature, as measured by the $\beta$ parameter 
in the log parabola model of paragraph 3.1 is $-$0.38 +/$-$ 0.1, a value somewhat steeper 
(possibly because of the energy dependant synchrotron cooling), but not too different, 
than the amount of curvature  found by Landau et al. 1986 ($-$0.22 to $-$0.09 ) in a 
sample of BL Lacs whose synchrotron 
power peaks at infra-red, optical frequencies. This similarity points to an intrinsically 
similar curvature in the spectrum of the emitting particles. The smoothly changing slope could be the 
spectral signature of a statistical acceleration mechanism where the acceleration process becomes 
less and less efficient as the particle's energy increases (Massaro 2002). 
In this scenario the widely different synchrotron \nupeak energies in LBL 
and HBL objects would be the result of the inefficiency in the acceleration 
process that sets off at different energies.

\bigskip

%\item[f-] A second synchrotron component, weak at low energies but dominating in the 
%hard X-ray band because of a higher synchrotron peak frequency, may be present in some 
%objects (e.g. 1ES2344+514, MKN 501, MKN 421, 1RXSJ123511.1-14033), especially when flaring. 

%\end{enumerate}

\acknowledgements

This research has made use of data retrieved from the ASI/ASDC-\sax public archive, 
the NASA/IPAC Extragalactic Database (NED), the NRAO VLA Sky Survey (NVSS), 
the Guide Star Catalog II (GSC2) and the Two Micron All Sky Survey (2MASS).

\end{document}